\RequirePackage{fix-cm}
\documentclass[twocolumn]{svjour3}          
\smartqed  
\usepackage{graphicx}
\usepackage{amsmath}
\usepackage{caption}
\usepackage{subcaption}
\usepackage{bm}
\usepackage{xfrac}
\usepackage{natbib}
\usepackage{blindtext}
\usepackage{amssymb}
\newcommand{\vecr}[1]{\bm{#1}}
\newcommand{\x}{\vecr{x}}
\newcommand{\uv}{\vecr{u}}

\newcommand{\bof}[1]{\left( #1 \right)}

\newcommand{\dpart}[2]{\frac{\partial #1}{\partial #2}}
\newcommand{\dpx}[1]{\frac{\partial #1}{\partial \x}}

\newcommand{\intO}[1]{\int_{\Omega} #1 \,dV}
\newcommand{\into}[1]{\int_{\omega} #1 \,dS}
\newcommand{\ui}{u_{i}}
\newcommand{\uj}{u_{j}}
\newcommand{\wi}{w_{i}}

\renewcommand{\xi}{x_{i}}
\newcommand{\xj}{x_{j}}
\newcommand{\xa}{x_{a}}
\newcommand{\xb}{x_{b}}

\newcommand{\uba}{\bar{u}_{a}}
\newcommand{\ubb}{\bar{u}_{b}}
\newcommand{\wba}{\bar{w}_{a}}

\usepackage{xcolor}

\usepackage[switch,pagewise,displaymath,mathlines]{lineno}

\let\oldequation\equation
\let\oldendequation\endequation
\let\oldalign\align
\let\oldendalign\endalign

\renewenvironment{equation}
  {\linenomathNonumbers\oldequation}
  {\oldendequation\endlinenomath}
 \renewenvironment{align}
  {\linenomathNonumbers\oldalign}
  {\oldendalign\endlinenomath}

\begin{document}


\title{
Revisiting the origin to bridge a gap between topology and topography optimisation of fluid flow problems
\thanks{This paper is dedicated to two pioneers of fluid flow topology optimisation: Joakim Petersson (1968-2002) and Allan Roulund Gersborg (1976-2020).}
}

\author{Joe Alexandersen}

\institute{J. Alexandersen \at
              Department of Mechanical and Electrical Engineering \\
              University of Southern Denmark \\
              Campusvej 55, DK-5230 Odense M \\
              Tel.: +45 65507465\\
              \email{joal@sdu.dk}
}

\date{Received: date / Accepted: date}

\maketitle

\begin{abstract}

This paper revisits the origin of topology optimisation for fluid flow problems, namely the Poiseuille-based frictional resistance term used to parametrise regions of solid and fluid.
The traditional model only works for true \textit{topology} optimisation, where it is used to approximate solid regions as areas with very small channel height and, thus, very high frictional resistance. It will be shown that if the channel height is allowed to vary continuously and/or the minimum channel height is relatively large and/or meaning is attributed to intermediate design field values, then the predictions of the traditional model are wrong.
To remedy this problem, this work introduces an augmentation of the mass conservation equation to allow for continuously varying channel heights. 
The proposed planar model accurately describes fully-developed flow between two plates of varying channel height. It allows for a significant reduction in the number of degrees-of-freedom, while generally ensuring a high accuracy for low-to-moderate Reynolds numbers in the laminar regime. The accuracy and limitations of both the traditional and proposed models are explored using in-depth parametric studies.
The proposed model is used to optimise the height of the fluid channel between two parallel plates and, thus, the \textit{topography} of the plates for a flow distribution problem.
Lastly, it is observed that the proposed model actually produces better \textit{topological} designs than the traditional model when applied to the \textit{topology} optimisation of a flow manifold.


\end{abstract}

\section{Introduction} \label{sec:intro}

\subsection{Motivation}

Plate heat exchangers represent a multiscale problem. The parallel plates will be stacked together with a relatively small spacing compared to the other dimensions of the plates and the full stack. Furthermore, it is well-known that the surface topography or corrugation plays an important role in flow distribution, pressure drop and heat transfer, see e.g. \citep{Kanaris2006,Tsai2009,Kilic2017,Li2021}. While modelling and optimisation of a single plate and fluid channel is feasible computationally, modelling an entire stack of plates and optimising either the stack or the surrounding chambers of the heat exchanger becomes infeasible. This is due to the large span in length scales, going from millimeters in terms of the spacing and corrugation all the way to tens of centimeters or even meters for the entire heat exchanger. Therefore, in order to optimise the macroscale of the heat exchangers, as well as the individual plate topographies simultaneously, it is necessary to develop simplified flow models for the flow between parallel plates of varying spacing and topography. This is an area of increasing interest in the field of topology optimisation for problems driven by fluid flow, as discussed in the recent review paper by \cite{Alexandersen2020}.

\subsection{Literature}

The seminal paper of topology optimisation for fluid flow problems is the work by \cite{Borrvall2003}. They presented a mathematical basis for topology optimisation of Stokes flow, using a design parametrisation based on Poiseuille flow. By introducing the frictional resistance from pressure-driven flow between parallel plates, they were able to introduce a design parametrisation where the solid domains are approximated by areas with vanishing channel height. 
This parametrisation was extended to Navier--Stokes flow by \cite{GersborgHansen2005}, also by basing the derivations on fully-developed flow between parallel plates.

Both sets of authors note the similarity of the out-of-plane frictional resistance with that of an idealised porous medium.
This is also the conceptualisation adhered to by the majority of papers thereafter \citep{Alexandersen2020}, starting with the work of \cite{Evgrafov2005,Evgrafov2006} and \cite{Olesen2006}. This also makes sense, since it is naturally extendable to three-dimensional problems, where the Poiseuille flow conceptualisation loses physical meaning.

One area of relevance of the out-of-plane viscous resistance is that of the so-called ``pseudo-3D'' models implemented for extruded heat sink design, initially conceptualised by \cite{McConnell2012}. However, for some reason, most of the work using such models do not include the out-of-plane viscous resistance, although they have small dimensions in the out-of-plane direction \citep{Haertel2018,Zeng2018,Zeng2019}. To the author's knowledge, the first model of the pseudo-3D type taking the out-of-plane resistance into account was the two-layer model by \cite{Yan2019}, which was subsequently extended to a three-layer model to improve thermal accuracy by \cite{Zhao2021}. 
In fact, according to the recent review \citep{Alexandersen2020}, only three out of seventeen papers \citep{Kobayashi2019,Yan2019,Behrou2019} actually include the viscous resistance from the friction due to the out-of-plane boundary layers.
Recently, \cite{Guo2020} examined the accuracy of plane two-dimensional topology optimisation compared to three-dimensional extruded versions of the microfluidic designs. However, instead of using a non-zero friction force in the fluid domains arising from the relatively small out-of-plane dimension, they argued for a maximum length scale constraint on the fluid channels to control accuracy of the two-dimensional approximation.

All of the above methodologies are perfectly well suited for \textit{topology} optimisation. But only when treating the \textit{topological} problem in terms of distributing discrete areas of fluid and solid, where the solid domain should not have any flow passing through (numerically very little). However, their accuracy fails as soon as physical meaning is attributed to intermediate design field values or these are utilised in a physical model to treat problems of continuously- and spatially-varying frictional resistance with non-infinite (or approximately so) maximum resistance - from either parallel plates or porous material. The missing physicality of the idealised porous media approach has recently been explored for flow in real porous media \citep{Phatak2021} and for varying porosity \citep{Bastide2018,Rakotobe2020}. Recently, an increase in accuracy was observed by using the Volume-Average Navier-Stokes for porous flow \citep{Theulings2021}. This paper will discuss similar issues and observations, but for flow between plates of spatially-varying spacing.

\subsection{Contributions}

The presented work stems from an effort to reduce the cost of simulating and optimising the flow through plate heat exchangers for building ventilation systems \citep{Veje2019}. While conditions in these are often transient and turbulent, the development of models began with steady and laminar flow. In order to model the flow between plates of spatially-varying spacing, it was initially observed that mass was not conserved using the original resistance terms \citep{Borrvall2003,GersborgHansen2005}. Thus, the presented model was derived taking the volumetric changes due to varying spacing into account.

This paper revisits the origin of topology optimisation for fluid flow problems in order to bridge a gap between \textit{topology} and \textit{topography} optimisation. The limitations of the traditional model will be discussed and the modified model will be introduced. The accuracy and limitations of the model will be explored using in-depth parametric studies and it will be shown that the accuracy is generally high for low and moderate Reynolds numbers in the laminar regime. Finally, the developed model will be used to optimise the topography of a single set of parallel plates and, thus, the height of the fluid channel/spacing between them.

\subsection{Paper layout}

The article is presented as follows:
Section \ref{sec:govequ} introduces the basic governing equations, as well as derivations, assumptions and limitations of the simplified model.
Section \ref{sec:anaexamples} discusses the implementation of the model and the simple analysis examples used for the parametric study of accuracy.
Section \ref{sec:optimisation} introduces the optimisation formulations and details of two example problems.
Section \ref{sec:results} presents and discusses the optimisation results for the two examples.
Finally, Section \ref{sec:conclusion} concludes on the presented work and presents future work.

\section{Governing equations and models} \label{sec:govequ}

\subsection{Navier-Stokes equations} \label{sec:govequ_ns}

This article restricts itself to steady-state laminar and incompressible flow, governed by the dimensional form of the Navier-Stokes equations:
\begin{subequations}
\begin{align}
    \rho u_{j}\dfrac{\partial u_i}{\partial x_j} - \mu \dfrac{\partial}{\partial x_j}\left( \dfrac{\partial u_i}{\partial x_j} + \dfrac{\partial u_j}{\partial x_i} \right) + \dfrac{\partial p}{\partial x_i} &= 0  \label{eq:navierstokes-a} \\
    \dfrac{\partial u_i}{\partial x_i} &= 0  \label{eq:navierstokes-b}
\end{align} \label{eq:navierstokes}
\end{subequations}
\hspace{-1.5ex} where $u_{i}$ is the $i$-th component of the velocity vector $\textbf{u}$, $p$ is the pressure, $\rho$ is the density, and $\mu$ is the dynamic viscosity.

\subsection{Original formulation} \label{sec:govequ_bp}

This section describes the derivation of the friction term arising in the Navier-Stokes equations from the out-of-plane viscous friction. This model was introduced by \citet{Borrvall2003} for Stokes flow and \citet{GersborgHansen2005} for Navier-Stokes flow.

\subsubsection{Basic assumption} \label{sec:govequ_bp_assump}

The model assumes a fully-developed and parallel flow profile which is derived on the basis of pressure-driven flow between two infinite parallel plates, known as\\ Poiseuille flow. For laminar flow, Stokes or Navier-Stokes, the governing equations lead to a parabolic flow profile between the two plates:
\begin{equation}
    u(x_3) = U_\mathrm{max}\left( 1 - \left( \frac{2x_3}{h} \right)^{2} \right) 
\end{equation}
where $U_\mathrm{max}$ is the maximum velocity at the midpoint ($x_3=0$) between the two plates at distance $h$ apart.
\begin{figure}
     \begin{subfigure}{\columnwidth}
         \centering
         \includegraphics[width=0.75\textwidth]{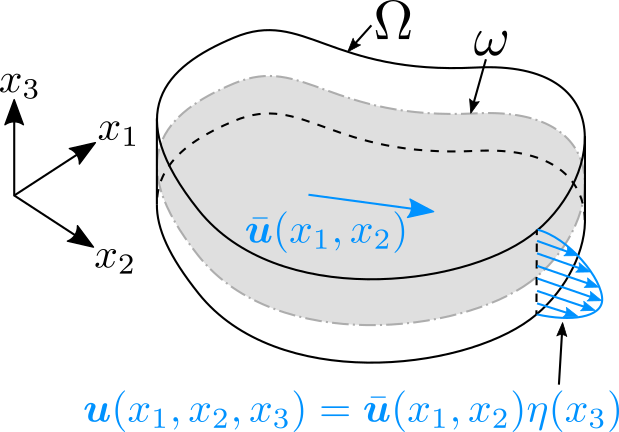}
         \caption{Three-dimensional geometry}
         \label{fig:simplifiedDomain-3D}
     \end{subfigure}
    \\
    \begin{subfigure}{\columnwidth}
         \centering \vspace{0.5cm}
         \includegraphics[width=0.725\textwidth]{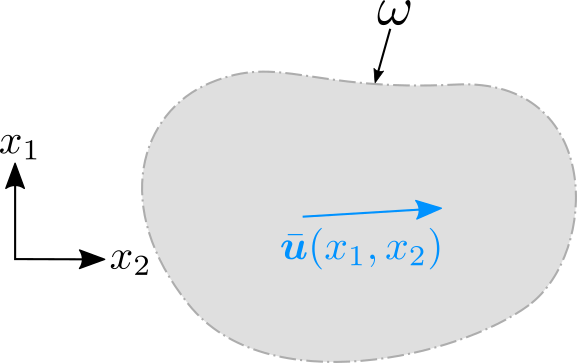}
         \caption{Two-dimensional mid-plane}
         \label{fig:simplifiedDomain-2D}
     \end{subfigure}
    \caption{Illustration of the mid-plane used to simplify a three-dimensional geometry to a two-dimensional representation.}
    \label{fig:simplifiedDomain}
\end{figure}
It is then assumed that the velocity at any point can be described as:
\begin{equation} \label{eq:u_sepVar}
    \uv(x_1,x_2,x_3) = \bar{\uv}(x_1,x_2) \eta(x_3)
\end{equation}
with:
\begin{equation} \label{eq:u_zDep}
    \eta(x_3) = \left( 1 - \left( \frac{2x_3}{h} \right)^{2} \right) 
\end{equation}
as is illustrated in Figure \ref{fig:simplifiedDomain-3D}. The vectors, $\uv(x_1,x_2,x_3)$ and  $\bar{\uv}(x_1,x_2)$, are both three-dimensional, but the velocity in the $x_3$-direction is assumed to be zero for both fields.

\subsubsection{Through-thickness resistance term} \label{sec:govequ_bp_augment}
With the separation of variables, the volumetric integration inherent to e.g. the finite element method can be decoupled and performed explicitly in the through-thickness direction, i.e. the $x_3$-direction, shown here for the three-dimensional case illustrated in Figure \ref{fig:simplifiedDomain}:
\begin{equation} \label{eq:int_sepVar}
    \int_{\Omega} \; dV = \int \int \left( \int_{-\frac{h}{2}}^{\frac{h}{2}} \; dx_3 \right) dx_1 \, dx_2 = \int_{\omega} \left( \int_{-\frac{h}{2}}^{\frac{h}{2}} \; dx_3 \right) dS
\end{equation}
The derivation is given in Appendix \ref{app:bp_derivation}, whereafter the final equivalent strong form of the conservation of momentum becomes:
\begin{equation} \label{eq:aug_navierstokes}
    \bar{\rho} \bar{u}_{b}\dfrac{\partial \bar{u}_a}{\partial x_b} - \mu \dfrac{\partial}{\partial x_b}\left( \dfrac{\partial \bar{u}_a}{\partial x_b} + \dfrac{\partial \bar{u}_b}{\partial x_a} \right) - \alpha \bar{u}_a + \dfrac{\partial \bar{p}}{\partial x_a} = 0
\end{equation}
where $a,b = 1,2$, $\bar{p} = \frac{5}{4}p$ is the scaled pressure, $\bar{\rho} = \frac{6}{7}\rho$ is the effective density, and $\alpha = \frac{10\mu}{h^2}$ is the through-thickness/out-of-plane viscous resistance factor\footnote{In the original work by \cite{Borrvall2003}, the factor is defined with respect to the half-thickness $\chi=\frac{h}{2}$, yielding $\alpha = \frac{5\mu}{2\chi^2}$.}.

\subsubsection{Limitations} \label{sec:govequ_bp_limits}

Equation \ref{eq:aug_navierstokes} correctly models the fully-developed flow between two parallel plates of \textit{constant spacing}, $h$. Applying the same procedure as in Appendix \ref{app:bp_derivation} to the conservation of mass, Equation \ref{eq:navierstokes-b}, leads to no change except going from three to two dimensions:
\begin{equation} \label{eq:bp_massconv}
    \dpart{\uba}{\xa} = 0
\end{equation}
As will be shown in Section \ref{sec:anaexamples}, when the spacing varies over the domain, a significant error is introduced mainly due to not including the height  change in the conservation of mass.

\subsection{Varying spacing height formulation}

The correct procedure to including the effects of a varying spacing height would be by letting $\eta$ depend on all spatial variables since the height depends on the location:
\begin{equation} \label{eq:u_fullyVar}
    \uv(x_1,x_2,x_3) = \bar{\uv}(x_1,x_2) \eta(x_1,x_2,x_3)
\end{equation}
with:
\begin{equation} \label{eq:u_zDep2}
    \eta(x_1,x_2,x_3) =  1 - \left( \frac{2x_3}{h(x_1,x_2)} \right)^{2}
\end{equation}
However, inserting this expression into the conservation of momentum and performing the procedure of Appendix \ref{app:bp_derivation} leads to an unnecessarily\footnote{Numerical experiments showed that the many additional terms do not significantly contribute to the results and accuracy of the model.} long and complex governing equation dependent on the gradient of the spacing height raised to several powers. 

Thus, the proposed model simply augments the conservation of mass by deriving it for a control volume of varying spacing height as detailed in the following.

\subsubsection{Mass balance}

The derivations for the correct mass balance will be shown for the two-dimensional case, but the results extend to three dimensions. 
\begin{figure}
    \centering
    \includegraphics[width=\columnwidth]{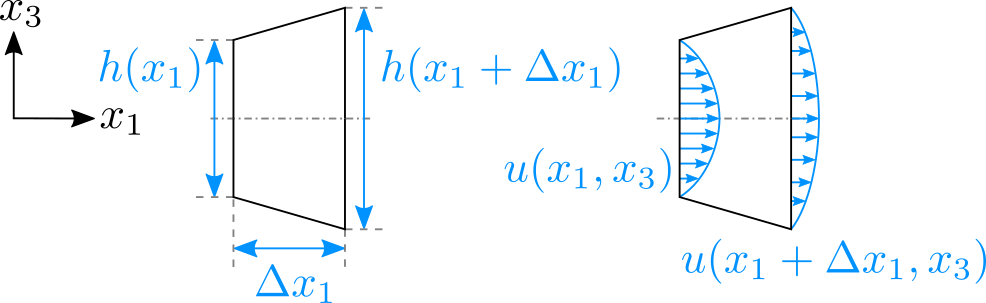}
    \caption{Illustration of control volume for mass balance calculation.}
    \label{fig:mass_balance}
\end{figure}
Figure \ref{fig:mass_balance} shows the control volume used for the mass balance calculations. The height is $h(x_1)$ at the left-hand side and $h(x_1+\Delta x_1)$ at the right-hand side, where $\Delta x_1$ is the width of the control volume. The velocity profile, at the control volume inlet and outlet, is of the same form, as originally assumed in Section \ref{sec:govequ_bp_assump}.

The mass balance is set up by equating the mass flow into and out of the control volume:
\begin{multline}
    \rho \int_{-\frac{1}{2}h(x_1)}^{\frac{1}{2}h(x_1)} \bar{u}(x_1) \eta(x_3) dx_3 = \\ \rho \int_{-\frac{1}{2}h(x_1+\Delta x_1)}^{\frac{1}{2}h(x_1+\Delta x_1)} \bar{u}(x_1+\Delta x_1) \eta(x_3) dx_3  
\end{multline}
Inserting the expression for $\eta$, Equation \ref{eq:u_zDep}, and integrating yields:
\begin{equation}
    \frac{2}{3} \bar{u}(x_1) h(x_1) = \frac{2}{3} \bar{u}(x_1+\Delta x_1) h(x+_1\Delta x_1)
\end{equation}
Rearranging and dividing by the width of the control volume gives:
\begin{equation}
    \frac{\bar{u}(x_1+\Delta x_1) h(x_1+\Delta x_1) - \bar{u}(x_1) h(x_1)}{\Delta x_1} = 0
\end{equation}
which can be changed to differential form by letting $\Delta x_1 \rightarrow 0$:
\begin{equation}
    \frac{d}{dx_1} \bof{\bar{u}(x_1) h(x_1)} = 0
\end{equation}
Finally, the product rule of integration yields:
\begin{equation} \label{eq:aug_continuity}
    h(x_1)\frac{d\bar{u}}{dx_1} + \frac{dh}{dx_1}\bar{u}(x_1) = 0
\end{equation}
where the first term is the standard divergence term weighted by the local height and the second term is non-standard, taking the change in the spacing height into account in the conservation of mass.

\subsubsection{Limitations}

The augmented continuity equation, Equation \ref{eq:aug_continuity}, extends the applicability of the order-reduced model to problems with varying spacing height. However, as will be shown, the model is still limited by the original assumption of a fully-developed flow profile to low-to-medium Reynolds numbers and/or relatively slowly varying spacing heights.

\section{Analysis examples} \label{sec:anaexamples}

In this section, both two- and three-dimensional examples are introduced to investigate the accuracy of the proposed models. The two- and three-dimensional problems are reduced to one- and two dimenions, respectively, using the previously described procedure of integrating the through-thickness direction \textit{a priori}.

The one-dimensional reduced problem highlights the significant error present in the traditional model. However, it also accentuates the drawbacks of the proposed model. The two-dimensional reduced problem is more forgiving for both models, but the proposed model will be shown to be superior.

\subsection{Simulation details}

For all simulations, COMSOL Multiphysics version 5.6 \citep{COMSOL} has been used. Second-order and first-order shape functions are used for the velocity and pressure fields, respectively, in order to avoid deriving stabilisation terms for the reduced-dimensional models. The geometries are meshed using line, triangle and tetrahedral elements in one, two and three dimensions, respectively.

The full dimensional models are implemented using the ``Laminar flow'' interface and the reduced-dimensional models are implemented using the ``Weak form boundary PDE'' interface applied to the mid-plane surface in order to ensure the exact same mesh for the comparison between full and approximate models.

To solve the systems of equations, a direct solver is used for the one- and two-dimensional models, whereas a geometric multigrid preconditioned FGMRES is used for the three-dimensional models. 

\subsection{Two-dimensional to one-dimensional} \label{sec:anaexamples_2D}

This example treats a two-dimensional flow problem, namely plane flow through an infinitely deep channel, which is further reduced to a one-dimensional problem following the process outlined previously.

\subsubsection{Problem setup}

\begin{figure}
     \begin{subfigure}{\columnwidth}
         \centering
         \includegraphics[width=\textwidth]{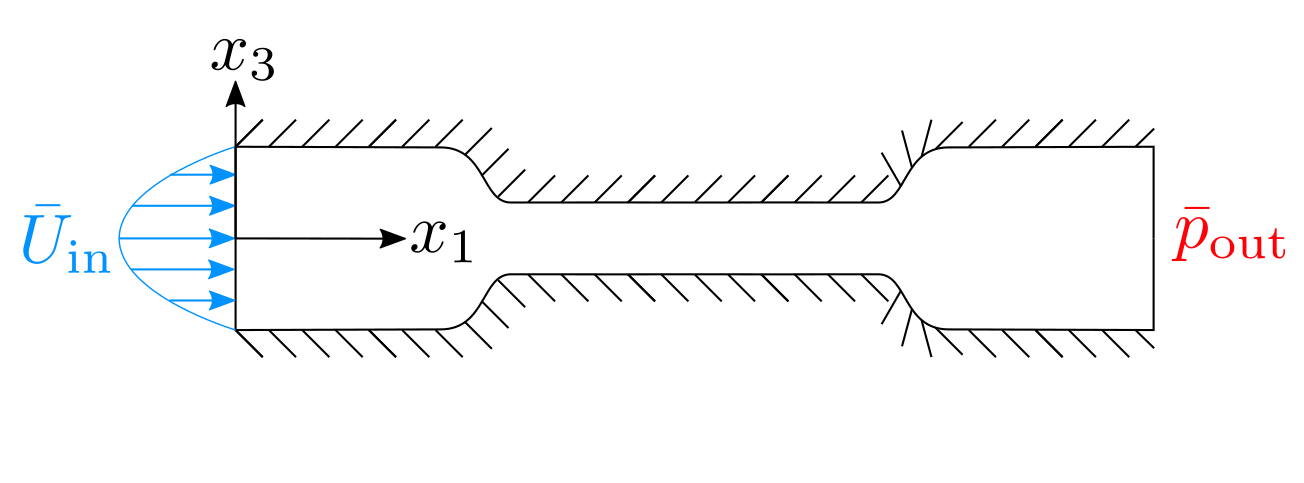}
         \caption{Boundary conditions}
         \label{fig:channel2D-bcs}
     \end{subfigure}
     \\
    \begin{subfigure}{\columnwidth}
         \centering
         \includegraphics[width=\textwidth]{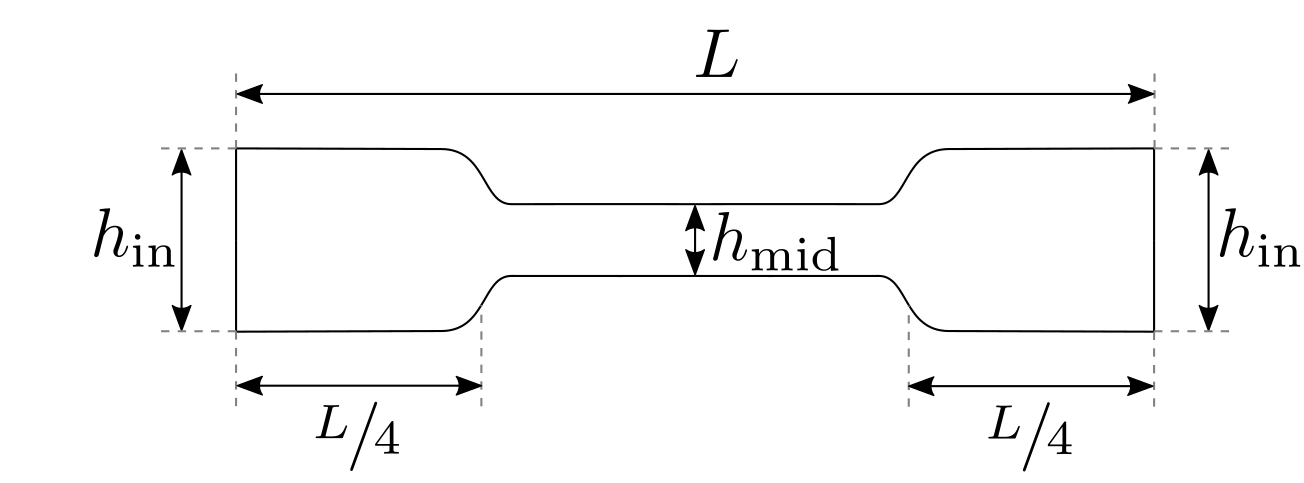}
         \caption{Dimensions}
         \label{fig:channel2D-dims}
     \end{subfigure}
    \caption{Two-dimensional channel of varying height.}
    \label{fig:channel2D}
\end{figure}
The dimensions and boundary conditions are shown in Figure \ref{fig:channel2D}.  The varying channel height shown in Figure \ref{fig:channel2D-dims} is represented as a function of the distance along the channel:
\begin{equation}
    h(x_1) = h_\text{mid} + (h_\text{in} - h_\text{mid})\gamma(x_1)
\end{equation}
where $\gamma(x_1) \in [0;1]$ is the function determining the height at a given point of the channel, $x_1$.
For the geometry under consideration, it is defined as:
\begin{equation}
    \gamma(x_1) = \frac{\tanh\left(\frac{\beta}{2}\right) + \tanh\left(\frac{\beta}{2} \cos\left(2\pi\frac{x_1}{L}\right)\right)}{2\tanh\left(\frac{\beta}{2}\right)}
\end{equation}
which gives a channel that contracts/expands in the middle, with $\beta > 1$ controlling the sharpness of the transition between the maximum and minimum heights.

The problem has been made dimensionless using the channel inlet height as the reference length, such that $h_\text{in} = 1$. The total length of the channel is set to $L=5$. As shown in Figure \ref{fig:channel2D-bcs}, a fully-developed flow with a maximum velocity of $\bar{U}_\text{in} = 1$ 
enters the inlet at the left-hand side. The top and bottom has no-slip and no-penetration conditions, while the outlet has a constant pressure condition, $\bar{p}_\text{out} = 0$. 
Due to symmetry, only the top half of the domain is simulated in the two-dimensional case.

\subsubsection{Parametric study values}

The two reduced one-dimensional models will be compared to the full two-dimensional model for a range of parameters. The Reynolds number, $Re=\frac{\bar{\rho} \bar{U}_\text{in} h_\text{in}}{\mu}$, will be varied to see the effect of the inertia in the system:
\begin{equation}
    Re \in [10^{-3}; 100]
\end{equation}
where $Re = 10^{-3}$ represents Stokes flow and $Re=100$ represents a moderate Reynolds number in the laminar regime.
\begin{figure*}
    \begin{subfigure}{\columnwidth}
         \centering
         \includegraphics[width=\textwidth]{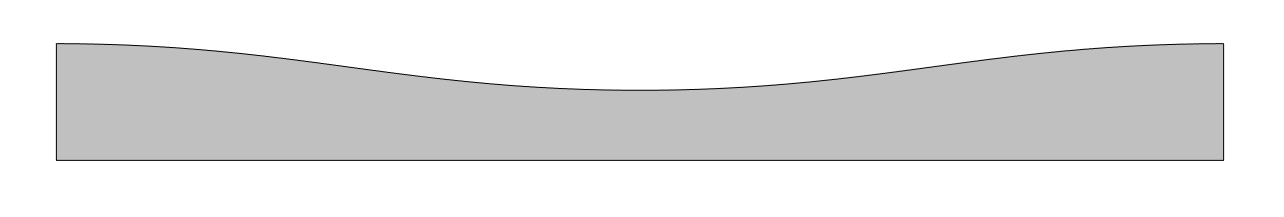}
         \caption{$\beta = 1$}
         \label{fig:channel2D-smooth}
     \end{subfigure}
     \hfill
     \begin{subfigure}{\columnwidth}
         \centering
         \includegraphics[width=\textwidth]{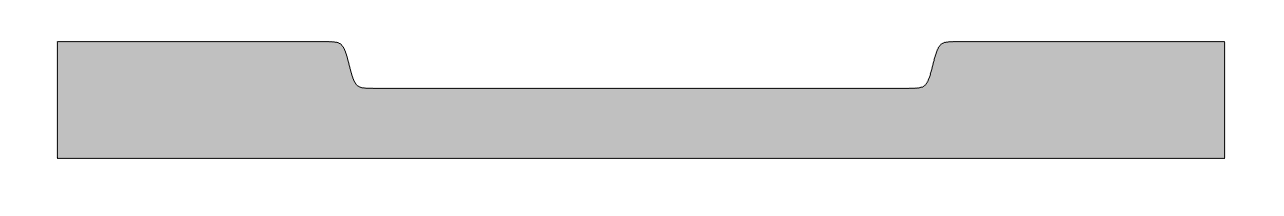}
         \caption{$\beta = 64$}
         \label{fig:channel2D-sharp}
     \end{subfigure}
    \caption{Two-dimensional channel geometries for different parameter settings.}
    \label{fig:channel2D_examples}
\end{figure*}
The height of the middle section, $h_\text{mid}$, will be varied to investigate both a contraction, $h_\text{mid} < h_\text{in}$, and an expansion, $h_\text{mid} > h_\text{in}$:
\begin{equation}
    h_\text{mid} \in [0.6; 1.4]
\end{equation}
Finally, the sharpness of the transition, $\beta$, is varied to investigate the effect of the height gradient:
\begin{equation}
    \beta \in [1; 64]
\end{equation}
Figure \ref{fig:channel2D_examples} shows the two-dimensional geometry with $h_\text{mid} = 0.6$ for $\beta = 1$ and $\beta = 64$. They show that when $\beta = 1$ the channel height changes very slowly and smoothly, while $\beta=64$ causes an abrupt and immediate change in the channel height. With $h_\text{mid} > h_\text{in}$, the geometry expands rather than contracts.

\subsubsection{Results}

\begin{figure}
    \begin{subfigure}{\columnwidth}
         \centering
         \includegraphics[height=0.7\textwidth]{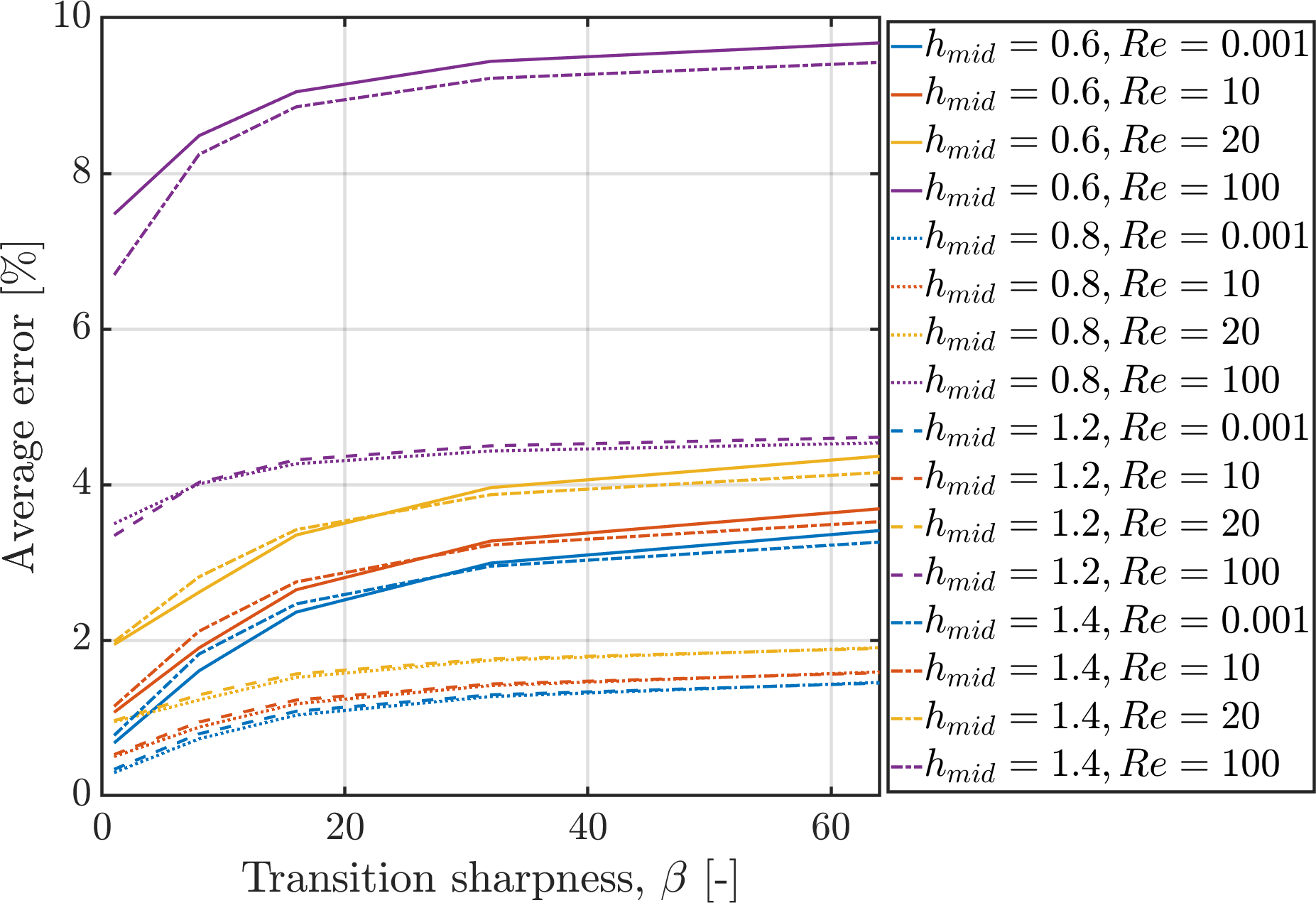}
         \caption{}
         \label{fig:channel2D_errors-a}
     \end{subfigure}
     \\
     \begin{subfigure}{\columnwidth}
         \centering
         \includegraphics[height=0.7\textwidth]{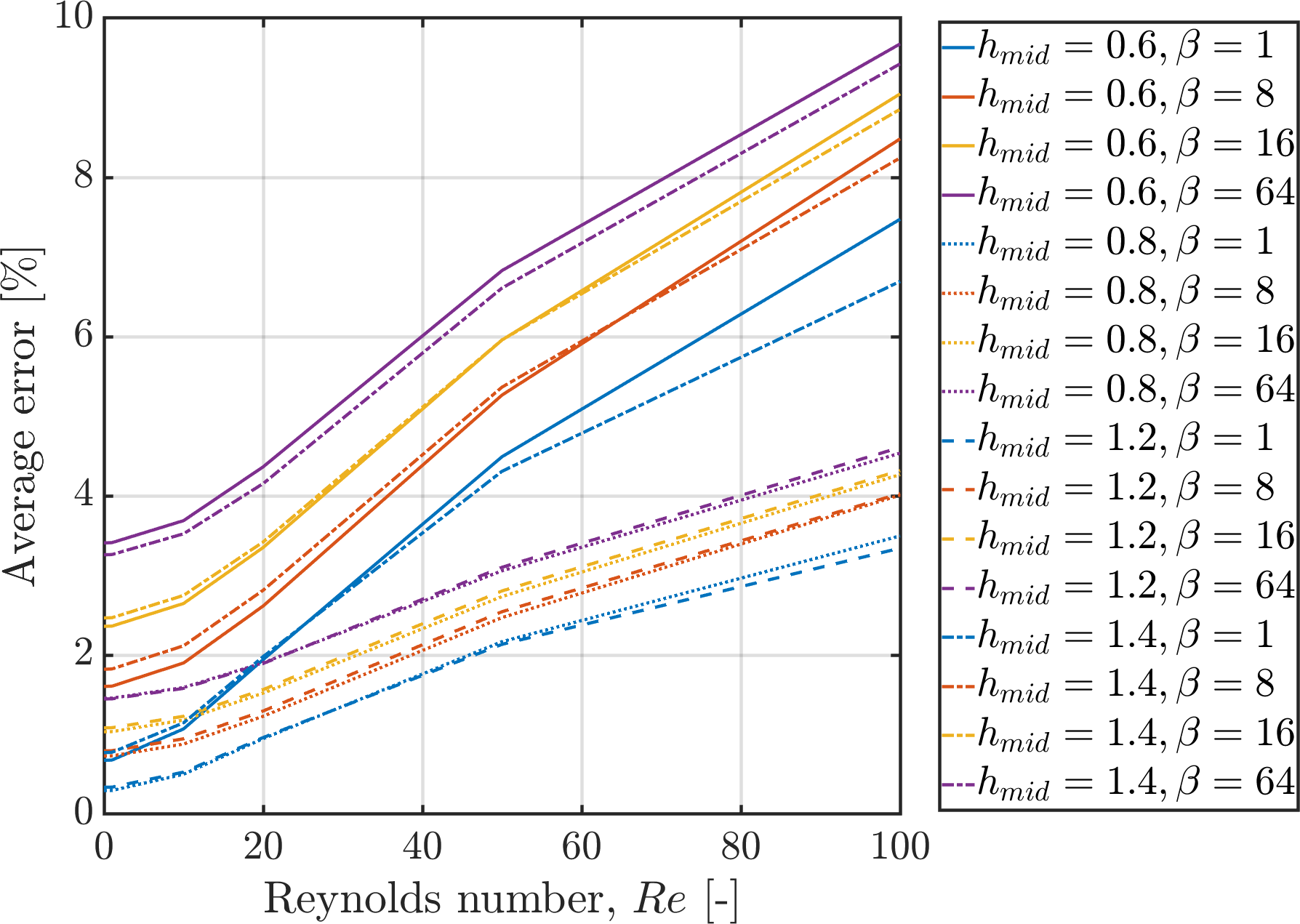}
         \caption{}
         \label{fig:channel2D_errors-b}
     \end{subfigure}
    \caption{Average relative error of the velocity along the channel for the presented one-dimensional model compared to the full two-dimensional channel. The error is show as a function of: (a) transition sharpness for a range of midpoint heights and Reynolds numbers; (b) Reynolds number for a range of midpoint heights and transition sharpness. Same line-style denotes the same midpoint heights, same colours denotes the same: (a) Reynolds number; (b) transition sharpness.}
    \label{fig:channel2D_errors}
\end{figure}
Figure \ref{fig:channel2D_errors} shows the results of the parametric study in terms of the average relative error of the velocity along the channel for the presented one-dimensional model compared to the mid-line velocity of the full two-dimensional model:
\begin{equation}
    e_\text{rel} = \frac{1}{L}\int^{L}_{0} \frac{\lvert \bar{U}_{2D} - \bar{U}_{1D} \rvert}{\bar{U}_{2D}} dx_1
\end{equation}
where $\bar{U}_{\square} = \| \bar{\uv} \|_{2}$ is the velocity magnitude for the given dimensional model.
Figure \ref{fig:channel2D_errors-a} shows that the average relative error depends only weakly on the transition sharpness, $\beta$. Once the transition is over a certain sharpness, the error seems to converge to a maximum value.
Figure \ref{fig:channel2D_errors-b} shows that the average relative error depends strongly on the Reynolds number, $Re$. This makes sense, since inertia becomes increasingly dominant and the constantly fully-developed flow assumption without separation begins to fail.
From both subfigures, it is seen that the average relative error is also strongly dependent on the change in channel height. The larger the difference, the larger the error. It is also observed that the average relative error is more or less the same whether there is a contraction or expansion of same size. This is likely because for both geometries, both one contraction and one expansion edge is present.

\begin{figure}
    \begin{subfigure}{\columnwidth}
         \centering
         \includegraphics[width=\textwidth]{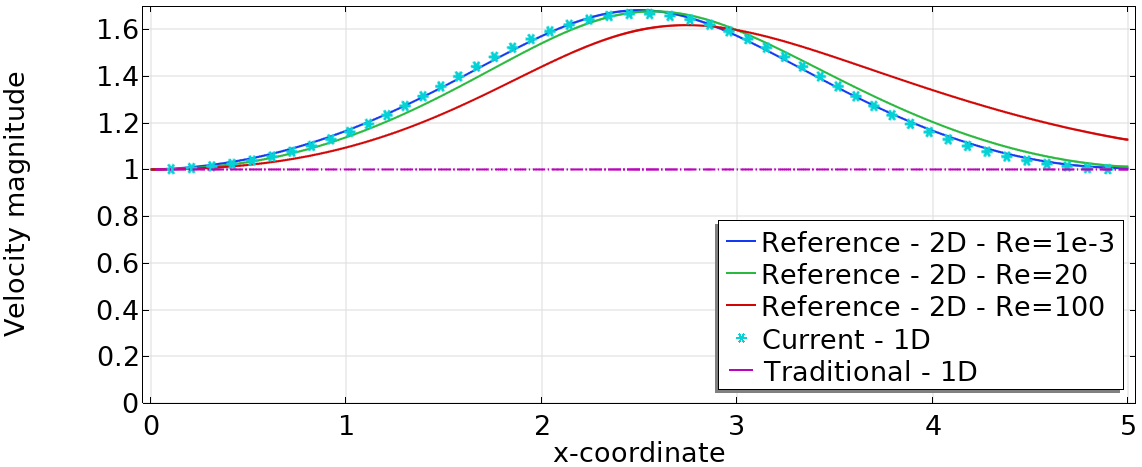}
         \caption{Contraction, $h_{mid} = 0.6,\, \beta = 1$}
         \label{fig:channel2D_midline-a}
    \end{subfigure}
    \\
    \begin{subfigure}{\columnwidth}
         \centering
         \includegraphics[width=\textwidth]{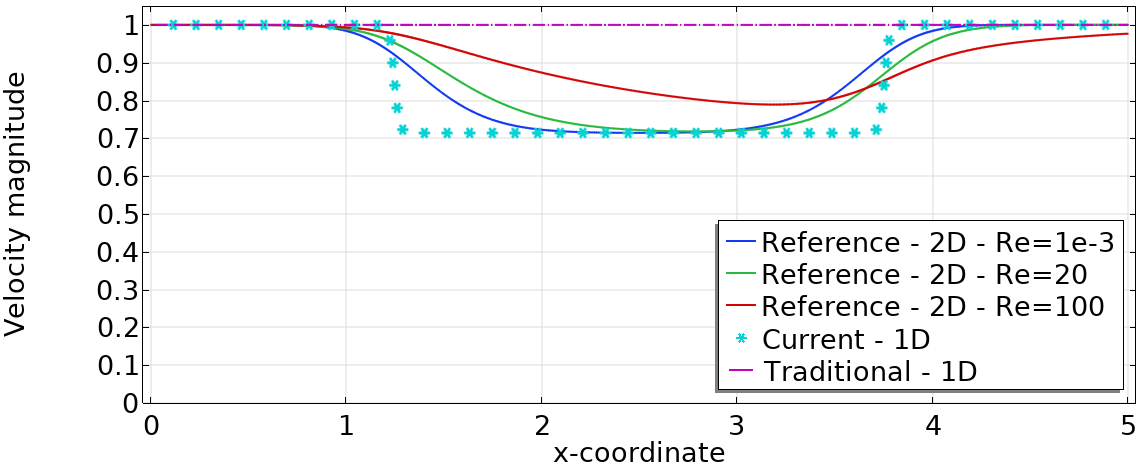}
         \caption{Expansion, $h_{mid} = 1.4,\, \beta = 64$}
         \label{fig:channel2D_midline-b}
    \end{subfigure}
    \caption{Representative examples of velocity magnitude comparison along the channel mid-line for the different models.}
    \label{fig:channel2D_midline}
\end{figure}
Figure \ref{fig:channel2D_midline} shows the velocity magnitude along the mid-line, comparing the traditional and proposed one-dimensional models with the full two-dimensional model for two representative contraction and expansion channel geometries.
Firstly, it is observed that the traditional one-dimensional model simply does not capture the change in channel height, since the conservation of mass is not adapted to accommodate this. So although a larger friction term exists when the channel height is reduced (giving a higher pressure drop), it has no effect on the velocity field when a prescribed inflow velocity for this one-dimensional case.
Secondly, for both one-dimensional models, the solution does not vary with a change in the Reynolds number. This is due to the restrictions that a one-dimensional problem presents together with the assumptions made. 

Further examples are shown and discussed in Appendix \ref{app:parameterStudies}. Overall a reasonably good accuracy is observed for many parameter values, in accordance with Figure \ref{fig:channel2D_errors}. The errors are mainly due to inaccuracies in the post-contraction and -expansion areas, especially for higher Reynolds numbers, higher height differences and higher transition sharpness. This is because both the traditional and proposed reduced-dimensional models exhibit instantaneous expansion and contraction, due to the assumption of a fully-developed flow profile at all points along the channel. This assumption allows for a reduction in the dimension of the problem, but like all assumptions it also introduces limitations and errors when outside these limitations. However, even for the upper bound of the parameter ranges investigated here, an average relative error below 10 percent is seen, which is acceptable.

\subsection{Three-dimensional to two-dimensional} \label{sec:anaexamples_3D}

The previous example is now extended to three dimensions, where the third dimension is now finite and must be taken into account. The flow domain is defined between two square plates with a circular dimple or protrusion in the center. The same function is used to describe the height of the channel in the $x_1 - x_3$ plane and the two-dimensional expansion or contraction profile is then revolved to form either a dimple or protrusion, respectively.

\begin{figure}
    \begin{subfigure}{\columnwidth}
         \centering
         \includegraphics[width=0.8\textwidth]{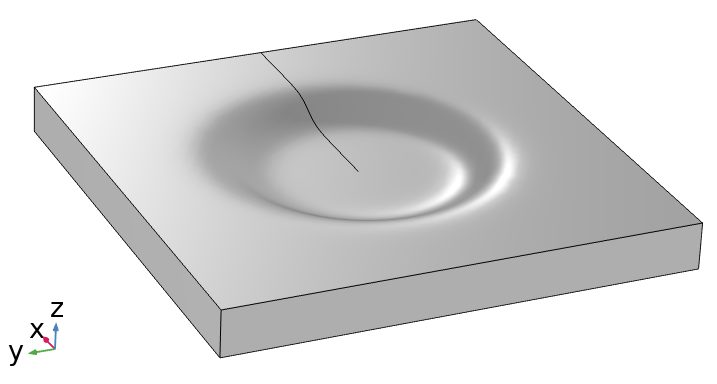}
         \caption{$h_\text{mid} = 0.6, \beta = 8$}
         \label{fig:channel3D_examples-a}
    \end{subfigure}
    \\
    \begin{subfigure}{\columnwidth}
         \centering
         \includegraphics[width=0.8\textwidth]{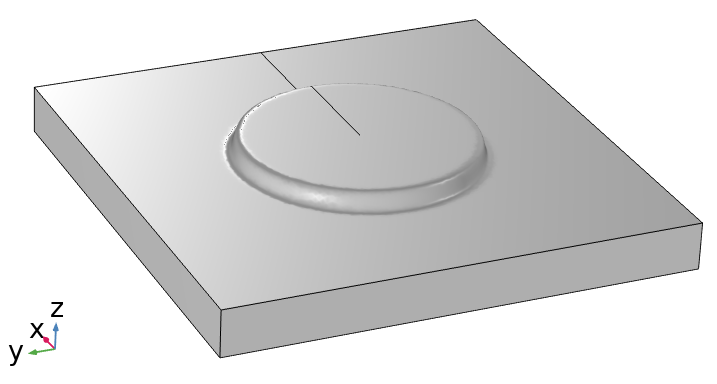}
         \caption{$h_\text{mid} = 1.4, \beta = 64$}
         \label{fig:channel3D_examples-b}
     \end{subfigure}
    \caption{Three-dimensional channel geometries for different parameter settings.}
    \label{fig:channel3D_examples}
\end{figure}
Figure \ref{fig:channel3D_examples} shows examples of the three-dimensional contraction/protrusion and expansion/dimple geometries. Figure \ref{fig:channel3D_examples-a} shows a contraction/protrusion geometry with $h_\text{mid} = 0.6$ and $\beta = 8$ and Figure \ref{fig:channel3D_examples-b} shows a expansion/dimple geometry with $h_\text{mid} = 1.4$ and $\beta = 64$. Similar to the two-dimensional case, due to symmetry only the top half of the domain is simulated for the three-dimensional case in order to save on computational time.

The same parametric study is carried out as for the two-dimensional channel problem, except $h_{mid}$ is only set to the bounds of $0.6$ and $1.4$.
The figures in the following subsections try to highlight the most important conclusions, since the parameter study produces a lot of data. For instance, at the lowest Reynolds number, $Re=10^{-3}$, when the geometries are smoothly-varying, the models are practically identical and are thus not shown here.

\subsubsection{Contraction geometry} \label{sec:anaexamples_3D_con}

\begin{figure}
    \begin{subfigure}{\columnwidth}
         \centering
         \includegraphics[width=0.64\textwidth]{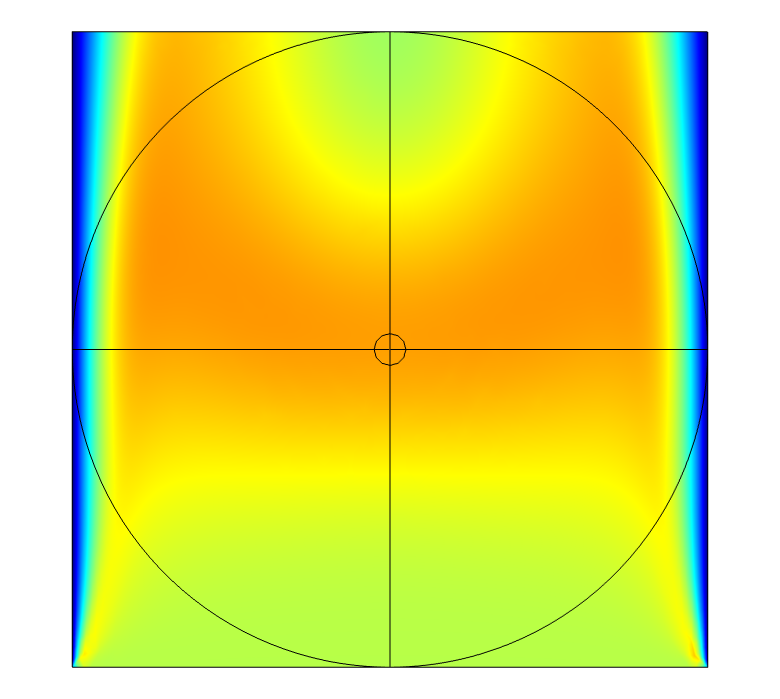}
         \caption{Full 3D model}
         \label{fig:channel3D_smoothContraction_highRe-a}
    \end{subfigure}
    \\
    \begin{subfigure}{\columnwidth}
         \centering
         \includegraphics[width=0.64\textwidth]{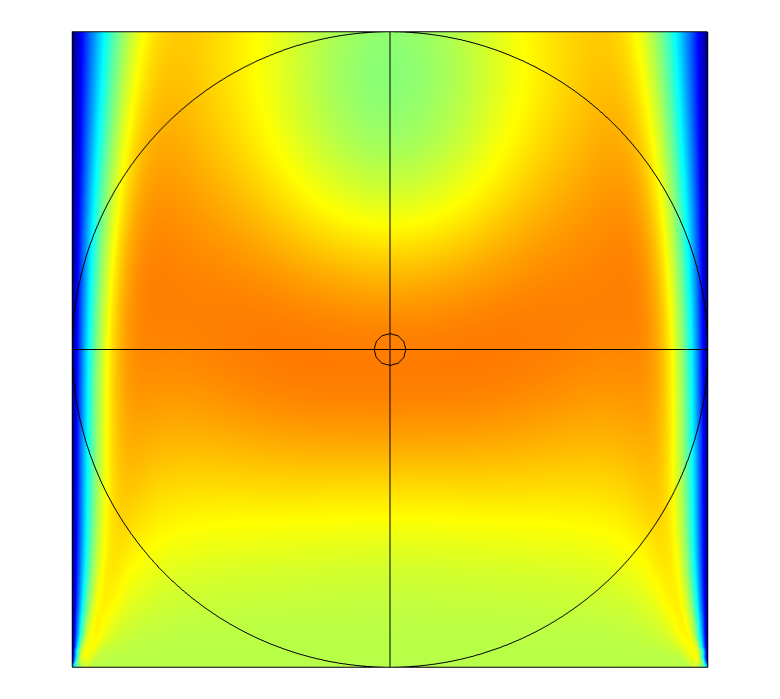}
         \caption{Proposed model}
         \label{fig:channel3D_smoothContraction_highRe-b}
    \end{subfigure}
    \\
    \begin{subfigure}{\columnwidth}
         \centering
         \includegraphics[width=0.64\textwidth]{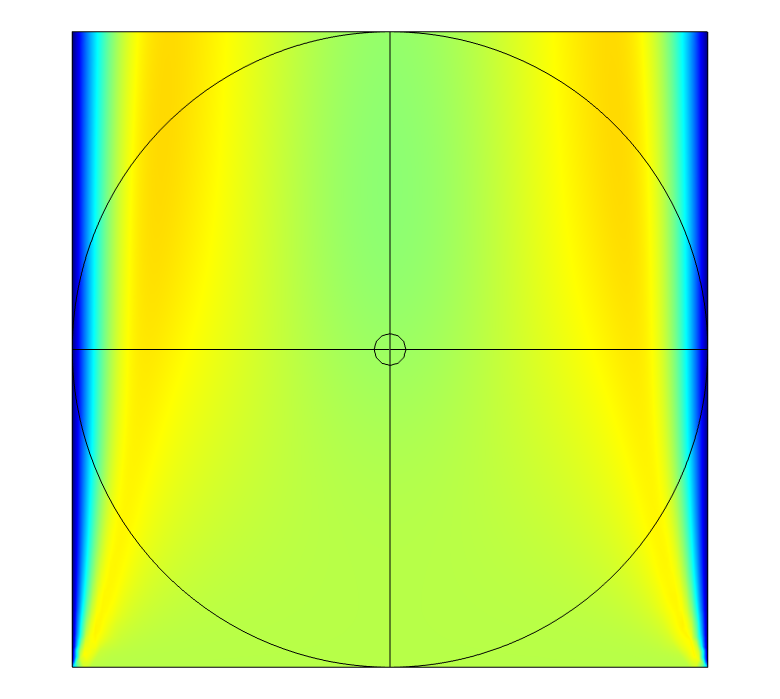}
         \caption{Traditional model}
         \label{fig:channel3D_smoothContraction_highRe-c}
    \end{subfigure}
    \caption{Velocity magnitude field on the mid-plane of the three-dimensional contraction channel geometry for different models with $h_\text{mid}=0.6$, $\beta = 1$ and $Re = 100$.}
    \label{fig:channel3D_smoothContraction_highRe}
\end{figure}
Figure \ref{fig:channel3D_smoothContraction_highRe} shows the mid-plane velocity magnitude field at $Re=100$ for the three different models applied to a smoothly-varying contraction channel geometry with $h_\text{mid}=0.6$ and $\beta = 1$. It can be seen that even at this moderately high Reynolds number, the agreement between the full three-dimensional model and the proposed planar model is very good. However, Figure \ref{fig:channel3D_smoothContraction_highRe-c} shows that the traditional model, without the augmentation of mass conservation, is way off. However, it is not as bad as for the two-dimensional case, due to the planar nature of the problem, where the flow is able to flow around the obstructions.

\begin{figure}
    \begin{subfigure}{\columnwidth}
         \centering
         \includegraphics[width=0.64\textwidth]{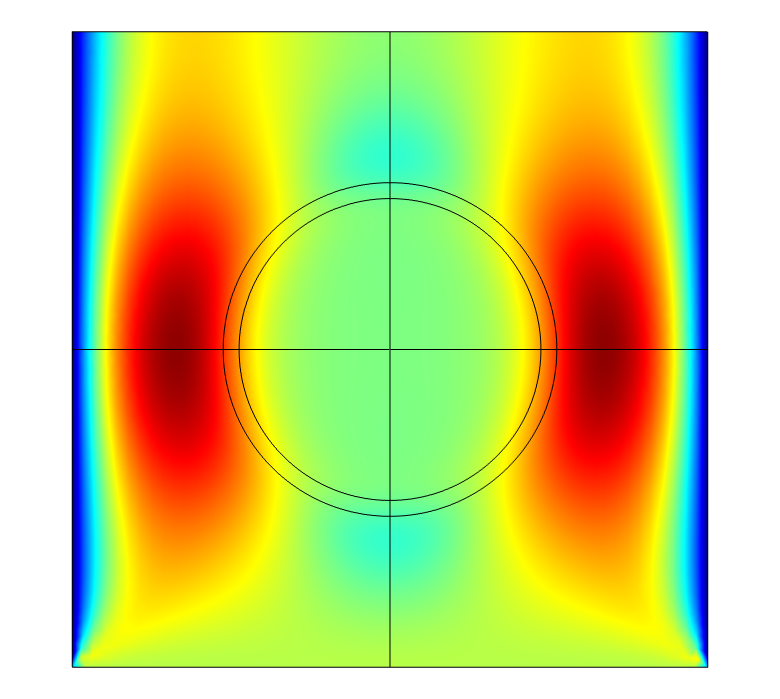}
         \caption{Full 3D model}
         \label{fig:channel3D_sharpContraction_lowRe-a}
    \end{subfigure}
    \\
    \begin{subfigure}{\columnwidth}
         \centering
         \includegraphics[width=0.64\textwidth]{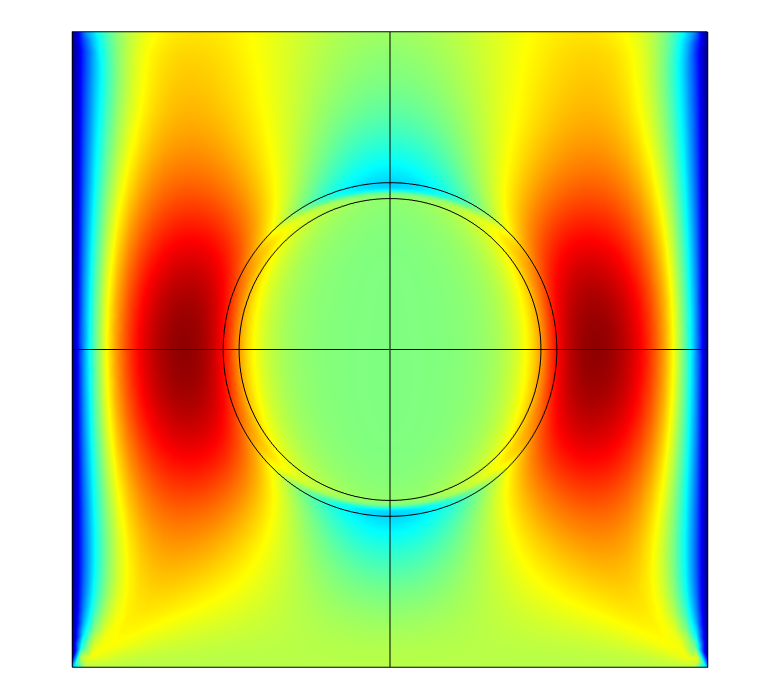}
         \caption{Proposed model}
         \label{fig:channel3D_sharpContraction_lowRe-b}
    \end{subfigure}
    \\
    \begin{subfigure}{\columnwidth}
         \centering
         \includegraphics[width=0.64\textwidth]{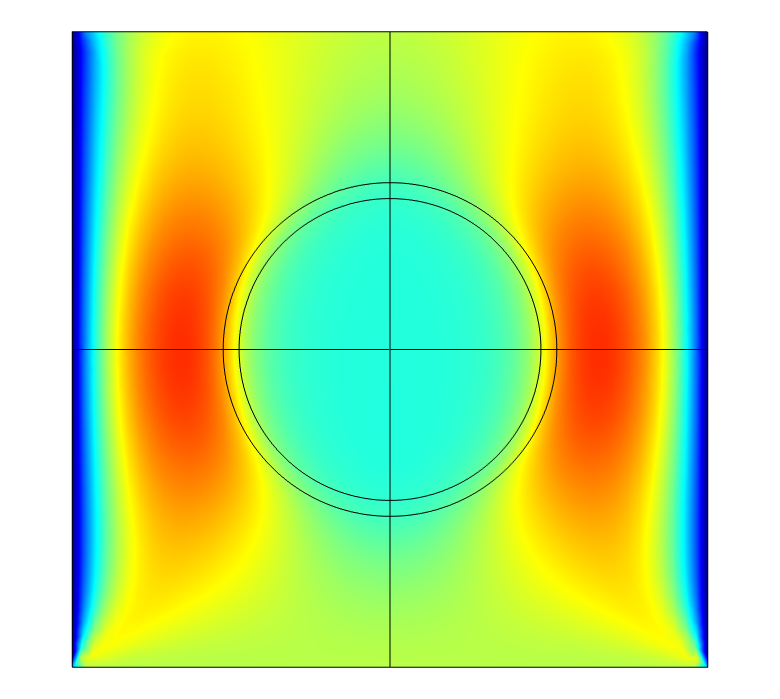}
         \caption{Traditional model}
         \label{fig:channel3D_sharpContraction_lowRe-c}
    \end{subfigure}
    \caption{Velocity magnitude field on the mid-plane of the three-dimensional contraction channel geometry for different models with $h_\text{mid}=0.6$, $\beta = 64$ and $Re = 10^{-3}$.}
    \label{fig:channel3D_sharpContraction_lowRe}
\end{figure}
Figure \ref{fig:channel3D_sharpContraction_lowRe} shows the mid-plane velocity magnitude field at $Re=10^{-3}$ for the different models applied to a sharply-varying contraction channel geometry with $h_\text{mid}=0.6$ and $\beta = 64$. Similar to before, the proposed planar model agrees very well with the full three-dimensional model. But it does appear that the proposed model over-predicts the stagnation effect of the protrusion compared to the full three-dimensional model. Figure \ref{fig:channel3D_sharpContraction_lowRe-c} shows that the prediction of the traditional model is not very good, with a significantly lower velocity and less fluid flow passing under the protrusion. This is because the model incorporates the resistance from the protrusion, but does not account for the reduction in flow volume. However, the flow outside the contraction is relatively acceptable. As it will be shown in Sections \ref{sec:topogToTopol_analysis} and \ref{sec:topogToTopol_optimisation}, this is why the model works for flow \textit{topology} optimisation where $h_{mid}\rightarrow0$. For the rest of this parameter study, the results of the traditional model will not be shown.

\begin{figure}
    \begin{subfigure}{\columnwidth}
         \centering
         \includegraphics[width=0.64\textwidth]{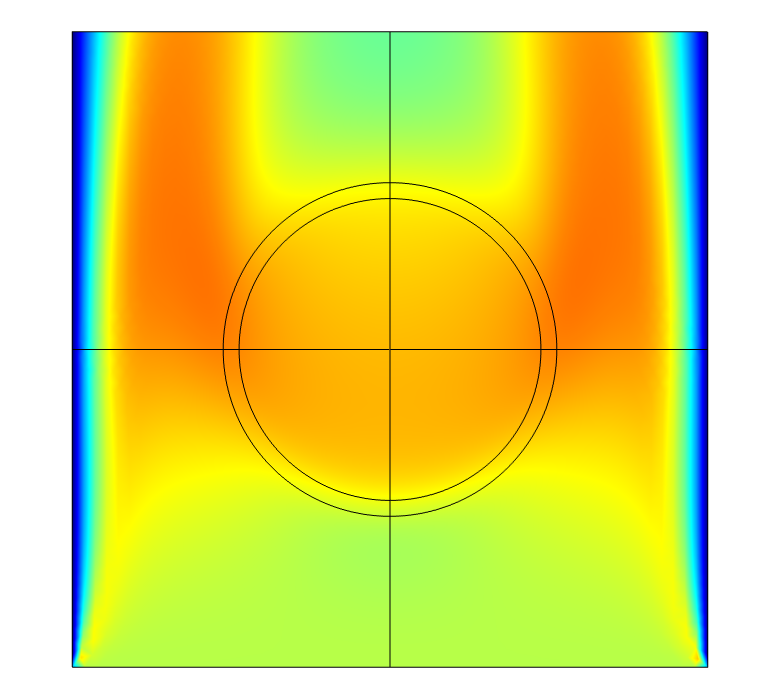}
         \caption{Full 3D model}
         \label{fig:channel3D_sharpContraction_highRe-a}
    \end{subfigure}
    \\
    \begin{subfigure}{\columnwidth}
         \centering
         \includegraphics[width=0.64\textwidth]{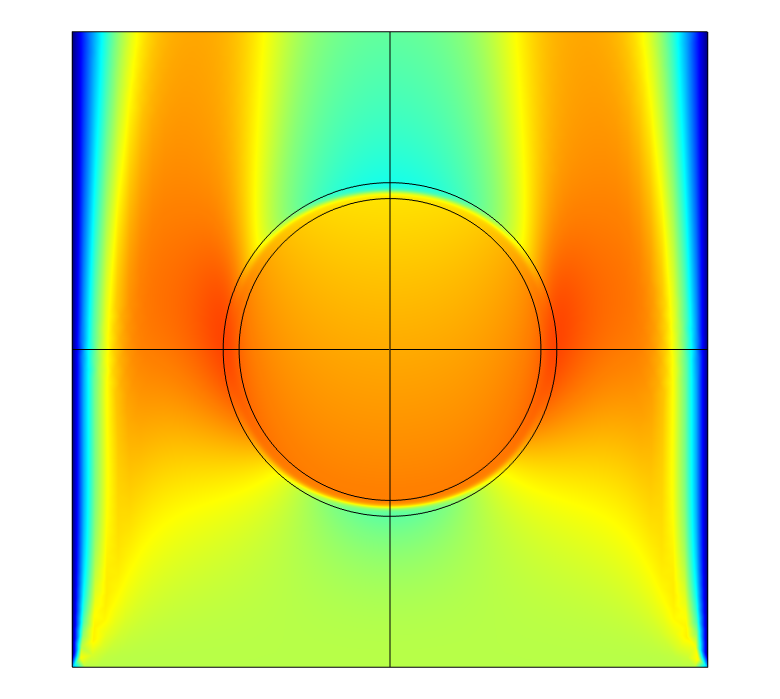}
         \caption{Proposed 2D model}
         \label{fig:channel3D_sharpContraction_highRe-b}
    \end{subfigure}
    \caption{Velocity magnitude field on the mid-plane of the three-dimensional contraction channel geometry for different models with $h_\text{mid}=0.6$, $\beta = 64$ and $Re = 100$.}
    \label{fig:channel3D_sharpContraction_highRe}
\end{figure}
Figure \ref{fig:channel3D_sharpContraction_highRe} shows the mid-plane velocity magnitude field at $Re=100$ for the reference and proposed models applied to a sharply-varying contraction channel geometry with $h_\text{mid}=0.6$ and $\beta = 64$. Even for this higher Reynolds numbers, the proposed planar model agrees very well with the full three-dimensional model. However, the instantaneous effect of contraction and expansion is clear for the proposed model in Figure \ref{fig:channel3D_sharpContraction_highRe-b}. The velocity increases as soon as the minimum height is reached and decreases as soon at the maximum height is reached, whereas inertia smoothens out the transition in the full model.

\subsubsection{Expansion geometry} \label{sec:anaexamples_3D_exp}

For the expansion case, corresponding to a dimple in the surface, generally the error is higher in the proposed model. This is because inertia plays a much larger role at an expansion of the channel height. Furthermore, the traditional model predicts results contrary to the physics of the problem, namely an increase in velocity under the expansion due to the lower flow resistance. However, the results are omitted here because an expansion is not relevant in \textit{topology} optimisation where the model is used.

\begin{figure}
    \begin{subfigure}{\columnwidth}
         \centering
         \includegraphics[width=0.64\textwidth]{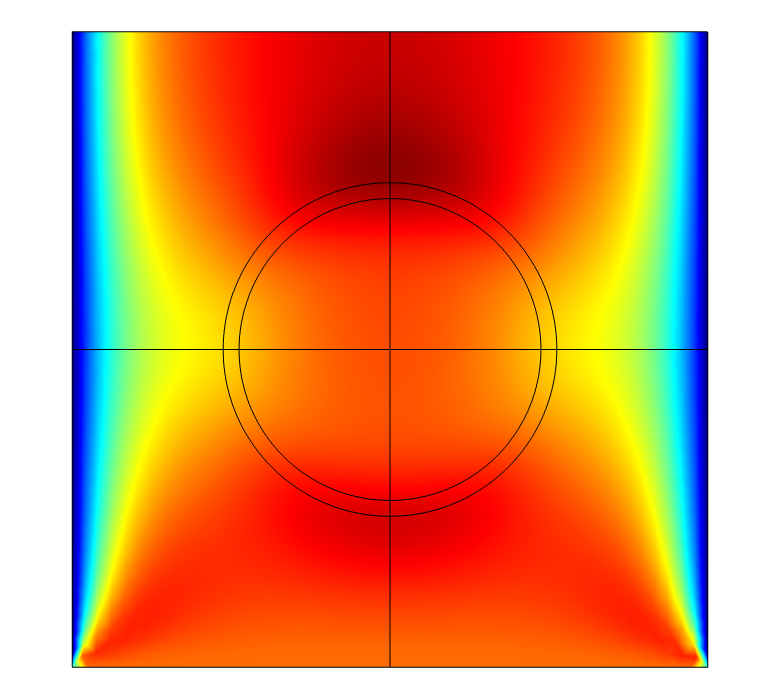}
         \caption{Full 3D model}
         \label{fig:channel3D_sharpExpansion_lowRe-a}
    \end{subfigure}
    \\
    \begin{subfigure}{\columnwidth}
         \centering
         \includegraphics[width=0.64\textwidth]{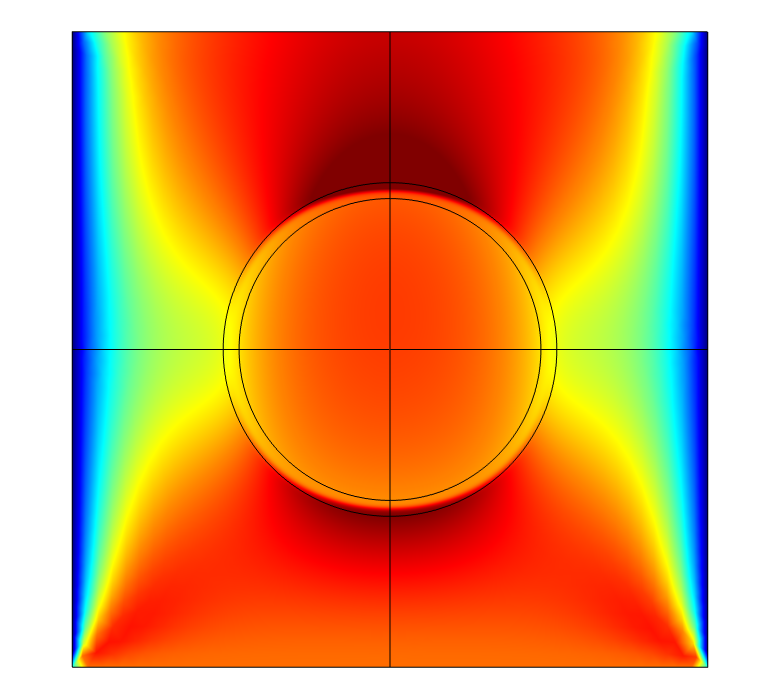}
         \caption{Proposed 2D model}
         \label{fig:channel3D_sharpExpansion_lowRe-b}
    \end{subfigure}
    \caption{Velocity magnitude field on the mid-plane of the three-dimensional expansion channel geometry for different models with $h_\text{mid}=1.4$, $\beta = 64$ and $Re = 10^{-3}$.}
    \label{fig:channel3D_sharpExpansion_lowRe}
\end{figure}
Figure \ref{fig:channel3D_sharpExpansion_lowRe} shows the mid-plane velocity magnitude field at $Re=10^{-3}$ for a sharply-varying expansion channel geometry with $h_\text{mid}=1.4$ and $\beta = 64$.
Figure \ref{fig:channel3D_sharpExpansion_lowRe-b} shows that the proposed model does not perform as well as for the contraction geometry. The problem arises from the instantaneous expansion of the flow felt by the model, due to the assumption of a fully-developed profile at all points in space.
\begin{figure}
    \begin{subfigure}{\columnwidth}
         \centering
         \includegraphics[width=0.64\textwidth]{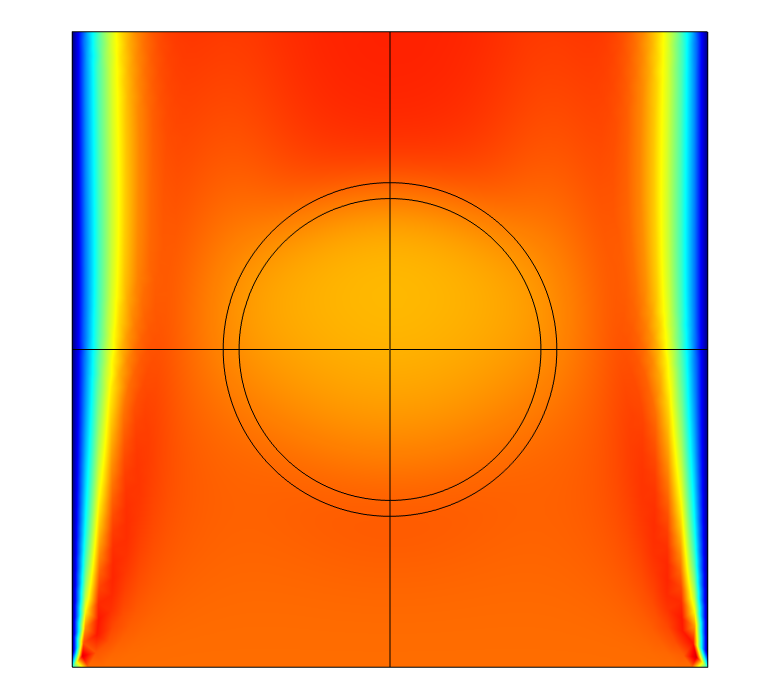}
         \caption{Full 3D model}
         \label{fig:channel3D_sharpExpansion_highRe-a}
    \end{subfigure}
    \\
    \begin{subfigure}{\columnwidth}
         \centering
         \includegraphics[width=0.64\textwidth]{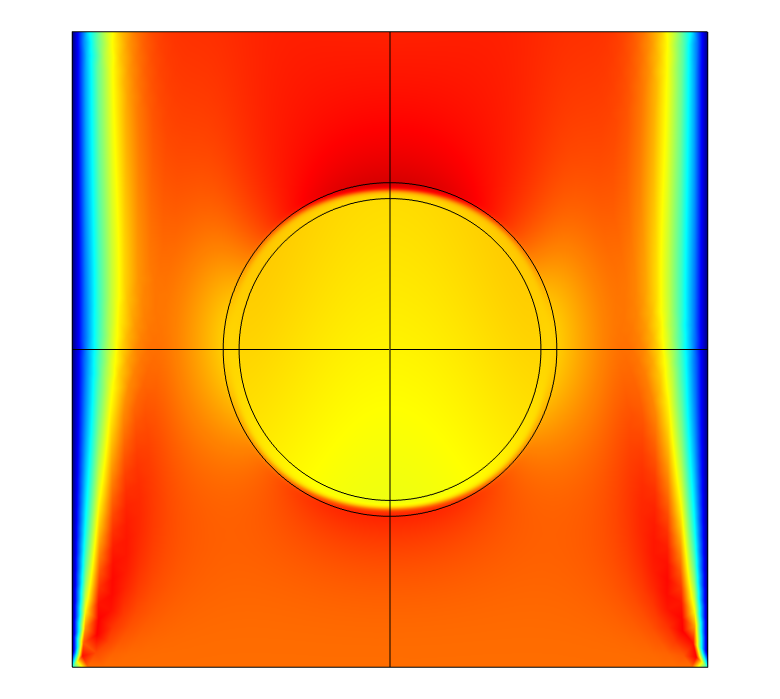}
         \caption{Proposed 2D model}
         \label{fig:channel3D_sharpExpansion_highRe-b}
    \end{subfigure}
    \caption{Velocity magnitude field on the mid-plane of the three-dimensional expansion channel geometry for different models with $h_\text{mid}=1.4$, $\beta = 64$ and $Re = 100$.}
    \label{fig:channel3D_sharpExpansion_highRe}
\end{figure}
This is further accentuated when increasing the Reynolds number, as seen in Figure \ref{fig:channel3D_sharpExpansion_highRe} which shows the mid-plane velocity magnitude field for the same channel geometry at $Re=100$. Because it is the mid-plane, the inertia of the fluid entering the expansion carries it forwards as in Figure \ref{fig:channel3D_sharpExpansion_highRe-a}, rather than expanding instantaneously as in Figure \ref{fig:channel3D_sharpExpansion_highRe-b}. This is the same as was observed for the two-dimensional channel in Figure \ref{fig:channel2D_midline}.

\subsubsection{Average relative error} \label{sec:anaexamples_3D_err}

\begin{figure}
    \begin{subfigure}{\columnwidth}
         \centering
         \includegraphics[height=0.7\textwidth]{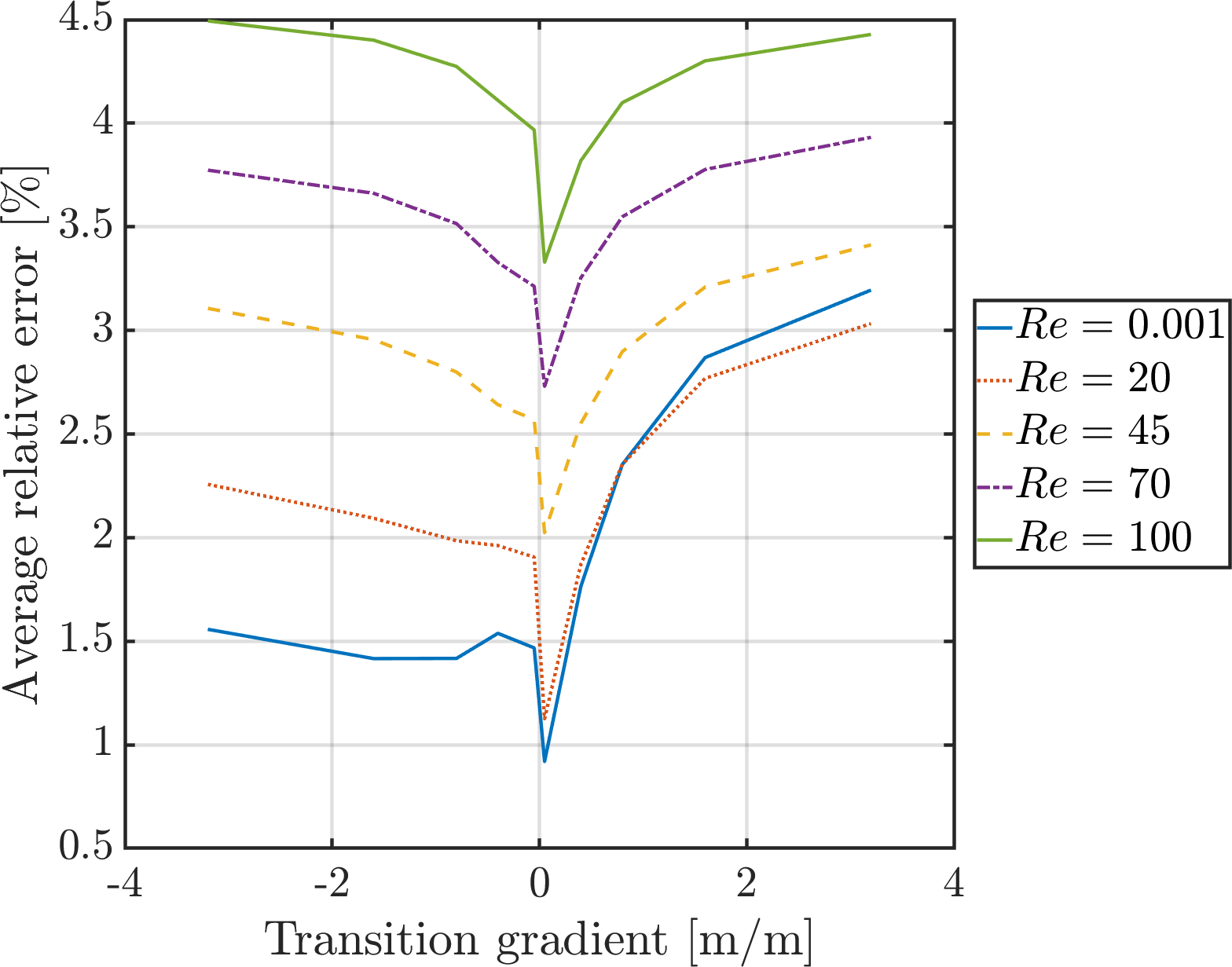}
         \caption{}
         \label{fig:channel3D_errors-a}
     \end{subfigure}
     \\
     \begin{subfigure}{\columnwidth}
         \centering
         \includegraphics[height=0.7\textwidth]{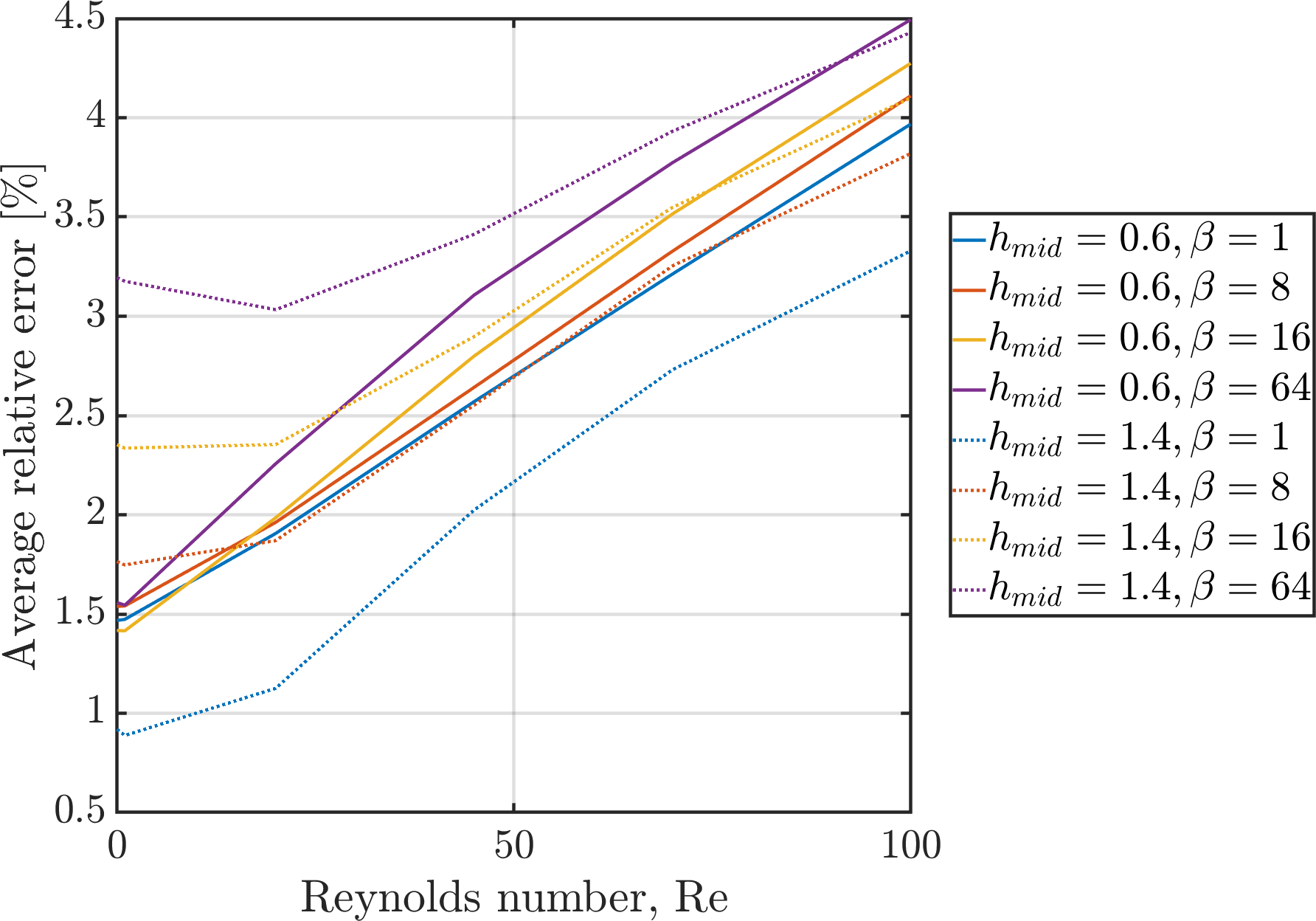}
         \caption{}
         \label{fig:channel3D_errors-b}
     \end{subfigure}
    \caption{Average relative error of the velocity magnitude on the mid-plane of  the channel for the presented two-dimensional model compared to the full three-dimensional channel. The error is show as a function of: (a) transition gradient for several Reynolds numbers; (b) Reynolds number for two midpoint heights and several transition sharpness. For (b) the same line-style denotes the same midpoint height, same colours denotes the same transition sharpness.}
    \label{fig:channel3D_errors}
\end{figure}
Figure \ref{fig:channel3D_errors} shows the average relative error of the proposed two-dimensional model compared to the full three-dimensional channel:
\begin{equation}
    e_\text{rel} = \int_{\omega} \frac{\lvert \bar{U}_{3D} - \bar{U}_{2D} \rvert}{\bar{U}_{3D}} dS
\end{equation}
Figure \ref{fig:channel3D_errors-a} shows the error as a function of the transition gradient. This has been computed based on the change in height and the width of the transition based on the transition sharpness. Figure \ref{fig:channel3D_errors-a} shows that the average relative error generally depends weaker on the transition gradient for contractions (negative gradient) rather than expansions (positive gradients). This difference seems to even out for higher Reynolds numbers.
As was also seen for the two-dimensional channel, Figure \ref{fig:channel3D_errors-b} shows that the average relative error depends strongly on the Reynolds number, $Re$. As before this makes sense, since inertia becomes increasingly dominant and the constantly fully-developed flow assumption without separation begins to fail.

\subsection{Transition from topography to topology} \label{sec:topogToTopol_analysis}

The transition from topography to topology will now be investigated by letting the minimum thickness of the contraction/protrusion geometry go towards 0. 
A height of $h_\text{mid} = 0$ would dictate a topology change, since there would be a hole introduced in the flow domain, rather than simply a protruding obstacle.
By setting $h_\text{mid}$ to something very small, this can be approximated numerically, which is the original idea behind the parametrisation by \citet{Borrvall2003} and \citet{GersborgHansen2005}.

Due to the topological change, flow separation around the cylinder is present with significant recirculation and low velocities behind the obstacle. Therefore, the domain is doubled in length in the flow direction to ensure this is captured by the models. 

\begin{figure}
    \centering
    \includegraphics[width=0.99\columnwidth]{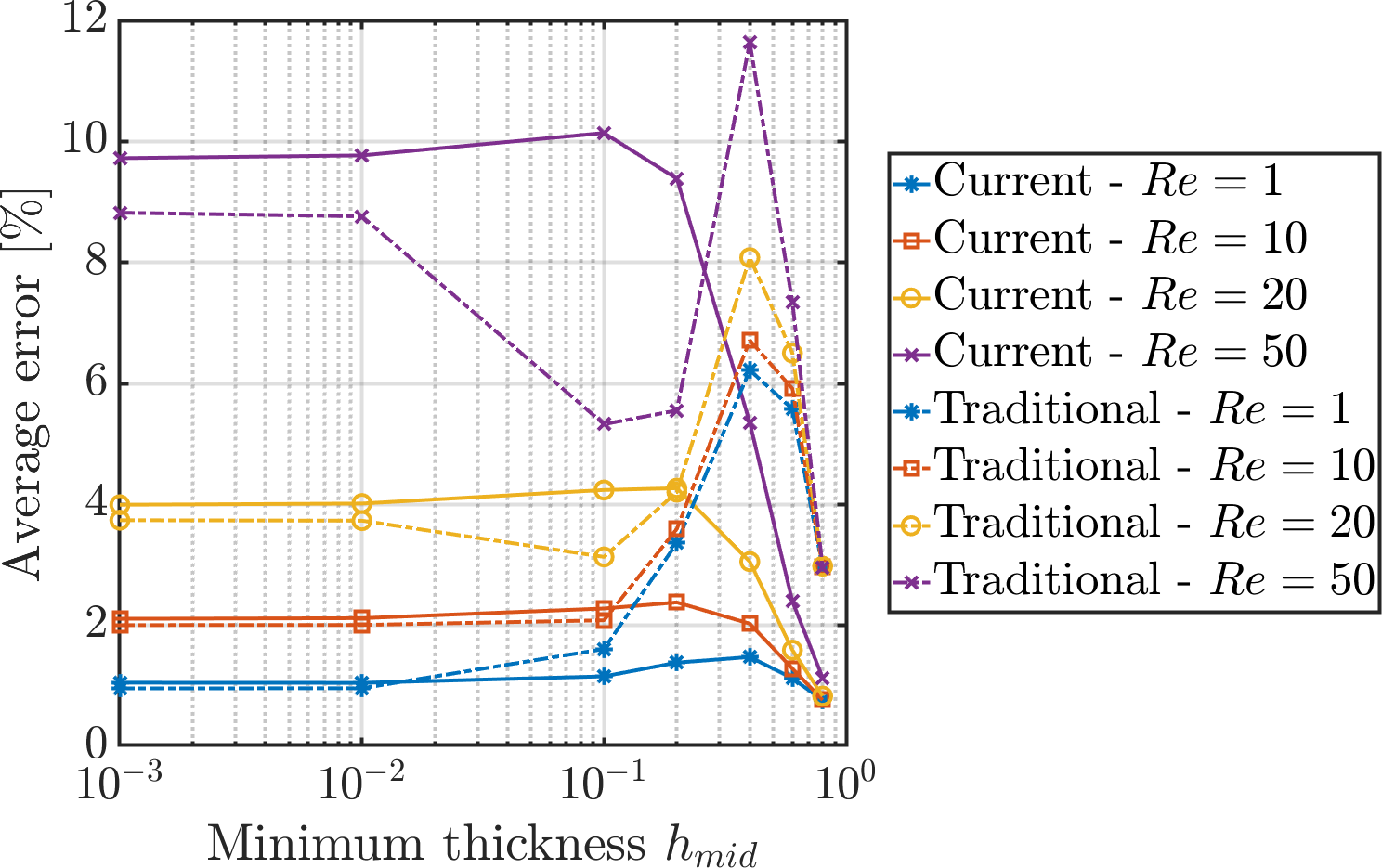}
    \caption{Average error of the velocity magnitude on the mid-plane of the channel for the presented and traditional two-dimensional models compared to the full three-dimensional channel.}
    \label{fig:channel3D_topoErrors}
\end{figure}
Figure \ref{fig:channel3D_topoErrors} shows the average error of the velocity magnitude for the two models compared to the three-dimensional model. The error is computed based on the inlet velocity as reference, since the regular relative error does not work for significant areas of low/almost-zero velocity:
\begin{equation}
    e_\text{rel} = \int_{\omega} \frac{\lvert \bar{U}_{3D} - \bar{U}_{2D} \rvert}{U_{in}} dS
\end{equation}
From the graphs it can be seen, that the proposed planar model with augmented mass conservation generally has a lower error for higher minimum thicknesses, than the traditional model with only the through-thickness resistance. This makes sense since the proposed model is developed for exactly that situation. As the minimum thickness is decreased, the two models appear to converge towards the same error levels. However, the traditional model actually exhibits better accuracy for higher Reynolds numbers already for a minimum height below $h_\text{mid} = 0.2$\footnote{This is probably not be a general conclusion, since it seems out-of-place.}. 

As predicted, it is seen that for infinitesimal minimum thicknesses, the traditional model is fine to use, since the accuracy is the same or sometimes even better than the proposed model with augmented mass conservation. This also makes sense as argued by \citet{GersborgHansen2005}, that the flow resistance term serves as ``merely an algorithmic device to implement a continuous transition between the limiting cases of viscous flow and zero flow'', which is also the concept chosen by a large majority of papers \citep{Alexandersen2020}. So as long as a \textit{topological} definition is taken, with regions of fluid flow and regions simulating solid regions, where a very small or infinitesimal minimum height is applied, the traditional model using the through-thickness viscous resistance only is sufficient.

However, the focus of this paper is problems where that is not true. The problems have surfaces of continuously varying height, where the minimum height is relatively large with a significant fluid flow through those areas. For these cases, it has been shown that the proposed model with the augmented mass conservation is strictly necessary.
Furthermore, in Section \ref{sec:results_manifold} it is actually shown that the proposed model seems to produce better performing topologies for similar settings.

\section{Optimisation formulation} \label{sec:optimisation}

The presented planar reduced-dimensional model will now be applied to the optimisation of the surface topography of the bounding plates. This is done by coupling the local channel height to a design field.

\subsection{Design parametrisation}

The local channel height is defined as:
\begin{equation} \label{eq:height}
    h(x_1,x_2) = h_\text{min} + (h_\text{max} - h_\text{min})\gamma(x_1,x_2)
\end{equation}
where $h_\text{min}$ is the minimum height, $h_\text{max}$ is the minimum height, and $\gamma(x_1,x_2) \in [0;1]$ is the design field to be determined.

The design field is discretised using nodal variables and linear shape functions. A reaction-diffusion filter \citep{Lazarov2011} is applied to the design field as a means to control the transition gradient to ensure accuracy, as discussed in Section \ref{sec:anaexamples_3D_err}. The filter also ensures continuity between the design domain and domains of prescribed channel height (i.e. inlets and outlets). The filter is controlled using the approximate radius of the filter kernel, $r_\text{min}$. This approximately ensures a transition width of $2 r_\text{min}$ and a maximum transition gradient of:
\begin{equation} \label{eq:maxgrad}
    \left\| \dpx{h} \right\|_{2} \leq \frac{h_\text{max}-h_\text{min}}{2 r_\text{min}}
\end{equation}

\subsection{Implementation details}

As for the simulations in Section \ref{sec:anaexamples}, COMSOL Multiphysics version 5.6 \citep{COMSOL} is used for the optimisation studies. The ``Laminar Flow'' interface is used, where the mass conservation has been augmented using a ``Weak Contribution'' node. The ``Optimization'' module is used to set up the design field, objective and constraint functionals. The sensitivities are automatically calculated using COMSOL's built-in adjoint sensitivity analysis and symbolic differentiation. 

To solve the optimisation problem, COMSOL's implementation of the GCMMA method \citep{Svanberg2002} is used with a move limit of 0.2, an optimality tolerance of 0.01, a maximum of 4 inner iterations, and otherwise default settings. The maximum number of outer iterations and model evaluations depends on the example and will be stated for each example separately.

\subsection{Flow distribution problem} \label{sec:formulation_distribution}

\begin{figure}
    \begin{subfigure}{\columnwidth}
         \centering
         \includegraphics[width=0.99\textwidth]{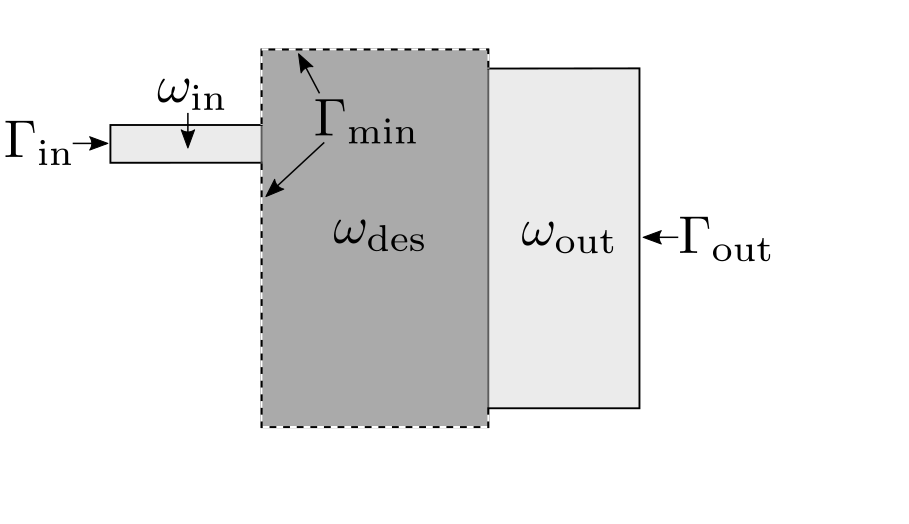}
         \caption{Domains and boundaries}
         \label{fig:example1_problem-a}
    \end{subfigure}
    \\
    \begin{subfigure}{\columnwidth}
         \centering
         \includegraphics[width=0.99\textwidth]{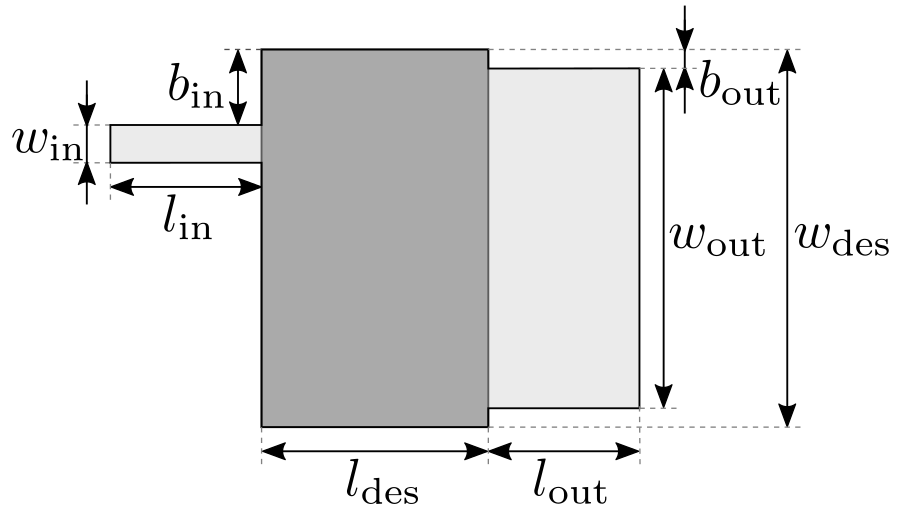}
         \caption{Dimensions}
         \label{fig:example1_problem-b}
    \end{subfigure}
    \caption{Problem setup for the flow distribution problem. Dimensions are given in Table \ref{tab:example1_dims}.}
    \label{fig:example1_problem}
\end{figure}
Figure \ref{fig:example1_problem} shows the setup for a flow distribution problem. The problem consists of a narrow inlet channel, where the flow enters at the left-most boundary, $\Gamma_\text{in}$. The inlet channel is connected to the flow distribution chamber, which also constitutes the design domain, $\omega_\text{des}$. Finally, the flow passes through the wide outlet channel and exits at the right-most edge, $\Gamma_\text{out}$. All other boundaries are no flow boundaries, $\bar{\uv} = 0$. The height in the design domain can vary between $h_\text{max}$ and $h_\text{min}$, whereas it is fixed to $h_\text{max}$ in the inlet and outlet domains, $\omega_\text{in}$ and $\omega_\text{out}$. Along the walls of the design domain, $\Gamma_\text{min}$ (dashed lines), the height is fixed to $h_\text{min}$. Figure \ref{fig:example1_problem-b} shows the dimensions of the domains, for which the values are listed in Table \ref{tab:example1_dims}.
\begin{table}
    \centering
    \begin{tabular}{c|c|c|c|c|c}
        $w_\text{in}$ & $l_\text{in}$ & $w_\text{des}$ & $l_\text{des}$ & $w_\text{out}$ & $l_\text{out}$ \\
        \hline
        5\text{ cm} & 20\text{ cm} & 50\text{ cm} & 30\text{ cm} & 5\text{ cm} & 45\text{ cm} \\
        \multicolumn{6}{c}{} \\
        \multicolumn{1}{c}{} & $b_\text{in}$ & $b_\text{out}$ & $h_\text{min}$ & \multicolumn{1}{c}{$h_\text{max}$} & \\
        \cline{2-5}
        \multicolumn{1}{c}{} & 10\text{ cm} & 2.5\text{ cm} & 3\text{ mm} & \multicolumn{1}{c}{5\text{ cm}} & 
    \end{tabular}
    \caption{Dimensions of the flow distribution problem shown in Figure \ref{fig:example1_problem-b}.}
    \label{tab:example1_dims}
\end{table}

The fluid enters the inlet, $\Gamma_\text{in}$, with a parabolic velocity distribution with a maximum velocity of $\bar{U}_\text{in}=5\text{ cm/s}$ and a zero reference pressure, $\bar{p}_\text{out} = 0\text{ Pa}$, is applied at the outlet, $\Gamma_\text{out}$. The fluid is considered to be air at $15^{o}\text{C}$ with a dynamic viscosity of $\mu = 1.802\text{ Pa\,s}$ and a density of $\rho = 1.225\text{ kg/m}^3$.

\subsubsection{Objective functional}

In order to improve the flow distribution at the outlet, the objective functional is defined as the standard deviation along the outlet boundary:
\begin{equation} \label{eq:objective}
    \Phi = \bar{u}_\text{std} = \sqrt{ \frac{1}{\lvert \Gamma_\text{out} \rvert} \int_{\Gamma_\text{out}} \bof{\bar{u}_{n} - \bar{u}_\text{avg}}^2 \,dL }
\end{equation}
where $\Gamma_\text{out}$ is the outlet boundary, $\bar{u}_n$ is the normal velocity and $\bar{u}_\text{avg}$ is the mean normal velocity:
\begin{equation} \label{eq:meanvelocity}
    \bar{u}_\text{avg} = \frac{1}{\lvert \Gamma_\text{out} \rvert} \int_{\Gamma_\text{out}} \bar{u}_{n} \,dL
\end{equation}
Due to the no-slip conditions at the edges of the outlet and the inherent boundary layers formed, the average and standard deviation is evaluated only for the middle 90\% in order to avoid the unchangeable boundary layer and its low velocities dominating the measures.

\subsubsection{Pressure drop constraint}

In order to control the pressure drop of the optimised solution, a constraint on the mean inlet pressure is applied:
\begin{equation} \label{eq:pressure_constraint}
    \Delta \bar{p} =  \frac{1}{\lvert \Gamma_\text{in} \rvert} \int_{\Gamma_\text{in}} \bar{p} \,dL
\end{equation}
This is possible because the outlet pressure is set to 0 for the model.

\subsubsection{Optimisation problem}

The final optimisation problem is posed as:
\begin{equation}\label{eq:optprob1}
\begin{split}
 \underset{\gamma}{\text{minimise:}} &\quad \Phi \\
\text{subject to:} & \quad \Delta \bar{p} \leq \bar{p}^{*} \\
& \quad 0 \leq \gamma \leq 1 \; \text{for } \x \in \omega_\text{des} \\
\text{with:}& \quad r(\vecr{s}(\gamma)),\gamma)= 0\\
\end{split}
\end{equation}
where $\bar{p}^{*}$ is the maximum pressure drop allowed, $\vecr{s} = \left\lbrace \vecr{u}, p \right\rbrace$ are the state variables and $r(\vecr{s}(\gamma),\gamma)$ is the weak form governing equations in residual form.

\subsection{Flow manifold problem} \label{sec:formulation_manifold}

This second will compare the presented \textit{topographical} model to the traditional \textit{topological} model for a \textit{topology} optimisation problem.

\begin{figure}
    \begin{subfigure}{\columnwidth}
         \centering
         \includegraphics[width=0.99\textwidth]{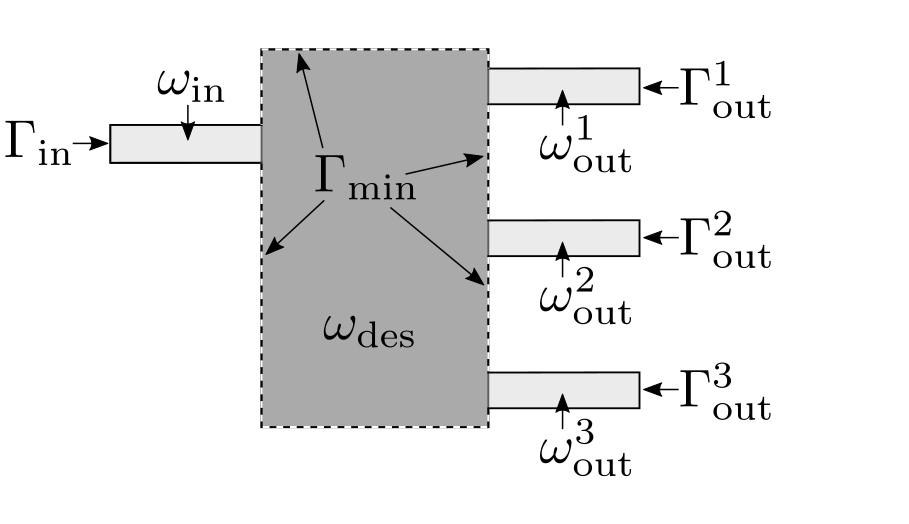}
         \caption{Domains and boundaries}
         \label{fig:example2_problem-a}
    \end{subfigure}
    \\
    \begin{subfigure}{\columnwidth}
         \centering
         \includegraphics[width=0.99\textwidth]{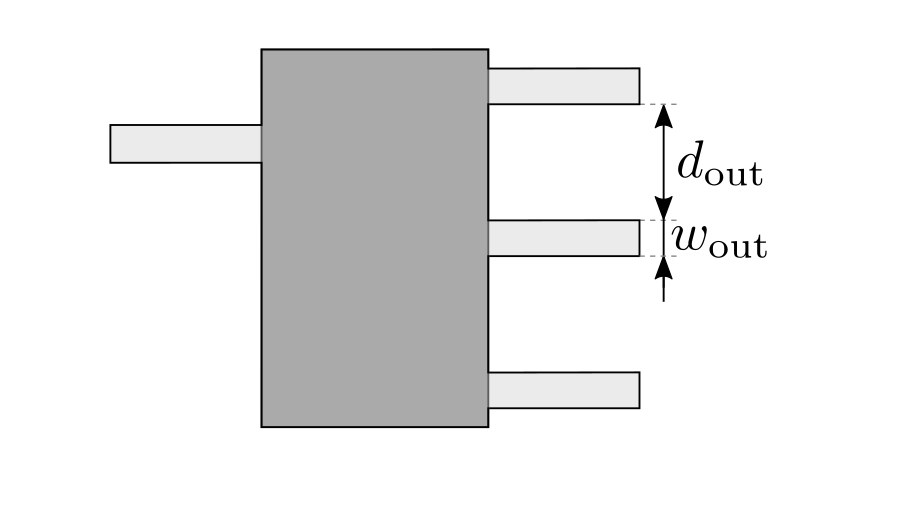}
         \caption{Dimensions}
         \label{fig:example2_problem-b}
    \end{subfigure}
    \caption{Problem setup for the flow manifold problem.}
    \label{fig:example2_problem}
\end{figure}
Figure \ref{fig:example2_problem} shows the setup for a flow manifold problem. The problem setup is identical to the previous example, except that the fluid now exits through three separate smaller outlets, $\Gamma^{i}_\text{out}$. The three outlets are identically dimensioned and are spaced equidistantly apart. Only the dimensions that vary from the previous example are shown in Figure \ref{fig:example2_problem-b} and are set to $w_\text{out} = 5\text{ cm}$; and $d_\text{out} = 15\text{ cm}$.

\subsubsection{Objective functional}

Contrary to the previous example, the optimisation will now seek to minimise the pressure drop of the optimised solution, as defined by Equation \ref{eq:pressure_constraint}: $\phi = \Delta \bar{p}$.

\subsubsection{Flow distribution constraints}

In order to control the distribution of the inlet flow among the three outlets of the manifold, constraints on the relative mass flow of each outlet is applied. The mass flow at the \textit{i}'th outlet is found using:
\begin{equation} \label{eq:distribution_constraints}
    \dot{m}^{i}_\text{out} = \int_{\Gamma^{i}_\text{out}} \rho \, \bar{u}_{n} \,dL
\end{equation}
where $\Gamma^{i}_\text{out}$ is the corresponding outlet. Each outlet will be restricted to be between a range of $\frac{1}{3}\dot{m}_\text{in} \pm \varepsilon$ where $\varepsilon$ is a small number and $\dot{m}_\text{in}$ is the mass flow of the inlet:
\begin{equation}
    \dot{m}_\text{in} = - \int_{\Gamma_\text{in}} \rho\, \bar{u}_{n} \,dL
\end{equation}

\subsubsection{Fluid area constraint}

In order to promote discrete \textit{topologies}, a constraint on the projected fluid area is introduced. This constraint is artificial, since it does not make sense physically in the context of the \textit{topographical} description. The projected used fluid area is defined as the integral of the design field over the design domain:
\begin{equation}
    A_{f} = \int_{\omega_\text{des}} \gamma \,dS
\end{equation}

\subsubsection{Optimisation problem}

The final optimisation problem is formally posed as:
\begin{equation}\label{eq:optprob2}
\begin{split}
 \underset{\gamma}{\text{minimise:}} &\quad \Delta \bar{p} \\
\text{subject to:} & \quad \dot{m}^{i}_{\text{out}} \leq \frac{1}{3}\dot{m}_\text{in} + \varepsilon \; \text{for } i \in \{ 1,2,3 \} \\
 & \quad \dot{m}^{i}_{\text{out}} \geq \frac{1}{3}\dot{m}_\text{in} - \varepsilon \; \text{for } i \in \{ 1,2,3 \} \\
 & \quad A_{f} \leq f_{a} A_\text{des} \\
 & \quad 0 \leq \gamma \leq 1 \; \text{for } \x \in \omega_\text{des} \\
\text{with:}& \quad r(\vecr{s}(\gamma)),\gamma)= 0\\
\end{split}
\end{equation}
where $f_a$ is the allowable fraction of the design domain area, $A_\text{des} = \int_{\omega_\text{des}} \,dS$.

\section{Optimisation results} \label{sec:results}

\subsection{Flow distribution problem} \label{sec:results_distribution}

The computational domain shown in Figure \ref{fig:example1_problem} is meshed using a regular quadratic mesh with elements of side length $2.5\text{ mm}$, with additional boundary layer refinement along the no-slip boundaries. This yields a total of 41,520 elements, 294,990 degrees-of-freedom and 25,197 design variables. The filter radius of the reaction-diffusion filter is set to $2.35\text{ cm}$, which should approximately satisfy a maximum thickness gradient of 1 as per Equation \ref{eq:maxgrad}. For the GCMMA optimiser, a maximum number of 150 outer iterations and 500 model evaluations is enforced\footnote{This optimisation problem seems to be quite difficult to solve, most likely due to the objective functional being very sensitive. For much of the optimisation history, 2-4 inner iterations are used. If less are taken, the optimisation procedure is very unstable and oscillatory.}.

\subsubsection{Single variable study}

In order to define some appropriate upper values for the pressure drop constraint, a single variable optimisation case is treated. The pressure drop and outlet velocity standard deviation is computed by varying the design field of the entire design domain, $\gamma_\text{des}$, simultaneously.
\begin{figure}
    \centering
    \includegraphics[width=0.95\columnwidth]{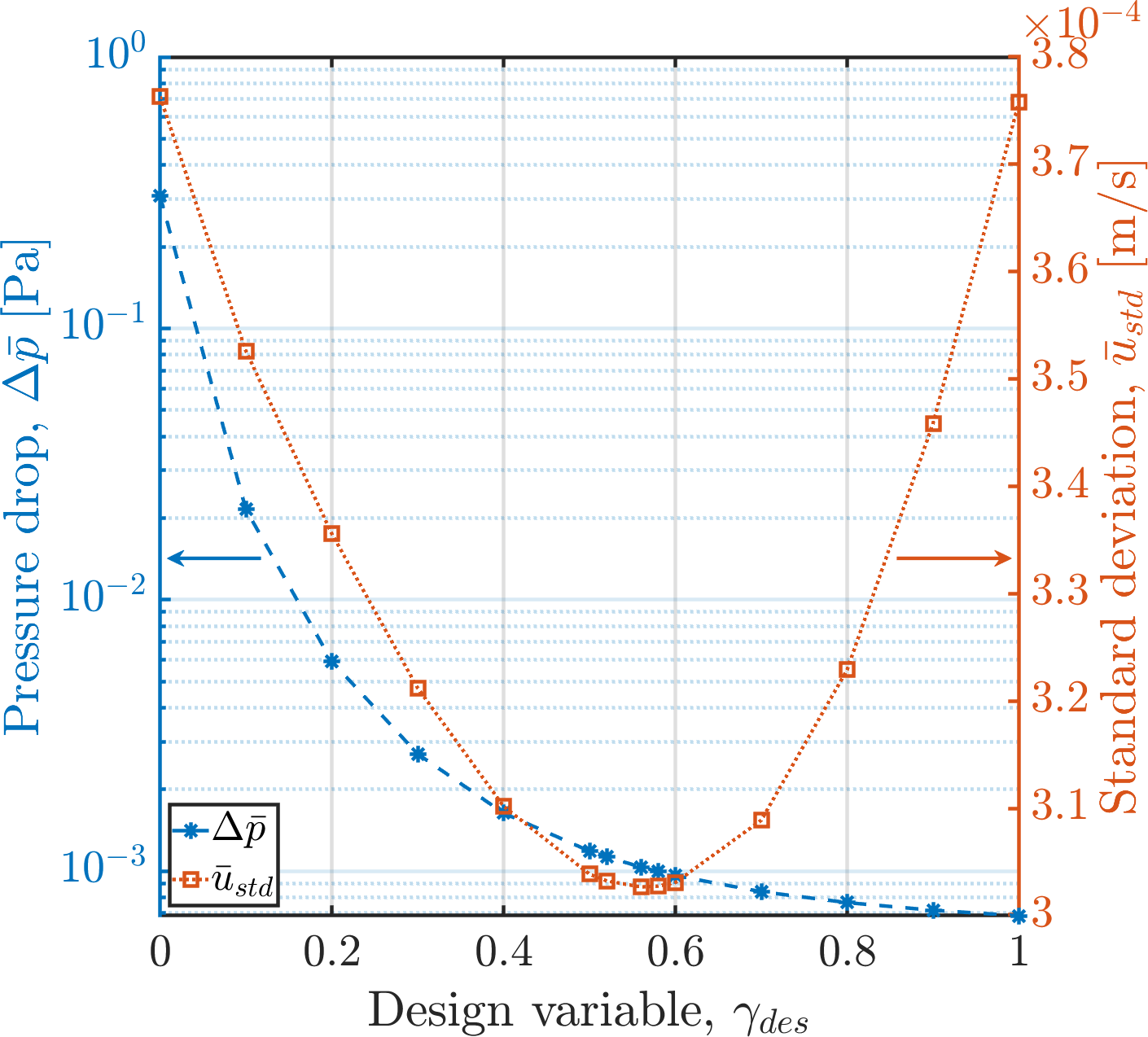}
    \caption{Results of the single variable case, where the spacing of the entire design domain is varied at once.}
    \label{fig:example1_singleVar}
\end{figure}
Figure \ref{fig:example1_singleVar} shows the two measures of interest for varying design field value. It can be seen that there is a clear optimal point due to the convex nature of the outlet velocity standard deviation with respect to the single design field value. This optimal point is around $\gamma_\text{des} = 0.56$ which is equivalent to $h(\gamma_\text{des}) = 2.932\text{ cm}$ over most of the design domain (except near the edges due to the boundary conditions enforced through the filtering process). On the other hand, as physically expected, the pressure drop decreases monotonously from the maximum value, $\Delta \bar{p}_\text{max} = 0.30822\text{ Pa}$, at the minimum height and the minimum value, $\Delta \bar{p}_\text{min} = 0.68437\text{ mPa}$, at the maximum height. The pressure drop at the approximate optimal point is $\Delta \bar{p}_\text{opt} = 1.0344\text{ mPa}$.

\subsubsection{Varying maximum pressure drop}

The optimisation problem, Equation \ref{eq:optprob1}, is now solved for a range of maximum allowable pressure drop, $\bar{p}^{*}$. The optimal single variable design is used as the initial design distribution, with the maximum pressure drop varying from the corresponding value down to the minimum value obtainable (maximum height everywhere, $\gamma_\text{des} = 1$). This is achieved by setting $\bar{p}^{*} = f_{\bar{p}} \Delta \bar{p}_\text{opt}$ using the following values for $f_{\bar{p}}$:
\begin{equation}
    f_{\bar{p}} \in \left\lbrace 1.01,0.9,0.8,0.7,0.66 \right\rbrace
\end{equation}

\begin{figure*}
    \begin{subfigure}{0.49\textwidth}
         \centering
         \includegraphics[width=0.99\textwidth]{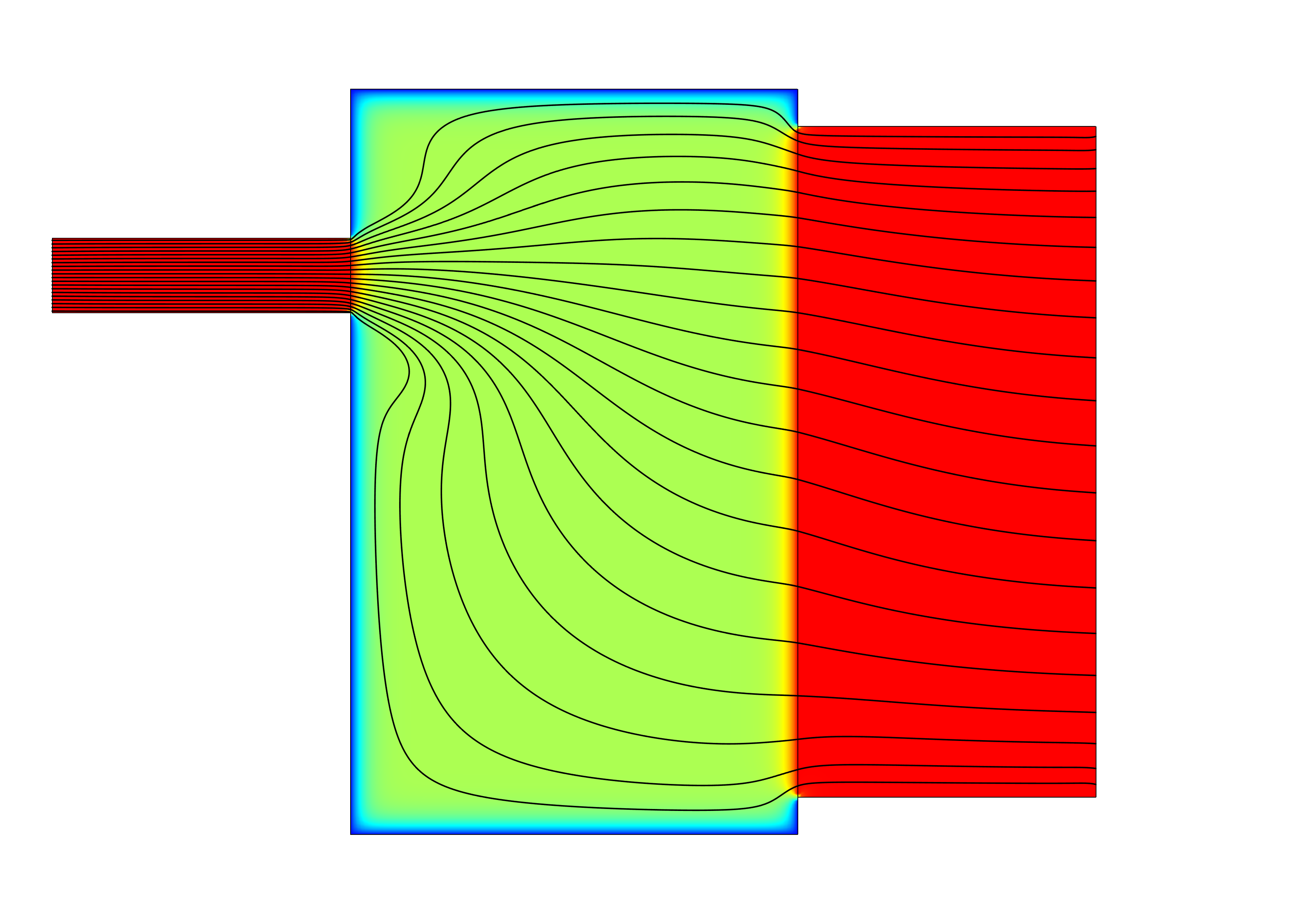}
         \caption{Initial}
         \label{fig:example1_thickOpt-000}
    \end{subfigure}
    \hfill
    \begin{subfigure}{0.49\textwidth}
         \centering
         \includegraphics[width=0.99\textwidth]{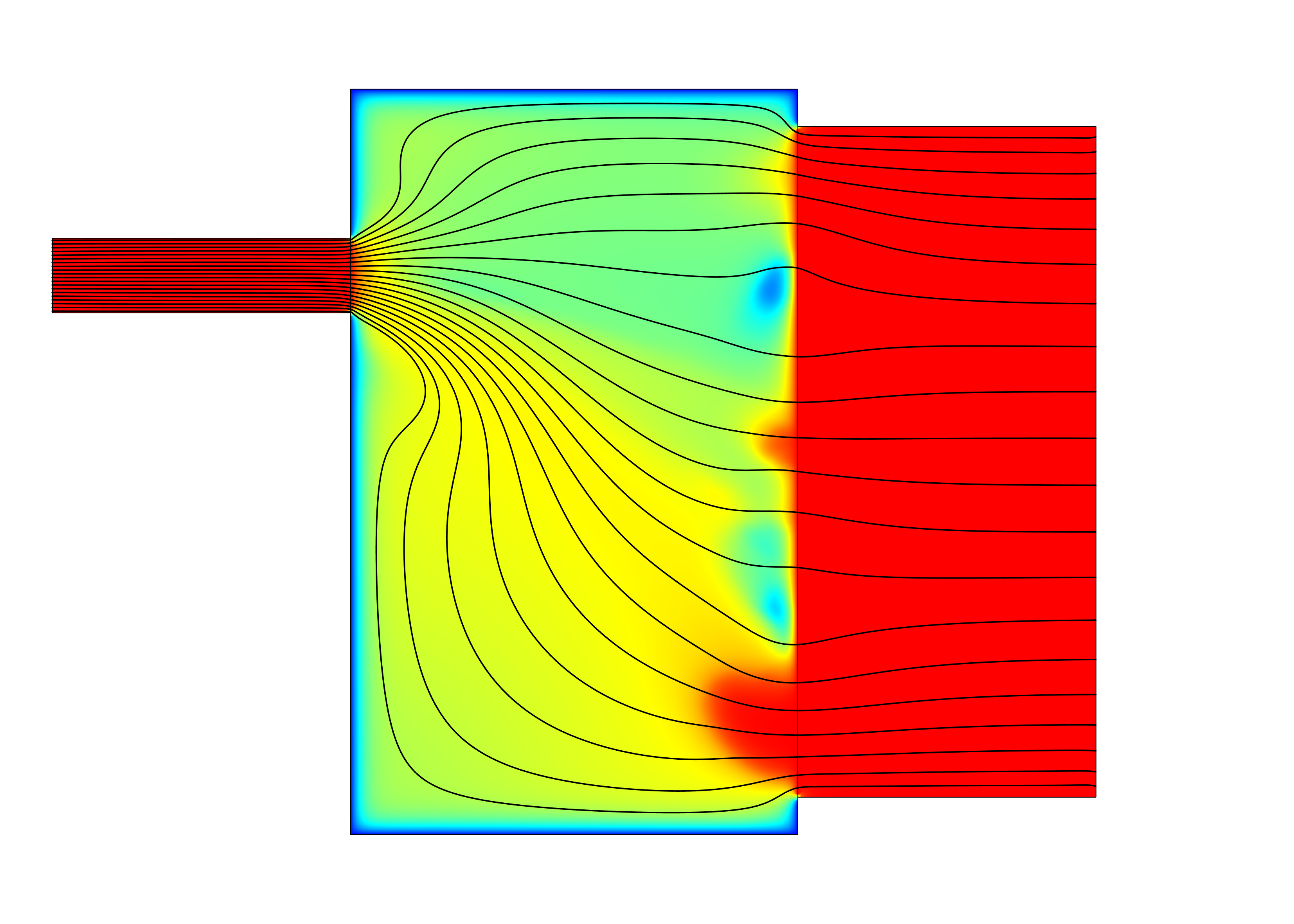}
         \caption{$f_{\bar{p}} = 1.01$}
         \label{fig:example1_thickOpt-101}
    \end{subfigure}
    \\
    \begin{subfigure}{0.49\textwidth}
         \centering
         \includegraphics[width=0.99\textwidth]{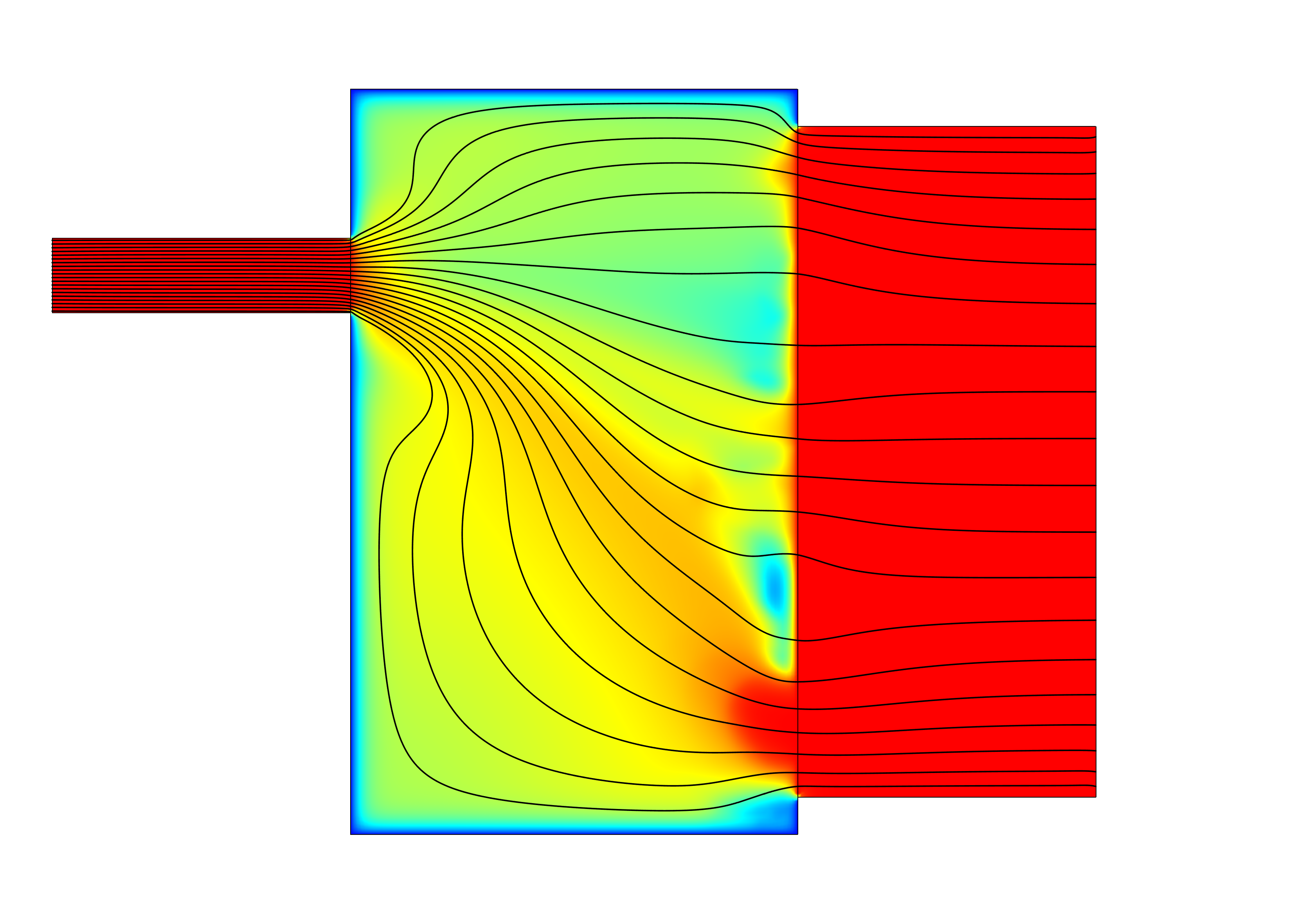}
         \caption{$f_{\bar{p}} = 0.9$}
         \label{fig:example1_thickOpt-090}
    \end{subfigure}
    \hfill
    \begin{subfigure}{0.49\textwidth}
         \centering
         \includegraphics[width=0.99\textwidth]{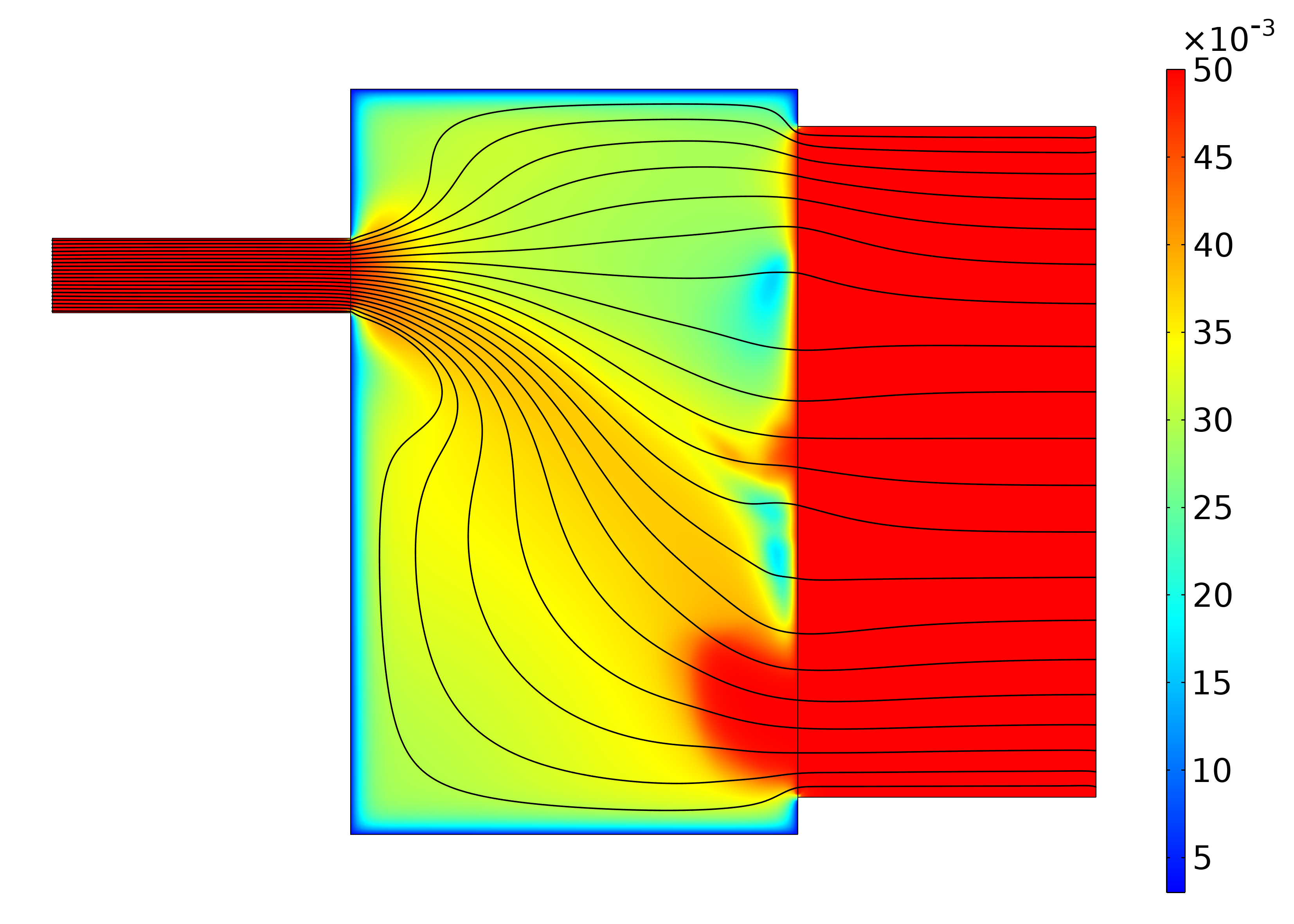}
         \caption{$f_{\bar{p}} = 0.8$}
         \label{fig:example1_thickOpt-080}
    \end{subfigure}
    \\
    \begin{subfigure}{0.49\textwidth}
         \centering
         \includegraphics[width=0.99\textwidth]{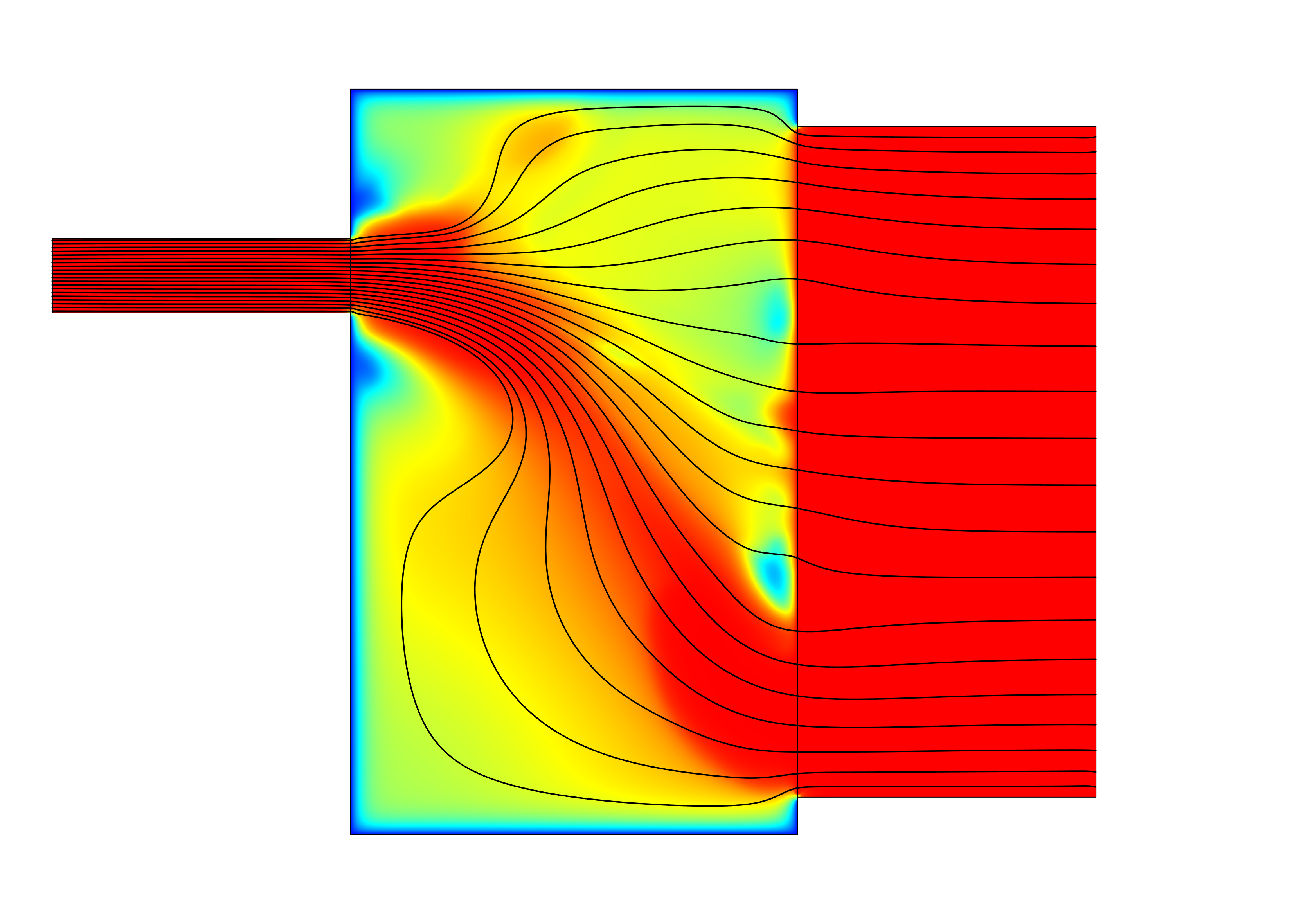}
         \caption{$f_{\bar{p}} = 0.7$}
         \label{fig:example1_thickOpt-070}
    \end{subfigure}
    \hfill
    \begin{subfigure}{0.49\textwidth}
         \centering
         \includegraphics[width=0.99\textwidth]{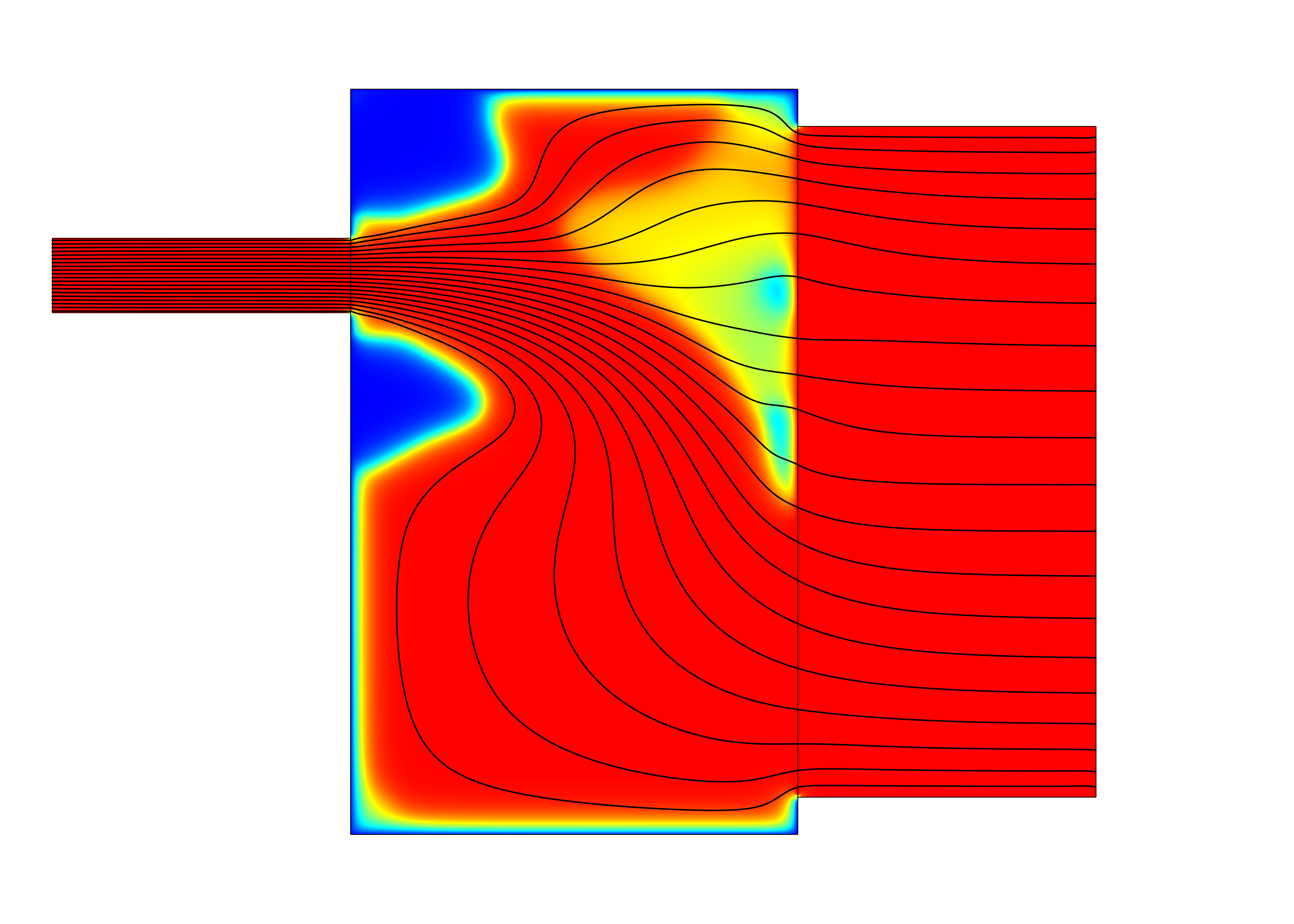}
         \caption{$f_{\bar{p}} = 0.65$}
         \label{fig:example1_thickOpt-065}
    \end{subfigure}
    \caption{Fluid channel height distributions for the flow distribution problem with different maximum allowable pressure drops. All subfigures use the same colour scale for the height in meters.}
    \label{fig:example1_thickOpt}
\end{figure*}
Figure \ref{fig:example1_thickOpt} shows the optimised topographies in terms of the fluid channel thickness. For all allowable pressure drops, it can be seen that the optimised designs have significantly varying spacing heights over the design domain. As the allowable pressure drop decreases, larger areas of the maximum height begin to appear. This is of course because the maximum height offers the lowest flow resistance. For the lowest pressure drop, the design is actually close to discrete, except some areas near the top right of the design domain, where intermediate heights are used to redistribute the flow.

\begin{figure}
    \centering
    \includegraphics[width=0.875\columnwidth]{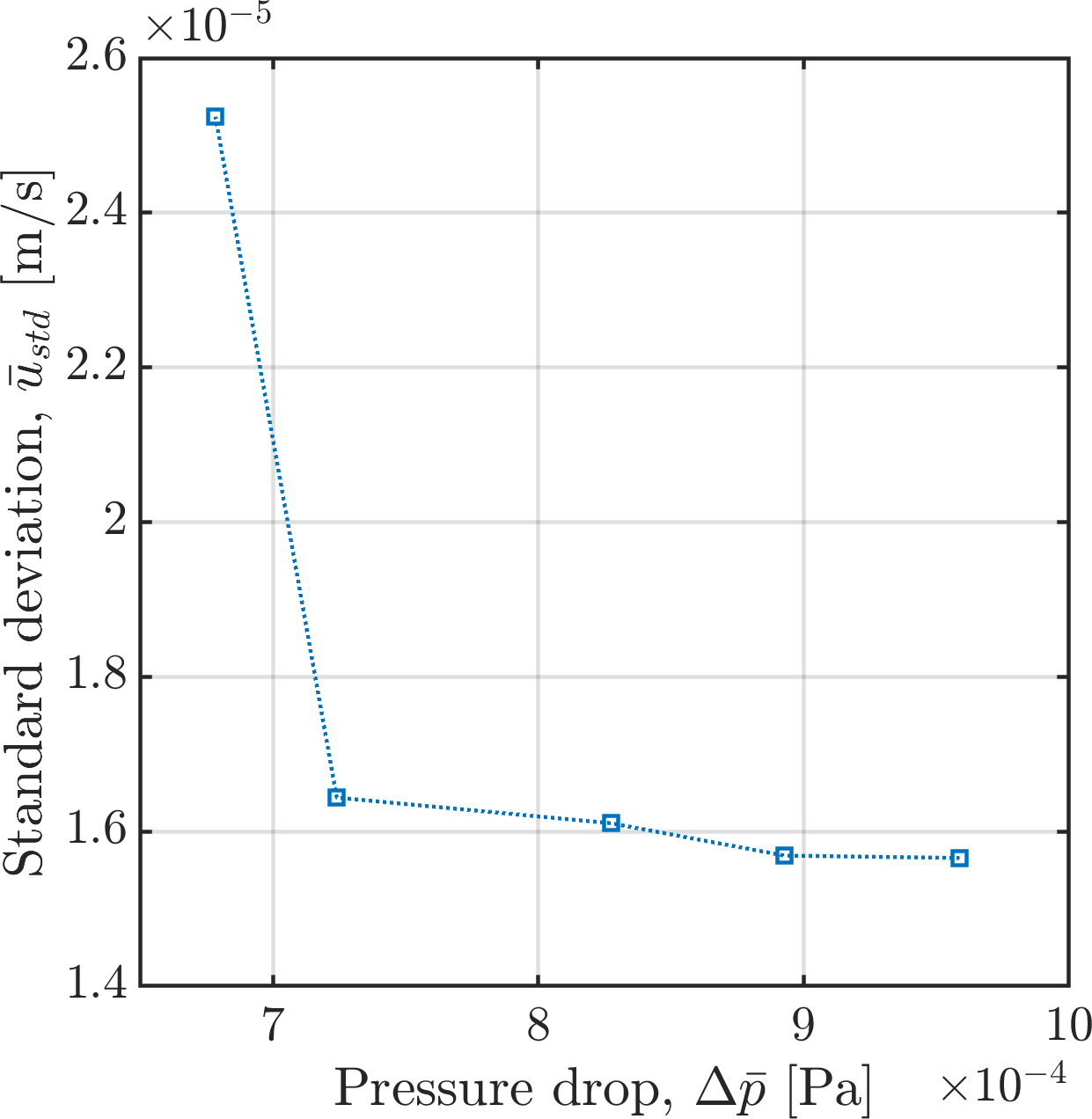}
    \caption{Performance of the optimised designs as a function of pressure drop.}
    \label{fig:example1_optPerf}
\end{figure}
Figure \ref{fig:example1_optPerf} shows the performance of the optimised designs as a function of their pressure drops. It can be seen that the lower the maximum allowable pressure drop, the worse the performance in terms of the standard deviation of the velocity distribution at the outlet. The effect is relatively weak with a significant spike for the lowest pressure drop. The performance of the optimised designs is up to 95\% better than the constant thickness case with a similar pressure drop, showing that a non-constant spacing and surface topography can have a huge effect on flow distribution.
The trend indicates that the higher the allowable pressure drop, the better the flow distribution. However, it has been observed (but not shown) that allowing an even higher pressure drop does not necessarily provide better performance.

\begin{figure*}
    \centering
    \,\,
    \begin{subfigure}{0.65\textwidth}
         \centering
         \includegraphics[width=0.99\textwidth]{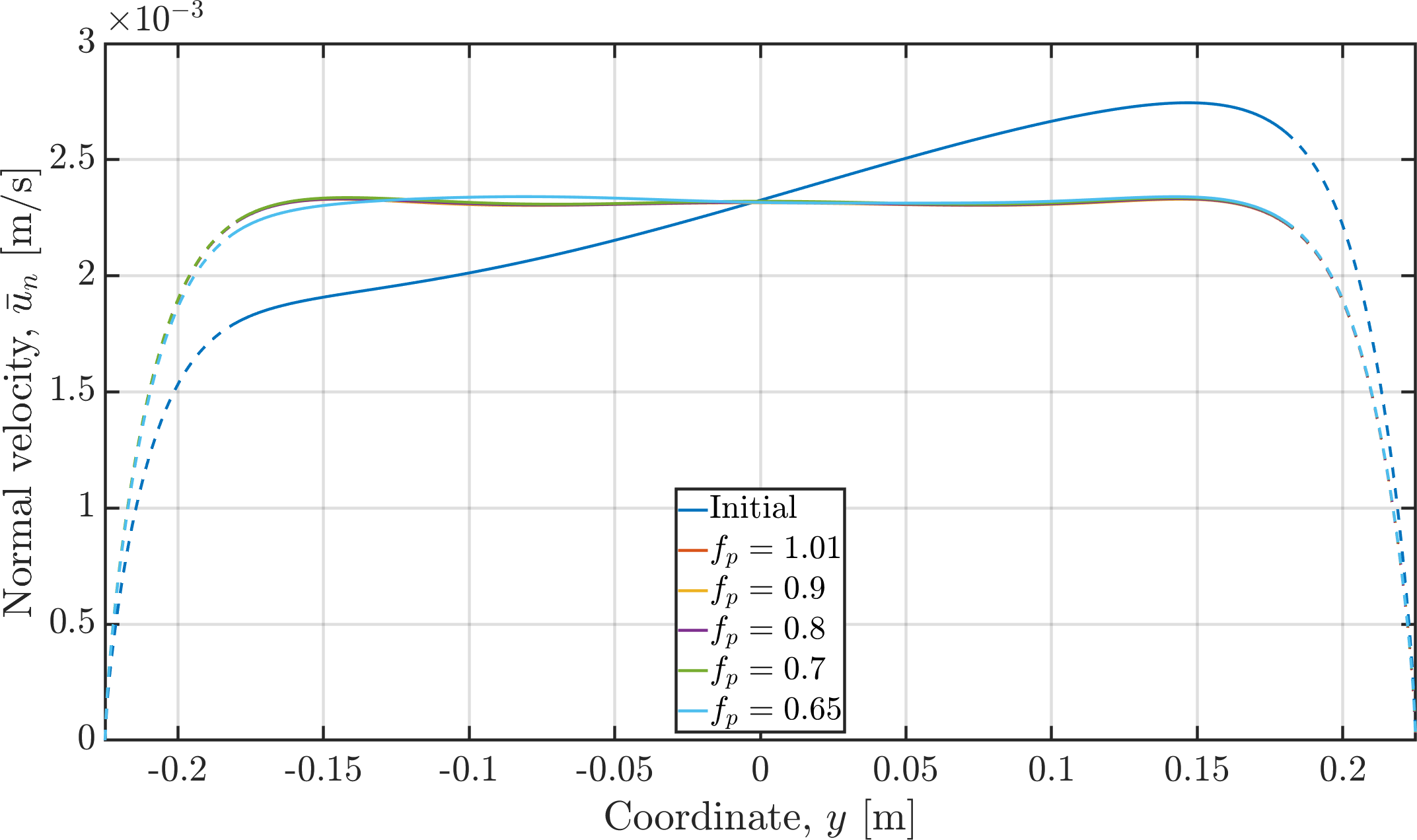}
         \caption{Full}
         \label{fig:example1_normvel-a}
    \end{subfigure}
    \\
    \begin{subfigure}{0.655\textwidth}
         \centering
         \includegraphics[width=0.999\textwidth]{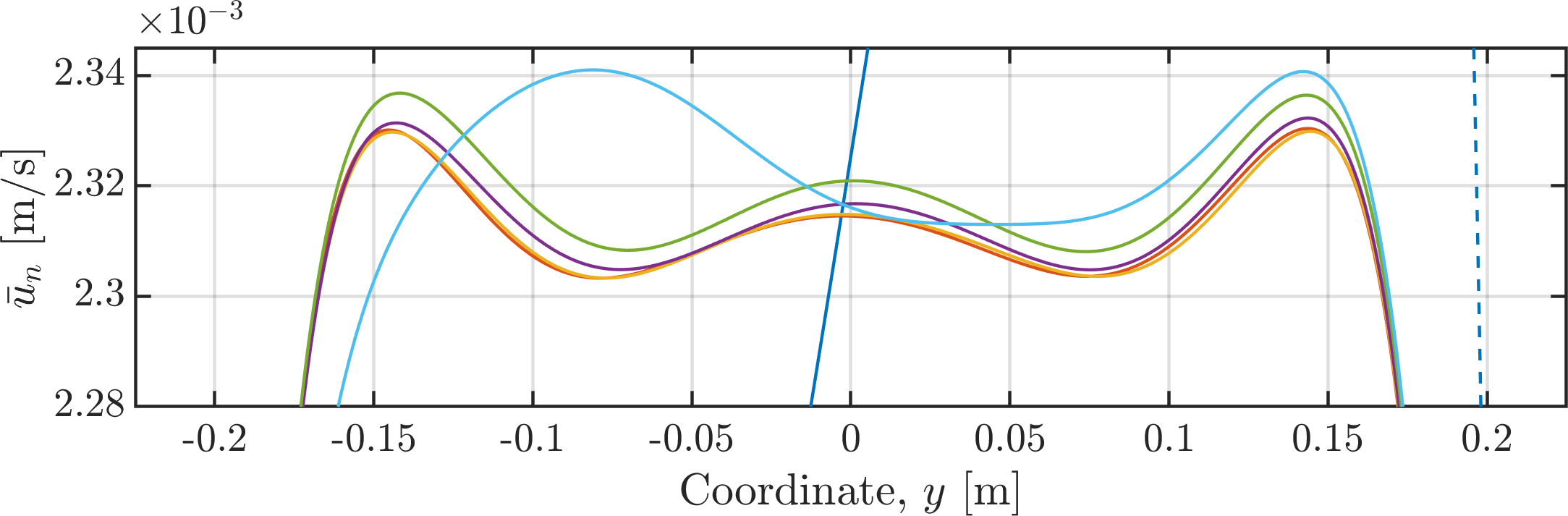}
         \caption{Zoom}
         \label{fig:example1_normvel-b}
    \end{subfigure}
    \caption{Normal velocity along the outlet boundary for the initial and optimised designs. The dotted lines represent the part of the outlet boundary not included in the evaluation of mean and standard deviation.}
    \label{fig:example1_normvel}
\end{figure*}
Figure \ref{fig:example1_normvel} shows the normal outlet velocity for the initial and optimised designs. From Figure \ref{fig:example1_normvel-a} it can be seen that all the optimised designs have a significantly more even flow distribution compared to the initial design with a constant thickness. Figure \ref{fig:example1_normvel-b} zooms in and shows the details for the optimised designs. Here it can be seen that the higher the pressure drop, the less evenly distributed the outlet flow - although the differences are marginal, as also seen from the numerical values in Figure \ref{fig:example1_optPerf}.

\subsubsection{Verification using full 3D model}

In order to further verify the accuracy of the plane two-dimensional model, the performance of the initial and two optimised designs are evaluated using a full three-dimensional model. For the optimised designs, $f_{\bar{p}} \in \left\lbrace 1.01, 0.7 \right\rbrace$ have been chosen as representative examples. It is important to note that the treated example is testing the limits of the formulated plane model, with a Reynolds number of $Re=100$ (see Figure \ref{fig:channel3D_errors-b}) and a height ratio of $\frac{h_\text{min}}{h_\text{max}} = 0.06$ (see Figure \ref{fig:channel3D_topoErrors}).

\begin{figure}
    \begin{subfigure}{\columnwidth}
         \centering
         \includegraphics[width=0.99\textwidth]{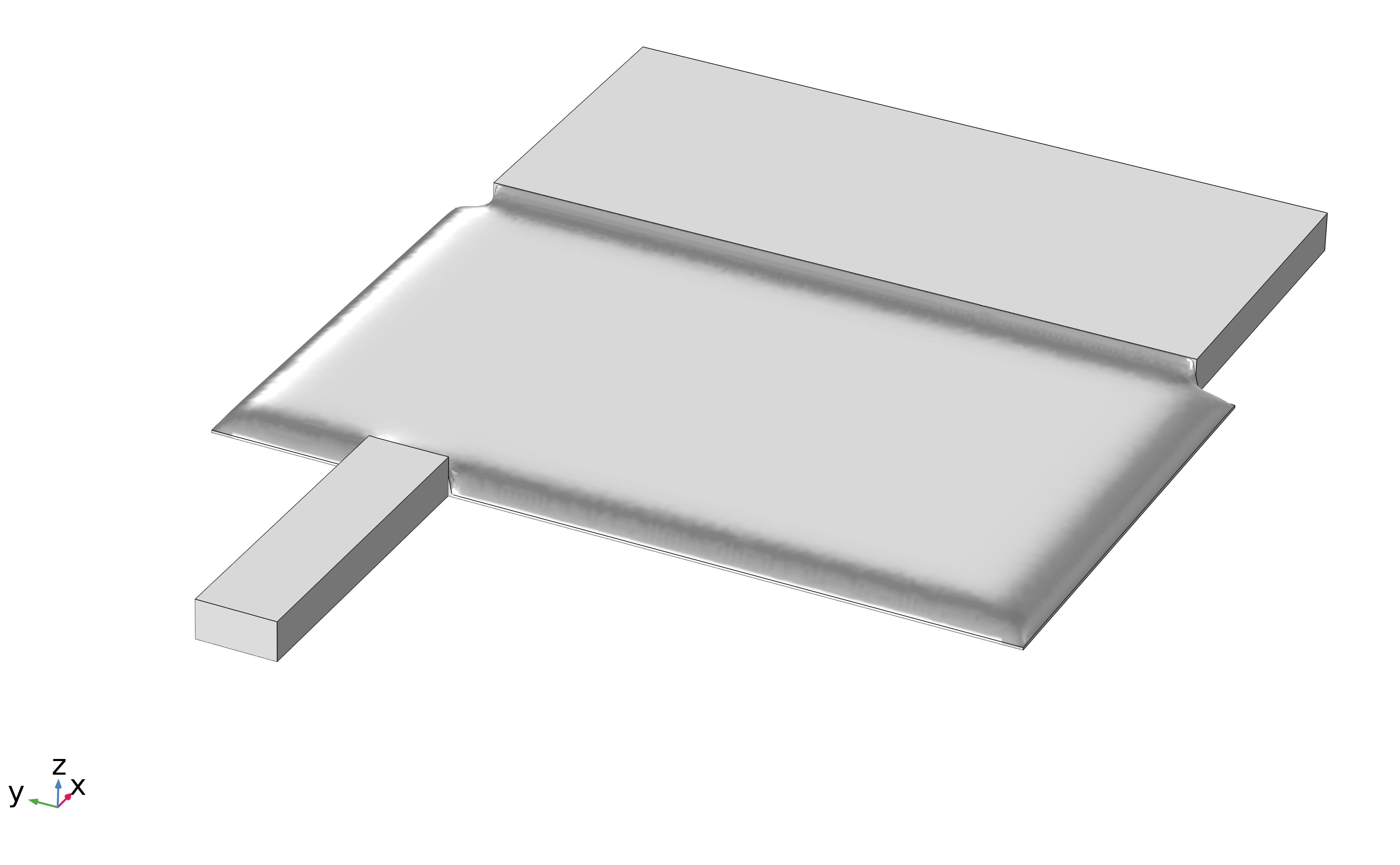}
         \caption{Initial}
         \label{fig:example1_geom3d-a}
    \end{subfigure}
    \\
    \begin{subfigure}{\columnwidth}
         \centering
         \includegraphics[width=0.999\textwidth]{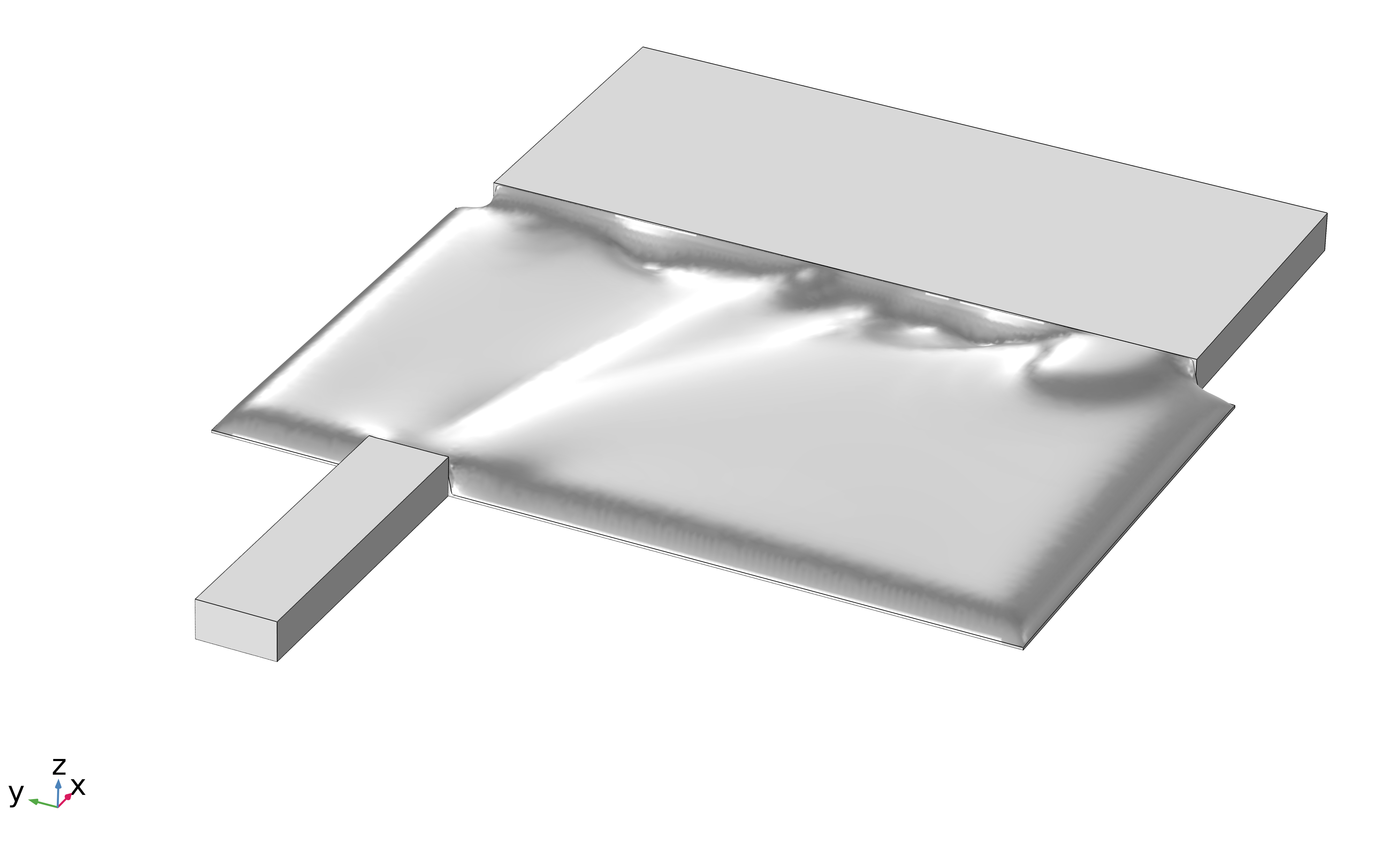}
         \caption{$f_{\bar{p}} = 1.01$}
         \label{fig:example1_geom3d-b}
    \end{subfigure}
    \\
    \begin{subfigure}{\columnwidth}
         \centering
         \includegraphics[width=0.999\textwidth]{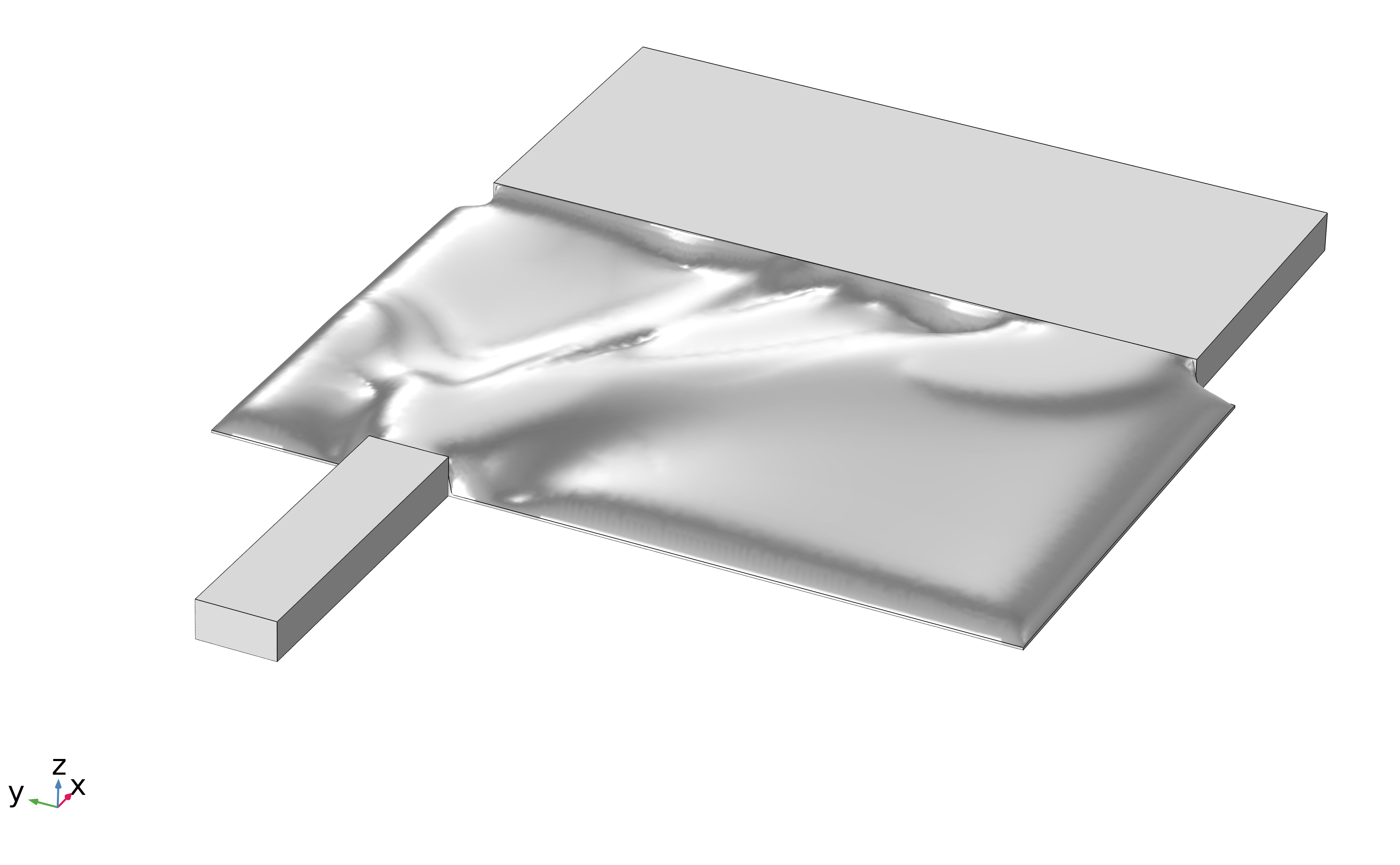}
         \caption{$f_{\bar{p}} = 0.7$}
         \label{fig:example1_geom3d-c}
    \end{subfigure}
    \caption{Three-dimensional geometries for the initial design and two optimised designs. Only the upper half of the total flow domain is shown, since the computational domain has been reduced to reduce computational cost.}
    \label{fig:example1_geom3d}
\end{figure}
Figure \ref{fig:example1_geom3d} shows the three-dimensional geometries for the initial design and the two chosen optimised designs. It is seen that the topography of the surface has been optimised by varying the fluid height over the design domain. For the lower maximum pressure drop it is evident that larger regions of maximum height and larger gradients are present.
The three-dimensional models are meshed with elements with a maximum size of 2 times that of the two-dimensional case. This results in between 700-800 thousand elements, primarily consisting of tetrahedral elements, with boundary layer meshes consisting of pyramids and prisms. This yields around 4.5 million degrees-of-freedom using second-order interpolation for velocity and first-order interpolation for pressure. The three-dimensional models are solved using the default algebraic multigrid preconditioned GMRES.

\begin{figure*}
    \centering
    \,\,
    \begin{subfigure}{0.65\textwidth}
         \centering
         \includegraphics[width=0.99\textwidth]{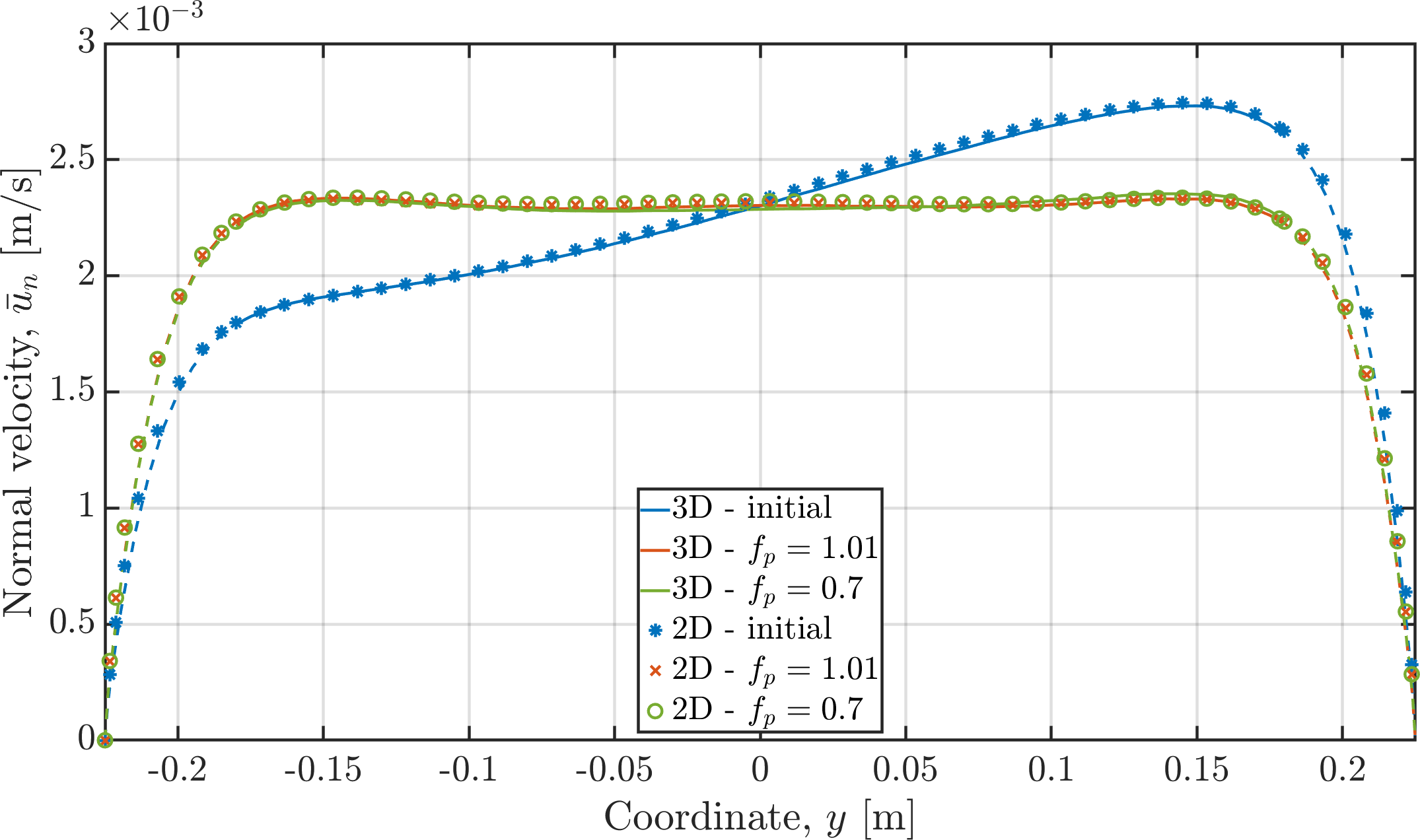}
         \caption{Full}
         \label{fig:example1_normvel3d-a}
    \end{subfigure}
    \\
    \begin{subfigure}{0.655\textwidth}
         \centering
         \includegraphics[width=0.999\textwidth]{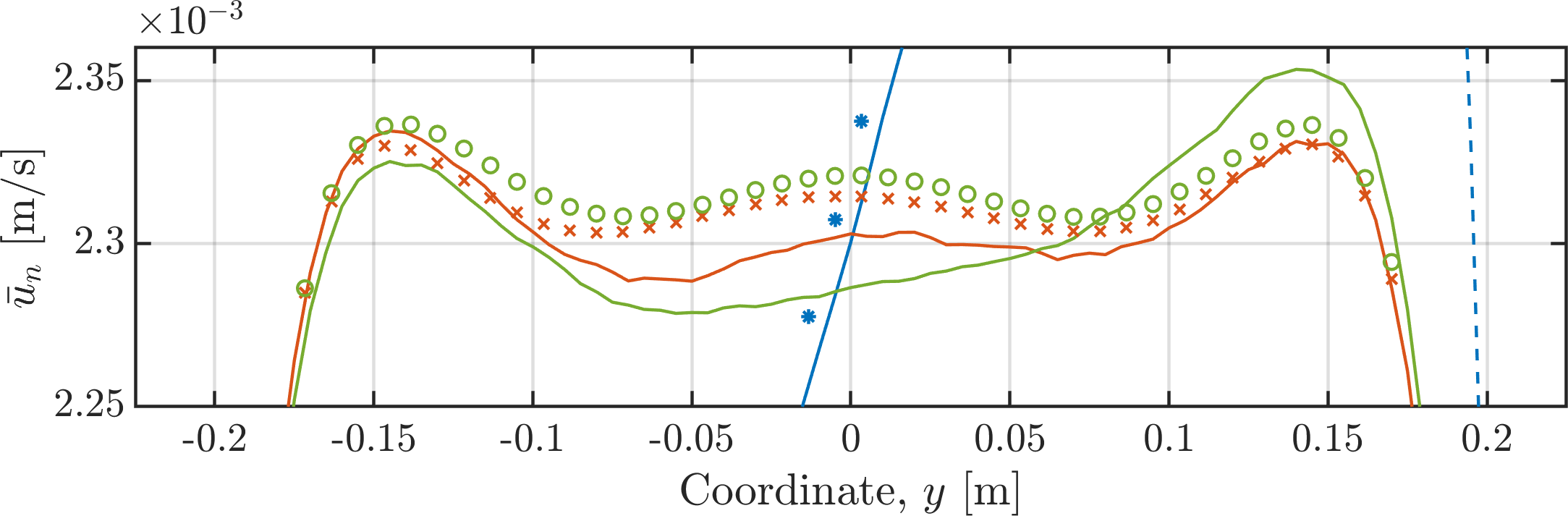}
         \caption{Zoom}
         \label{fig:example1_normvel3d-b}
    \end{subfigure}
    \caption{Normal velocity along the outlet boundary for the initial and two optimised designs comparing the full three-dimensional model to the plane two-dimensional model. The dotted lines represent the part of the outlet boundary not included in the evaluation of mean and standard deviation.}
    \label{fig:example1_normvel3d}
\end{figure*}
Figure \ref{fig:example1_normvel3d} shows the normal outlet velocity for the initial and two optimised designs. From Figure \ref{fig:example1_normvel3d-a} it can be seen that overall the agreement between the two models are quite good. However, Figure \ref{fig:example1_normvel3d-b} shows the details for the optimised designs and here some differences can be observed. Most significant is the difference for $f_{\bar{p}} = 0.7$, where the central peak is not present in the full three-dimensional result.

\begin{table}
    \centering
    \begin{tabular}{c||c|c|c||c|c|c}
        \multicolumn{1}{c||}{} & \multicolumn{3}{c||}{$\bar{u}_\text{std} [\times10^{-5}\text{ m/s}]$} & \multicolumn{3}{c}{$\Delta p [\times10^{-4}\text{ Pa}]$} \\
        Design & 2D & 3D & \% & 2D & 3D & \% \\
        \hline
        Initial & 30.27 & 29.63 & 2.2 & 8.275 & 8.768 & -5.6 \\
        $f_{\bar{p}} = 1.01$ & 1.566 & 1.761 & -11.1 & 7.671 & 7.993 & -4.0 \\
        $f_{\bar{p}} = 0.7$ & 1.644 & 2.462 & -33.2 & 5.792 & 5.899 & -1.8
    \end{tabular}
    \caption{Measures of interest computed using both a full three-dimensional model and the plane two-dimensional model for the flow distribution problem for the initial and two optimised designs.}
    \label{tab:example1_3d}
\end{table}
To further assess the accuracy, Table \ref{tab:example1_3d} shows the measures of interest of the optimisation problem, namely the standard deviation of the normal velocity along the central line of the outlet, $\bar{u}_\text{std}$, and the pressure drop in the three-dimensional scale\footnote{Please note that $\bar{p} = \frac{5}{4}p$ is the scaled pressure arising from the derivation process for Equations \ref{eq:aug_navierstokes} and, therefore, $\Delta p = \frac{4}{5} \Delta \bar{p}$.}, $\Delta p$. 
Before discussing the accuracy, it should be repeated that the treated example is testing the limits of the formulated plane model.
The error for the pressure drop prediction is rather small with the largest deviation of $-5.6\%$ for the initial design. Unfortunately, the error for the velocity standard deviation is rather large, with a $-33.2\%$ error for the lowest pressure drop design. This large error is due to the peak in the normal velocity for the two-dimensional model as seen in Figure \ref{fig:example1_normvel3d-b}. This peak reduces the standard deviation along the outlet, whereas the full three-dimensional model has a rather large deviation from the mean at the centre due to the large valley.

\begin{figure}
    \begin{subfigure}{\columnwidth}
         \centering
         \includegraphics[width=0.99\textwidth]{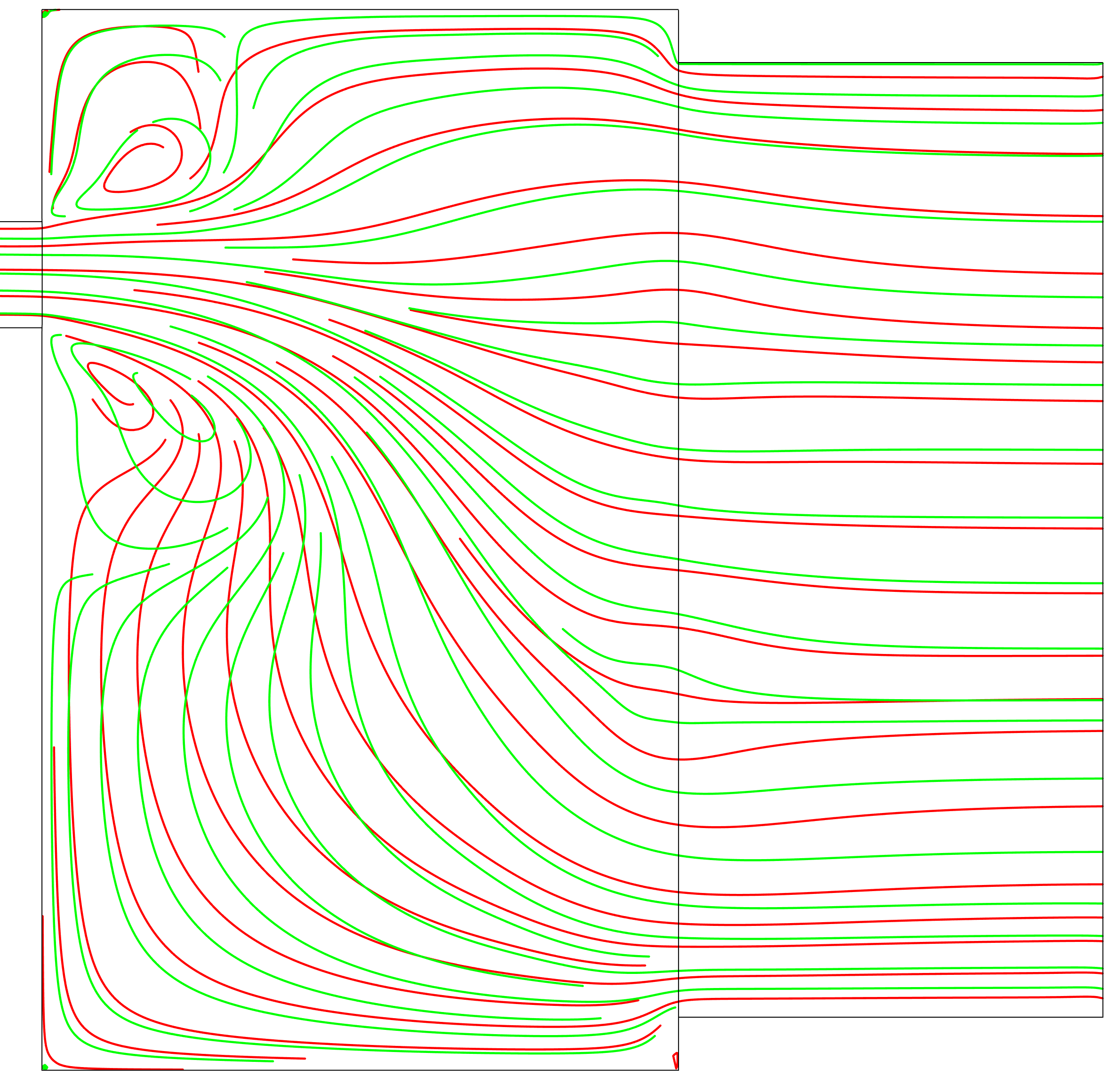}
         \caption{Central plane}
         \label{fig:example1_2Dvs3D-2D}
    \end{subfigure}
    \\
    \begin{subfigure}{\columnwidth}
         \centering
         \includegraphics[width=0.999\textwidth]{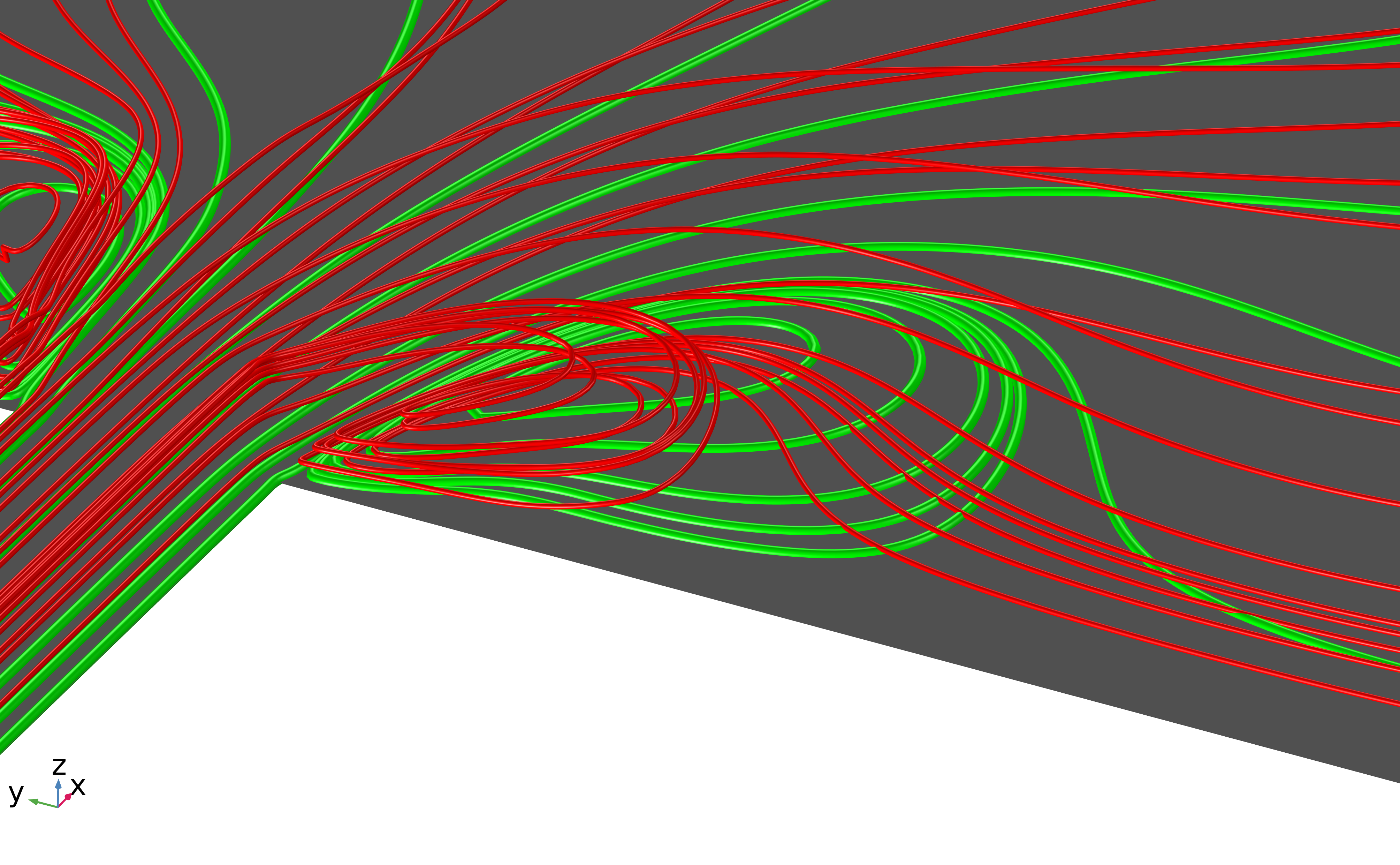}
         \caption{3D zoom}
         \label{fig:example1_2Dvs3D-3D}
    \end{subfigure}
    \caption{Streamline comparison for the full three-dimensional model (red) and the plane two-dimensional model (green): (a) streamlines along the central plane only; (b) three-dimensional streamlines.}
    \label{fig:example1_2Dvs3D}
\end{figure}
Figure \ref{fig:example1_2Dvs3D} compares the streamlines of the two- and three-dimensional models. Figure \ref{fig:example1_2Dvs3D-2D} shows the streamlines in the central plane only, for the three-dimensional model the out-of-plane velocities have been ignored and the two-dimensional model naturally exists in this plane. Overall a strong agreement is observed when comparing the central plane streamlines. However, a significant difference is observed just below where the inlet channel meets the design domain. The in-plane re-circulation zone is predicted to be much larger in the two-dimensional model than compared to the three-dimensional model. On the other hand, Figure \ref{fig:example1_2Dvs3D-3D} shows the three-dimensional streamlines in this region and it can be seen that there are three-dimensional flow effects in this region. From Figure \ref{fig:example1_geom3d-c}, it can be seen that this region is one of significant height change and therefore a large change in geometry. Since the two-dimensional model is not capable of capturing such three-dimensional effects due to the basic assumption of in-plane flow only, this is likely why it predicts a larger degree of separation and re-circulation.

\subsection{Flow manifold problem} \label{sec:results_manifold}

The computational domain shown in Figure \ref{fig:example2_problem} is meshed using a regular quadratic mesh with elements of side length $2.5\text{ mm}$, with additional boundary layer refinement along the no-slip boundaries. This yields a total of 32,800 elements, 331,732 degrees-of-freedom and 25,433 design variables. The filter radius of the reaction-diffusion filter is set to $6.0\text{ mm}$, which is the minimum allowable to ensure a stable solution (2.4 times element size). This minimum size is imposed since a topological solution is sought for this problem, with as clear a definition of the boundary as possible. The allowable fluid area is determined by the fraction $f_{a} = 0.30$. For the GCMMA optimiser, a maximum number of 100 outer iterations and 150 model evaluations is enforced\footnote{This problem seems smoother than the previous, since only 1 inner iteration is consistently used throughout the optimisation procedure for all parameter cases.}.

\subsubsection{Decreasing minimum thickness}  \label{sec:topogToTopol_optimisation}

The flow manifold problem will be optimised for a range of decreasing minimum height:
$h_\text{min} \in \left\lbrace 2.5, 1.0, 0.5, 0.1,\right.$ $\left. 0.01 \right\rbrace \text{cm}$.
This is done in order to observe the convergence of the \textit{topographical} model towards a \textit{topological} problem, both in the design distribution and the design performance. 

\begin{figure}
    \centering
    \includegraphics[width=0.9\columnwidth]{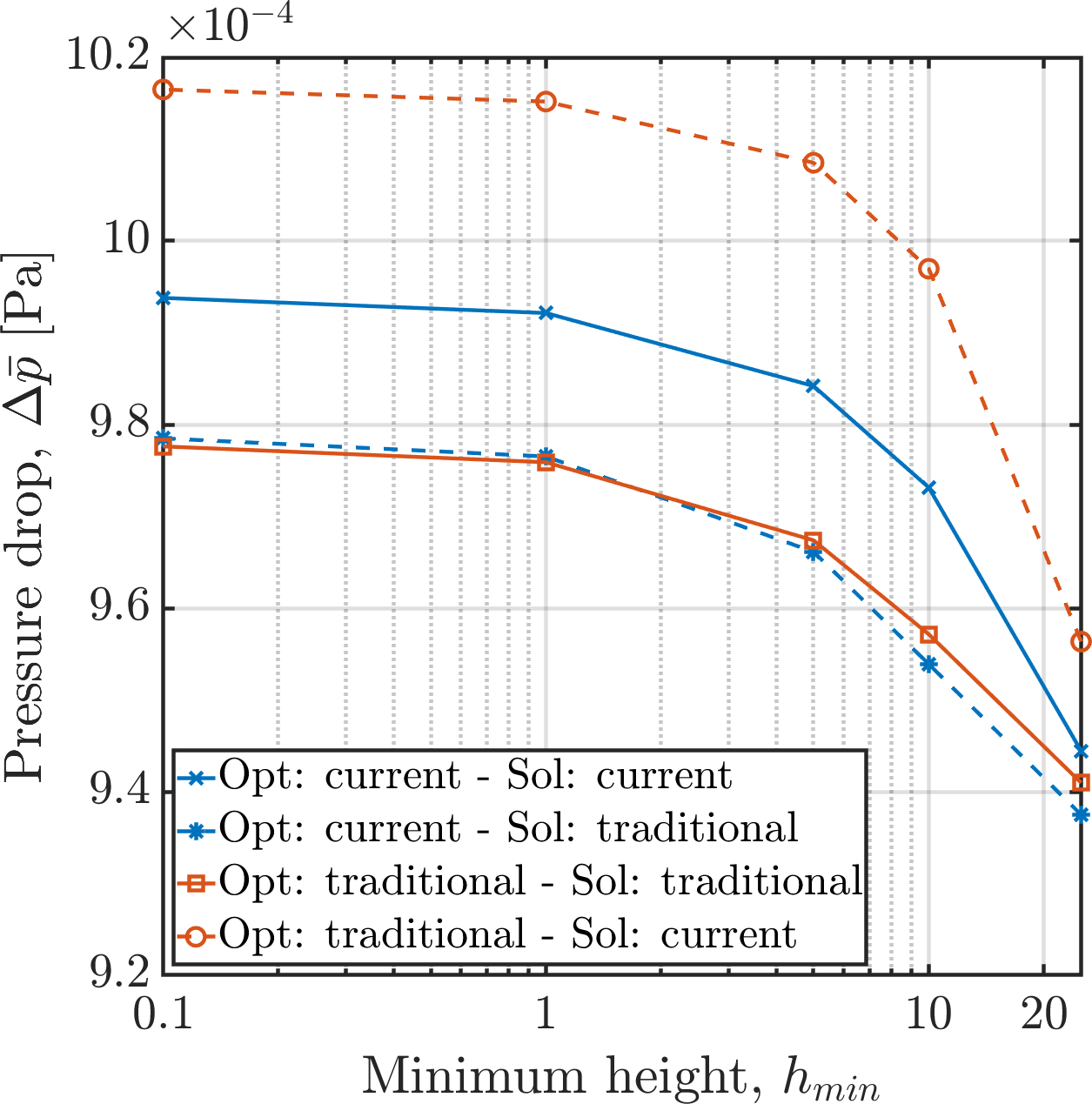}
    \caption{Pressure drop for optimised designs with varying minimum height using both the proposed model and the traditional model, evaluated using both models in a cross-check. The blue lines are optimised using the proposed model and the orange lines using the traditional model. Full lines show the performance evaluated using the same model as optimised with. Dashed lines show the performance evaluated using the other model than optimised with.}
    \label{fig:example2_opt_pres}
\end{figure}
Figure \ref{fig:example2_opt_pres} shows the optimised pressure drop as a function of the minimum fluid channel height. It can be seen that generally the traditional model predicts lower pressure drops than the proposed model. Furthermore, when evaluated using the proposed model, it is clear that designs optimised using the same model performs significantly better than when optimised using the traditional model. Lastly, it is observed that when evaluated using the traditional model, the designs optimised using the traditional model actually only begin to (marginally) outperform those optimised using the proposed model for very small minimum heights, $h_\text{min} \leq 1\text{ mm}$. That means that the proposed topographical model outperforms the traditional model for relatively large minimum heights, even when evaluated using the traditional model. This may well be due to the significant difference in the amount of intermediate design field values present.

\begin{figure*}
    \begin{subfigure}{0.19\textwidth}
         \centering
         \includegraphics[height=1.5\textwidth]{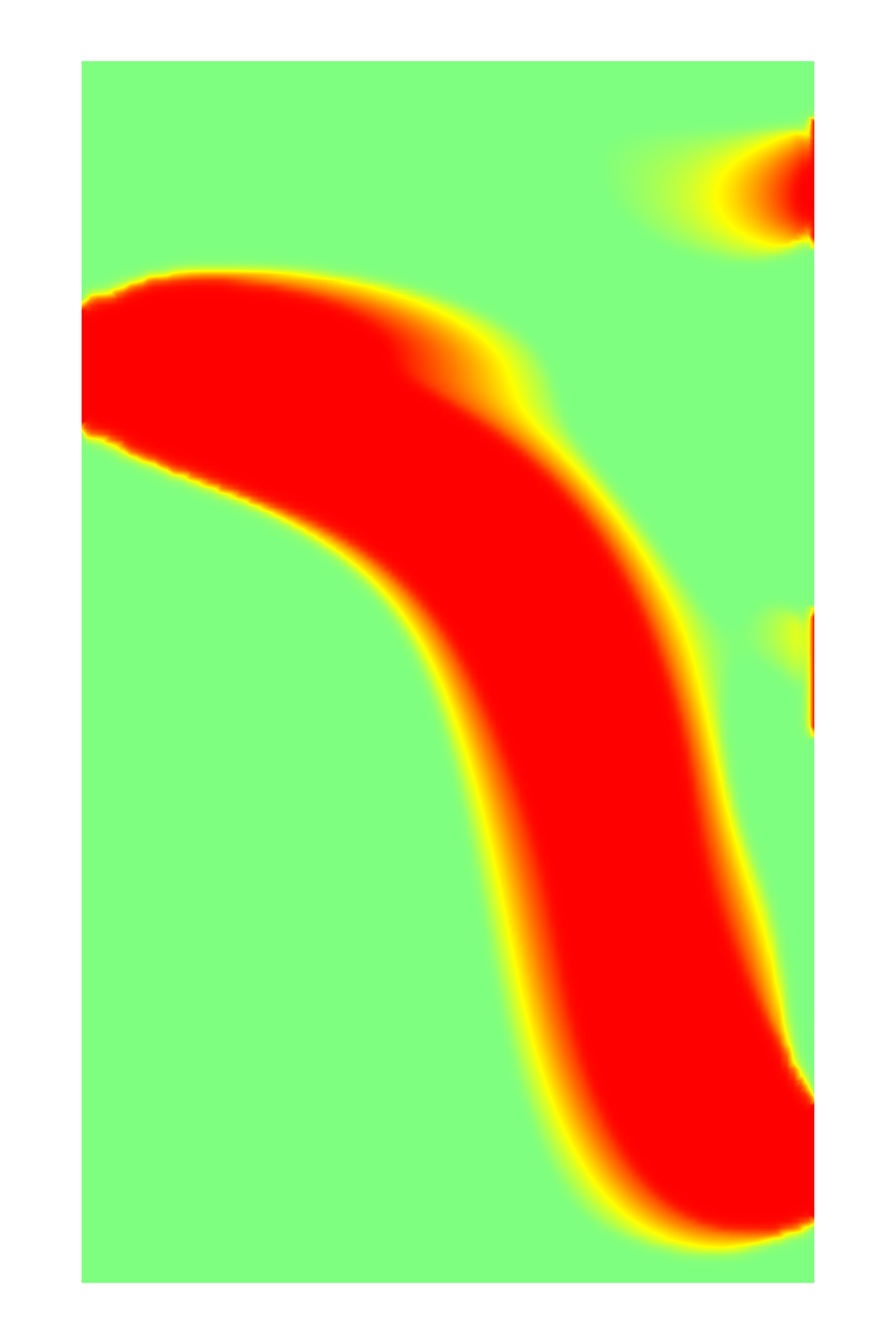}
         \caption{}
         \label{fig:example2_optDes-a}
    \end{subfigure}
    \begin{subfigure}{0.19\textwidth}
         \centering
         \includegraphics[height=1.5\textwidth]{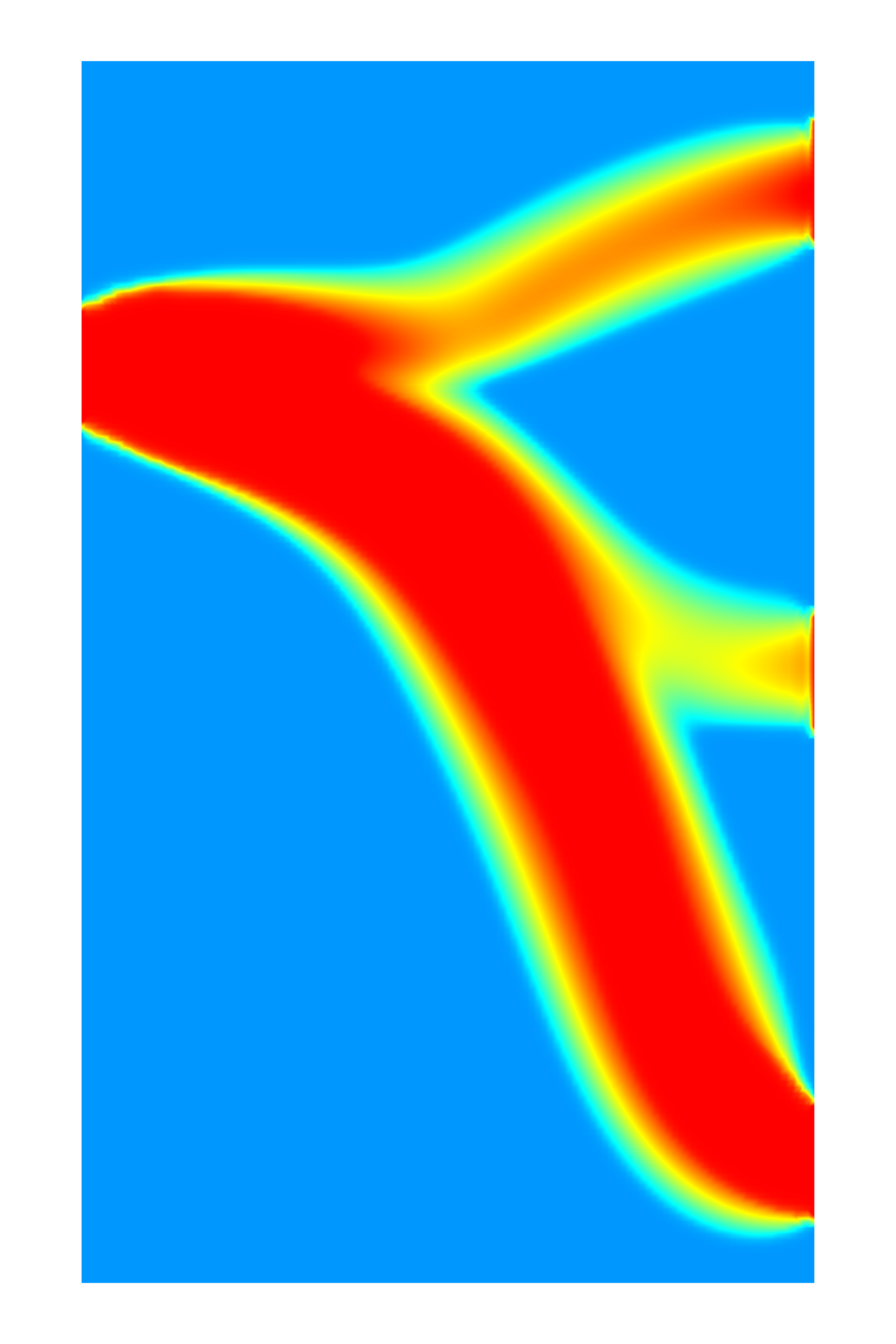}
         \caption{}
         \label{fig:example2_optDes-b}
    \end{subfigure}
    \begin{subfigure}{0.19\textwidth}
         \centering
         \includegraphics[height=1.5\textwidth]{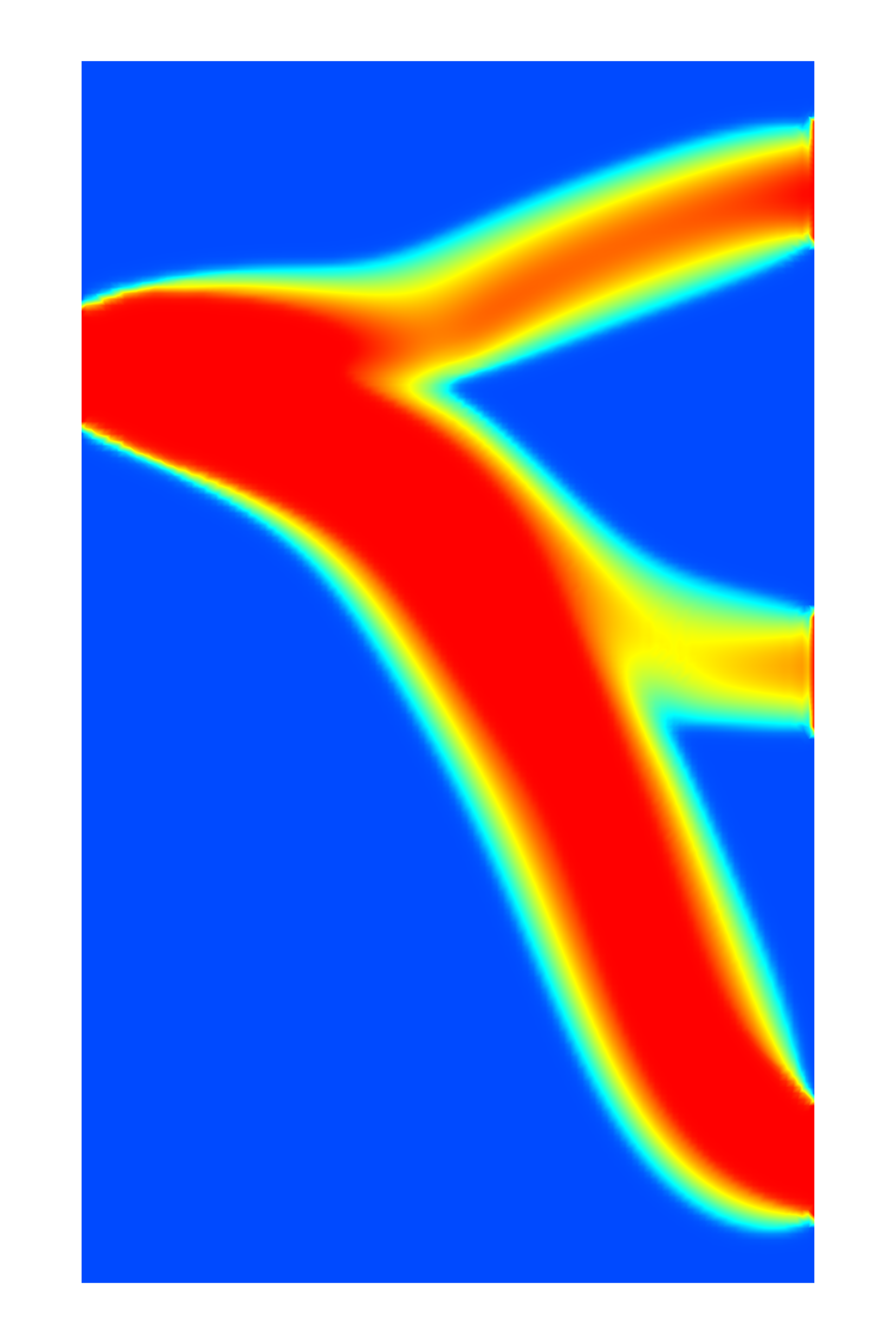}
         \caption{}
         \label{fig:example2_optDes-c}
    \end{subfigure}
    \begin{subfigure}{0.19\textwidth}
         \centering
         \includegraphics[height=1.5\textwidth]{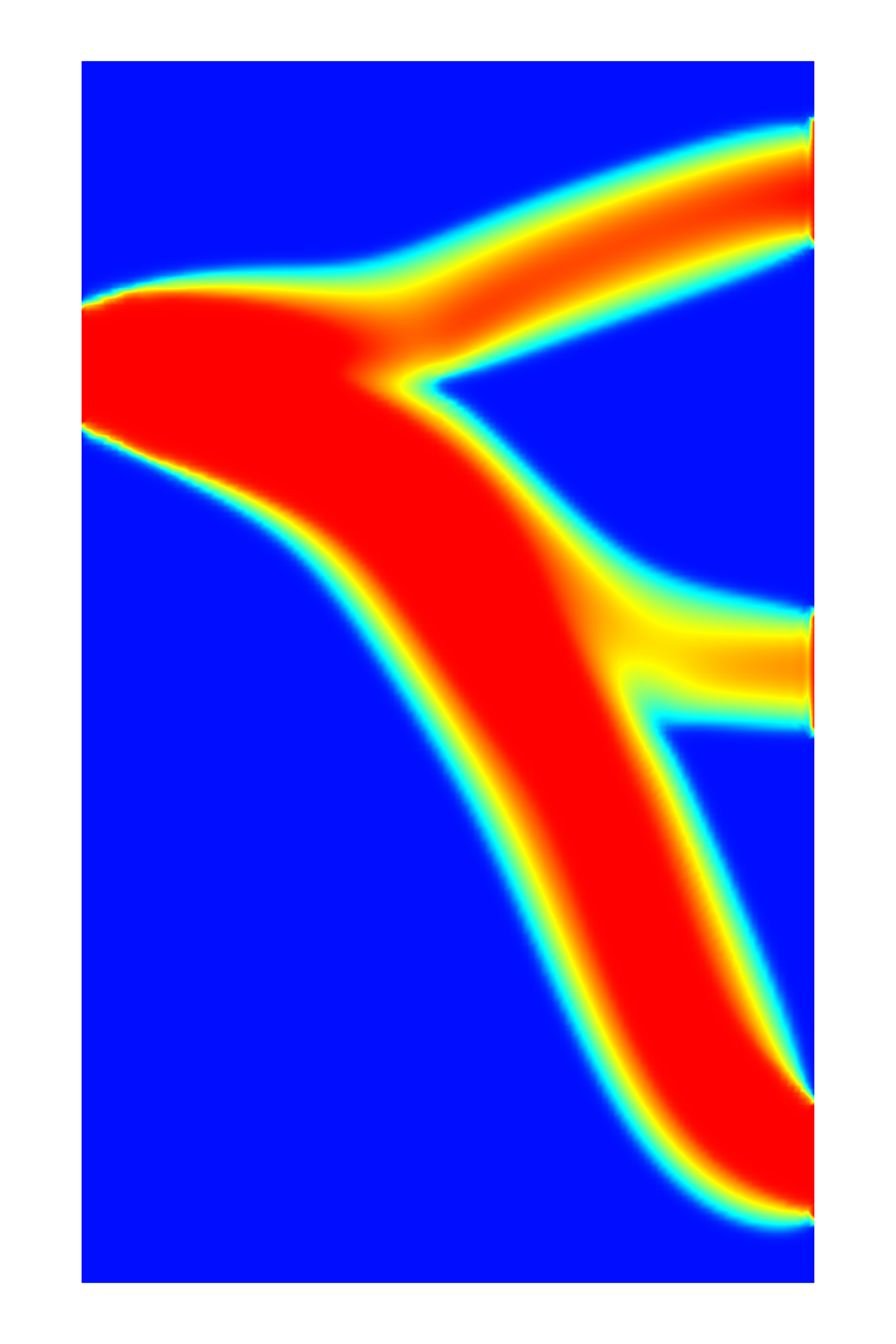}
         \caption{}
         \label{fig:example2_optDes-d}
    \end{subfigure}
    \begin{subfigure}{0.19\textwidth}
         \centering
         \includegraphics[height=1.5\textwidth]{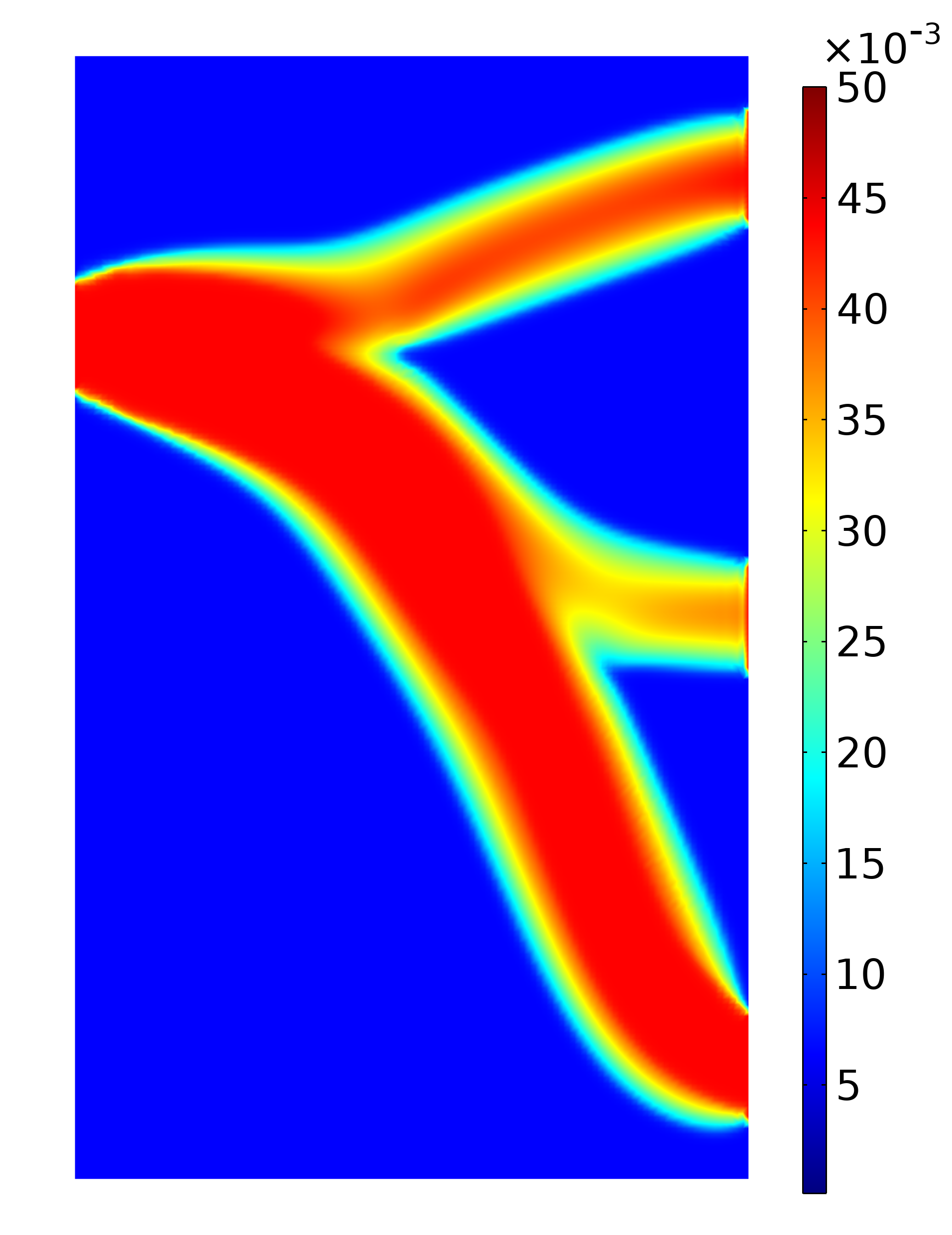}
         \caption{}
         \label{fig:example2_optDes-e}
    \end{subfigure}
    \\ \vspace{0.5cm}
    \begin{subfigure}{0.19\textwidth}
         \centering
         \includegraphics[width=\textwidth]{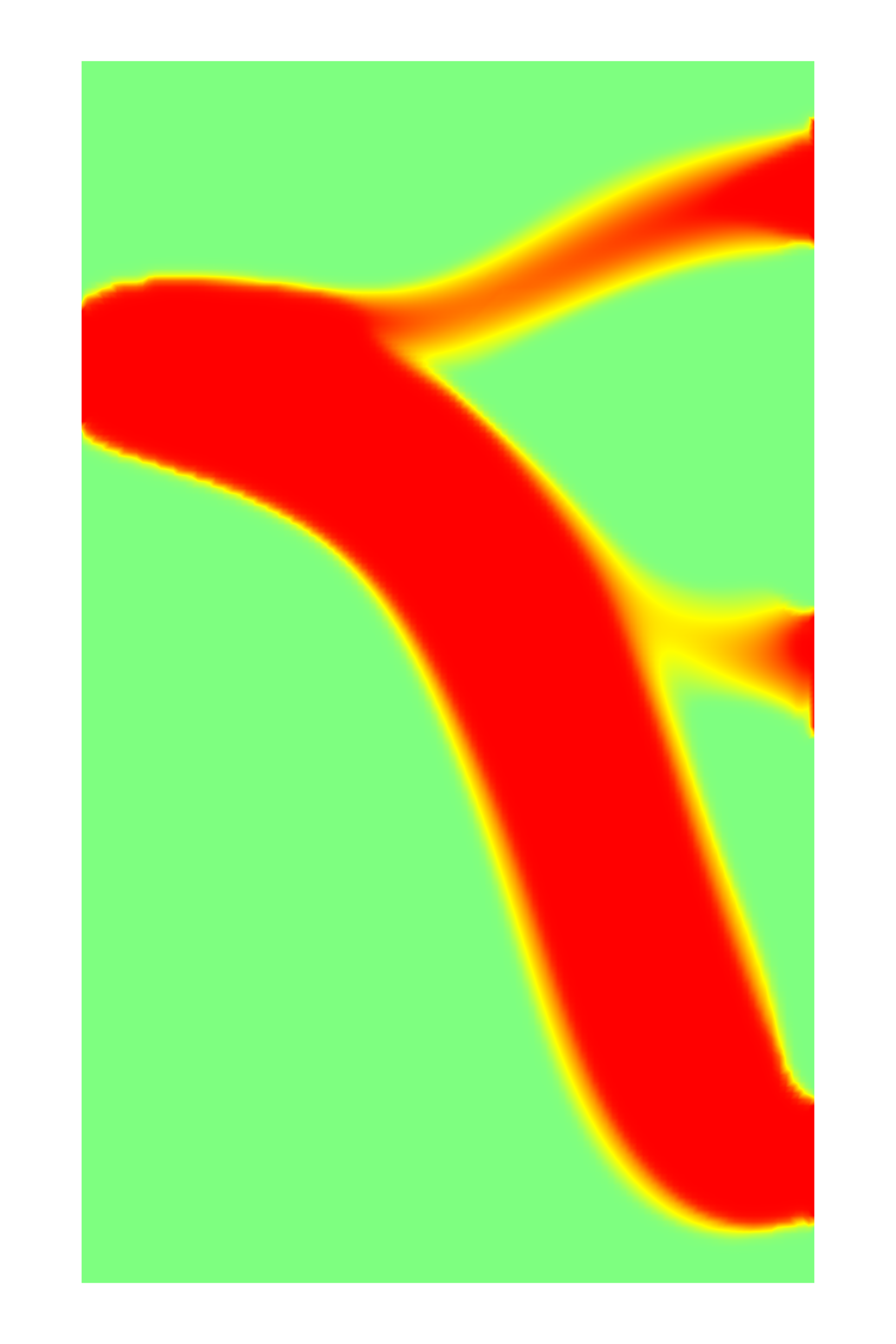}
         \caption{}
         \label{fig:example2_optDes-f}
    \end{subfigure}
    \begin{subfigure}{0.19\textwidth}
         \centering
         \includegraphics[width=\textwidth]{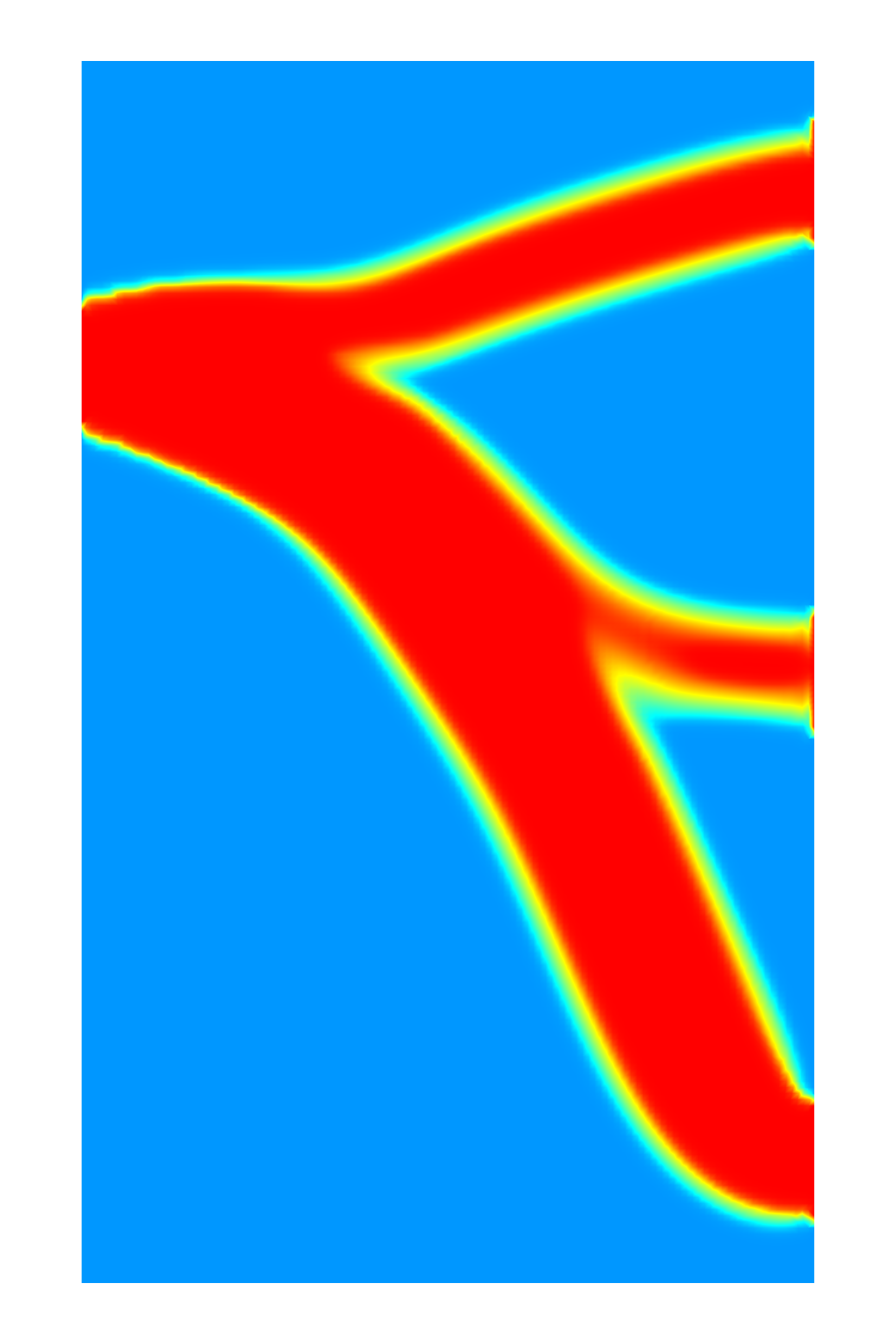}
         \caption{}
         \label{fig:example2_optDes-g}
    \end{subfigure}
    \begin{subfigure}{0.19\textwidth}
         \centering
         \includegraphics[width=\textwidth]{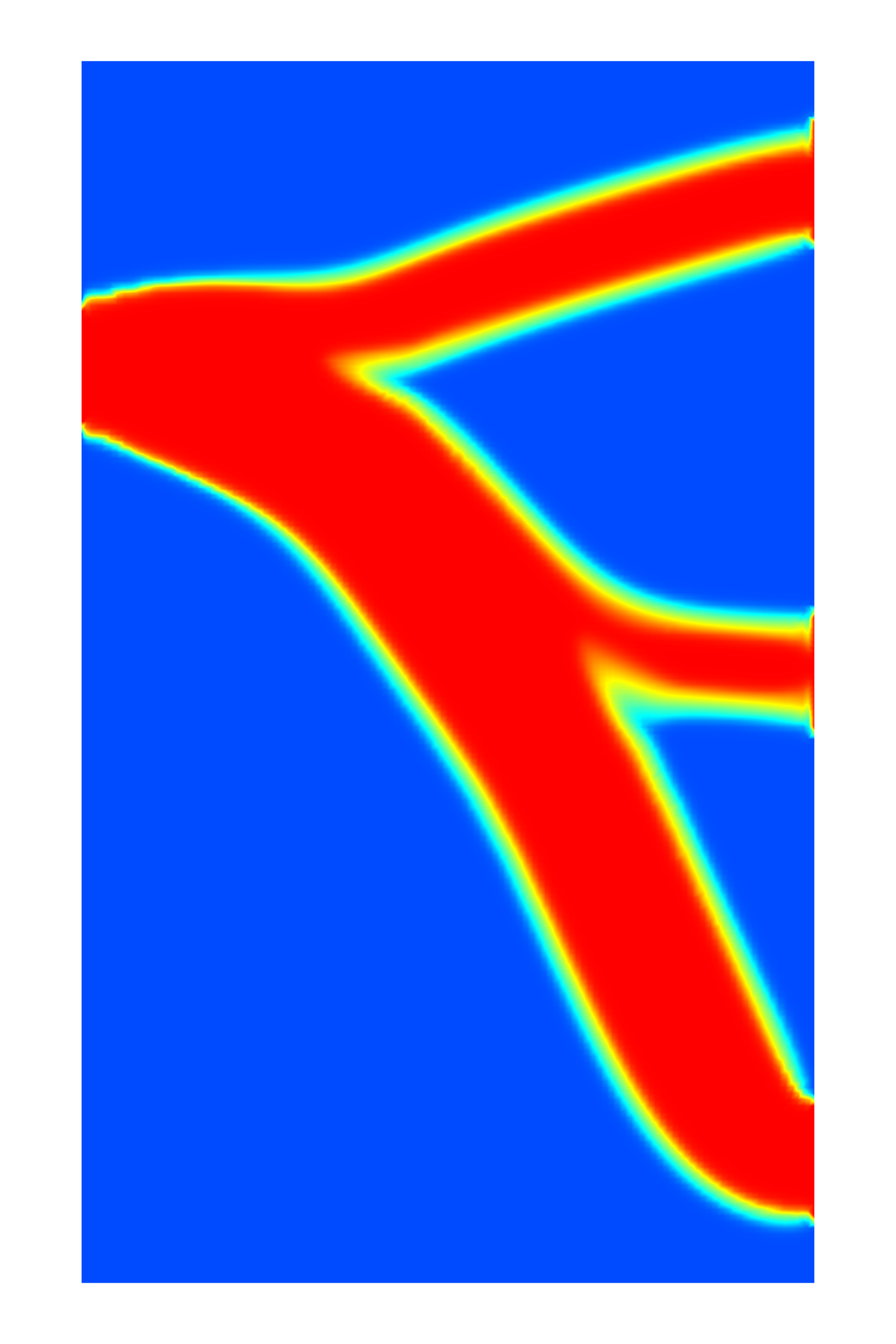}
         \caption{}
         \label{fig:example2_optDes-h}
    \end{subfigure}
    \begin{subfigure}{0.19\textwidth}
         \centering
         \includegraphics[width=\textwidth]{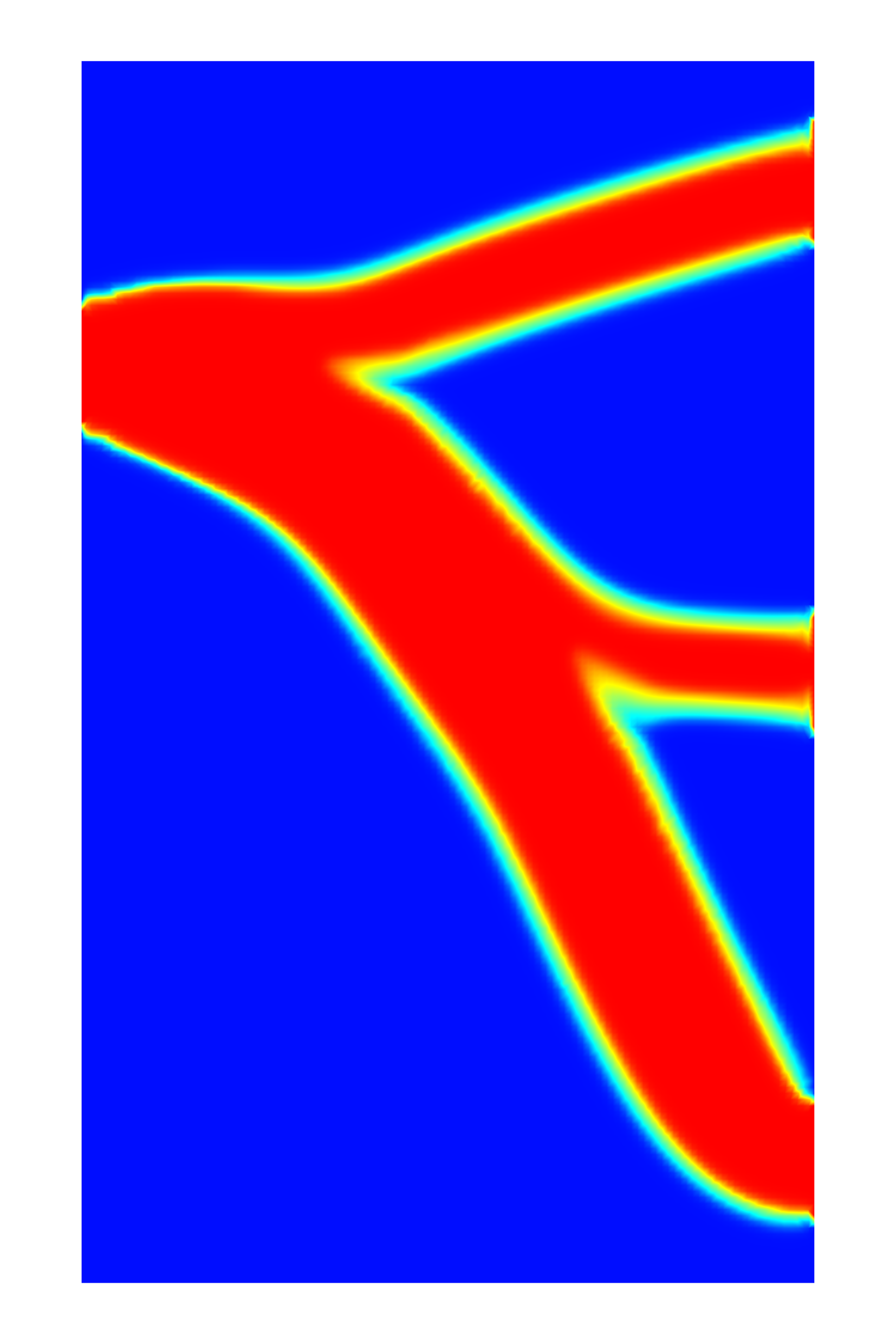}
         \caption{}
         \label{fig:example2_optDes-i}
    \end{subfigure}
    \begin{subfigure}{0.19\textwidth}
         \centering
         \includegraphics[width=\textwidth]{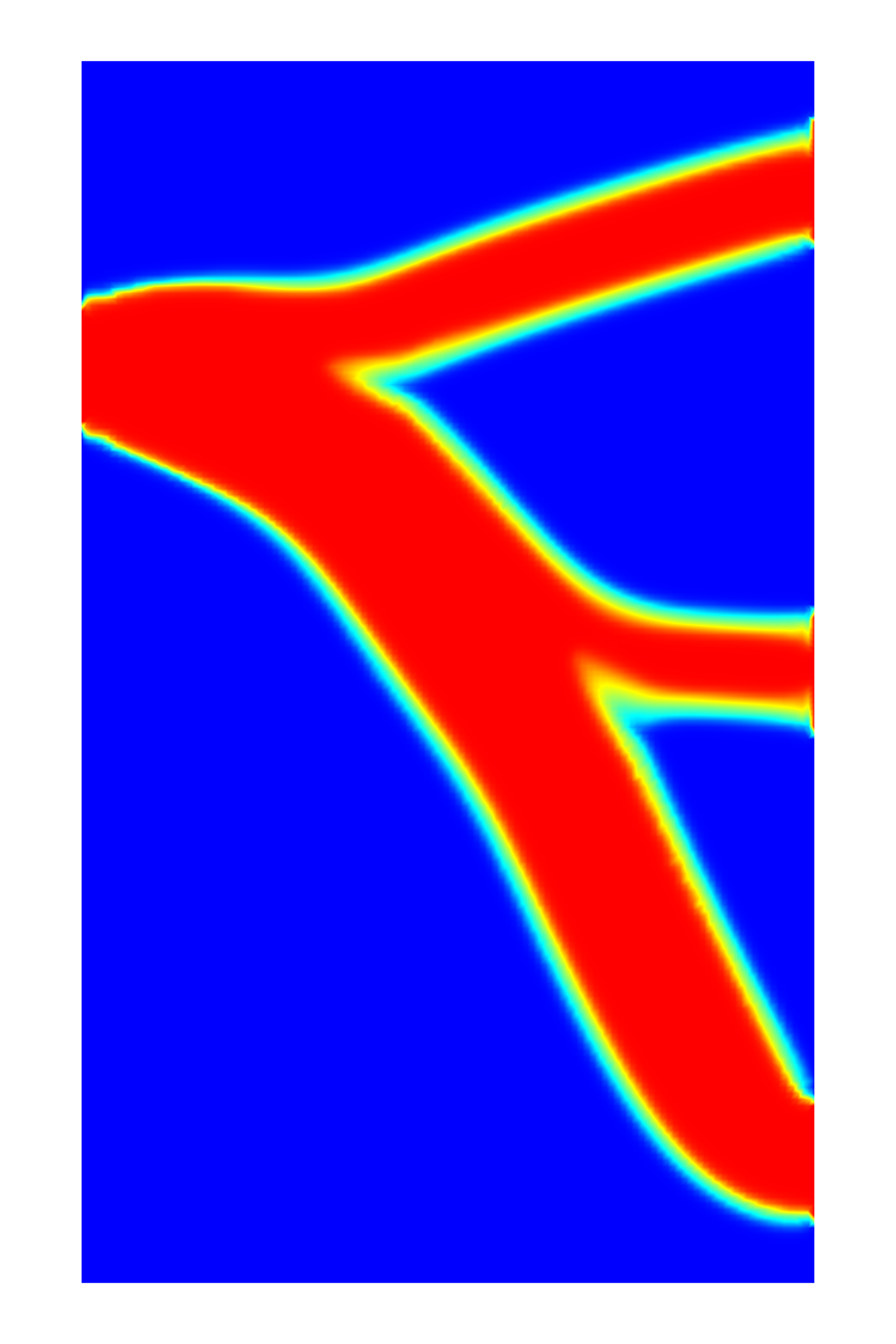}
         \caption{}
         \label{fig:example2_optDes-j}
    \end{subfigure}
    \caption{Optimised height fields for the flow manifold problem using decreasing minimum heights and the two different plane models: (a-e) traditional model; (f-j) proposed model; (a,f) $h_\text{min} = 25\text{ mm}$; (b,g) $h_\text{min} = 10\text{ mm}$; (c,h) $h_\text{min} = 5\text{ mm}$; (d,i) $h_\text{min} = 1\text{ mm}$; (e,j) $h_\text{min} = 0.1\text{ mm}$. All figures use the same colour scale.}
    \label{fig:example2_optDes}
\end{figure*}
Figure \ref{fig:example2_optDes} shows the optimised height fields using the two models for a range of decreasing minimum heights. It appears that when using the proposed model, the topology seems to have already stabilised at around $h_\text{min} = 10\text{ mm}$, whereas the traditional model does not stabilise until around $h_\text{min} = 1\text{ mm}$. Even more evident is the fact that the penalisation of the height-based interpolation, Equation \ref{eq:height}, does not encourage discrete solutions for the traditional model, where larger areas of intermediate design field values and heights remain in the final design. On the contrary, the final designs using the proposed model have significantly less intermediate design field values and heights, with only the transition due to the filter remaining.

It is evident that using a linear height-based interpolation for the topology, there is only a slight penalisation of intermediate design field values and, thus, does not yield fully discrete solutions - even without a filter. However, using a linear interpolation on the penalty term, as was discussed originally by \cite{Borrvall2003}, does strongly penalise intermediate values and yields near discrete solutions - without a filter.

\subsubsection{Verification using body-fitted models}

\begin{figure*}
    \begin{subfigure}{0.42\textwidth}
         \centering
         \includegraphics[width=\textwidth]{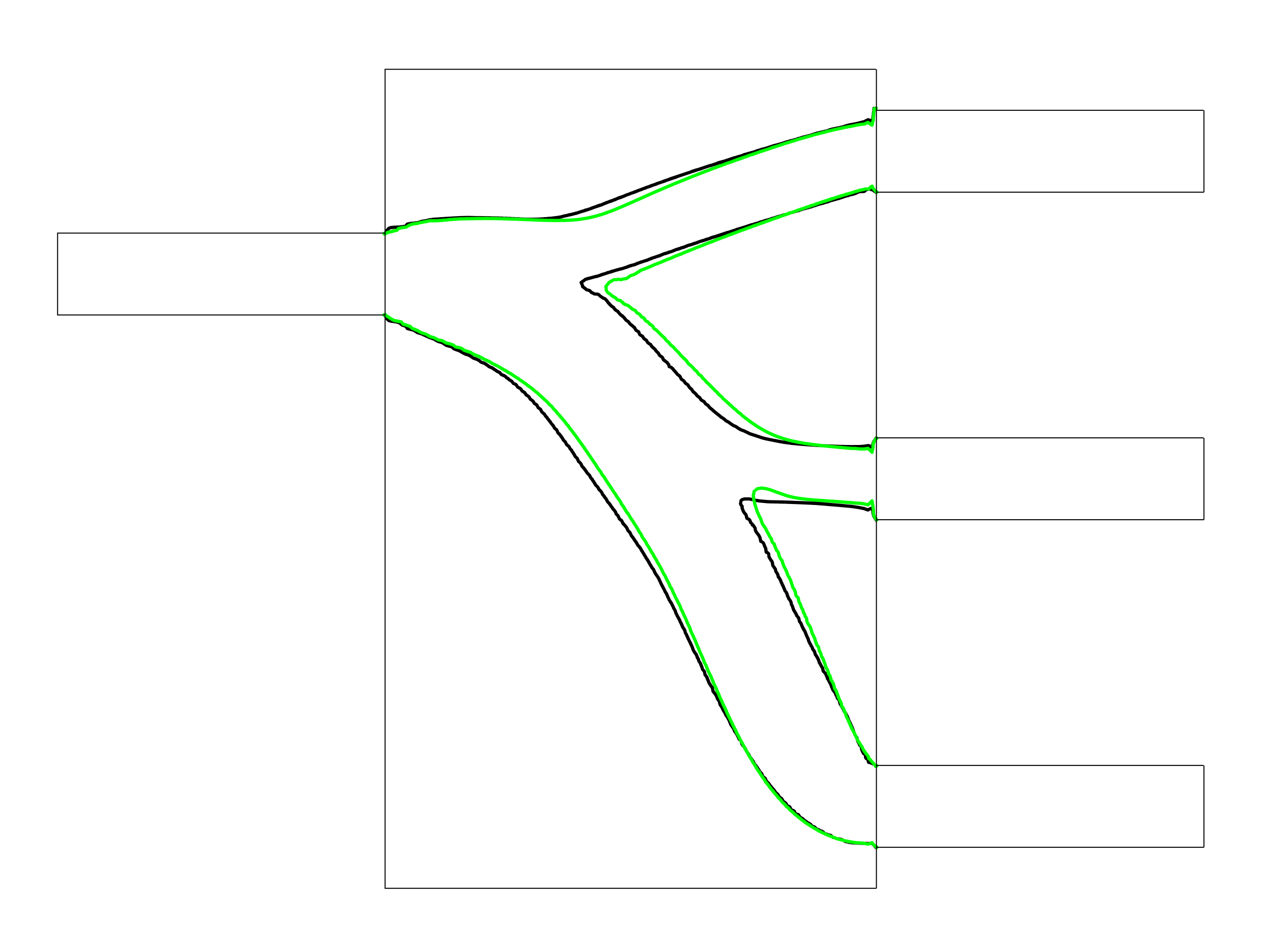}
         \caption{Contours}
         \label{fig:example2_isovol-a}
    \end{subfigure}
    \begin{subfigure}{0.28\textwidth}
         \centering
         \includegraphics[width=\textwidth]{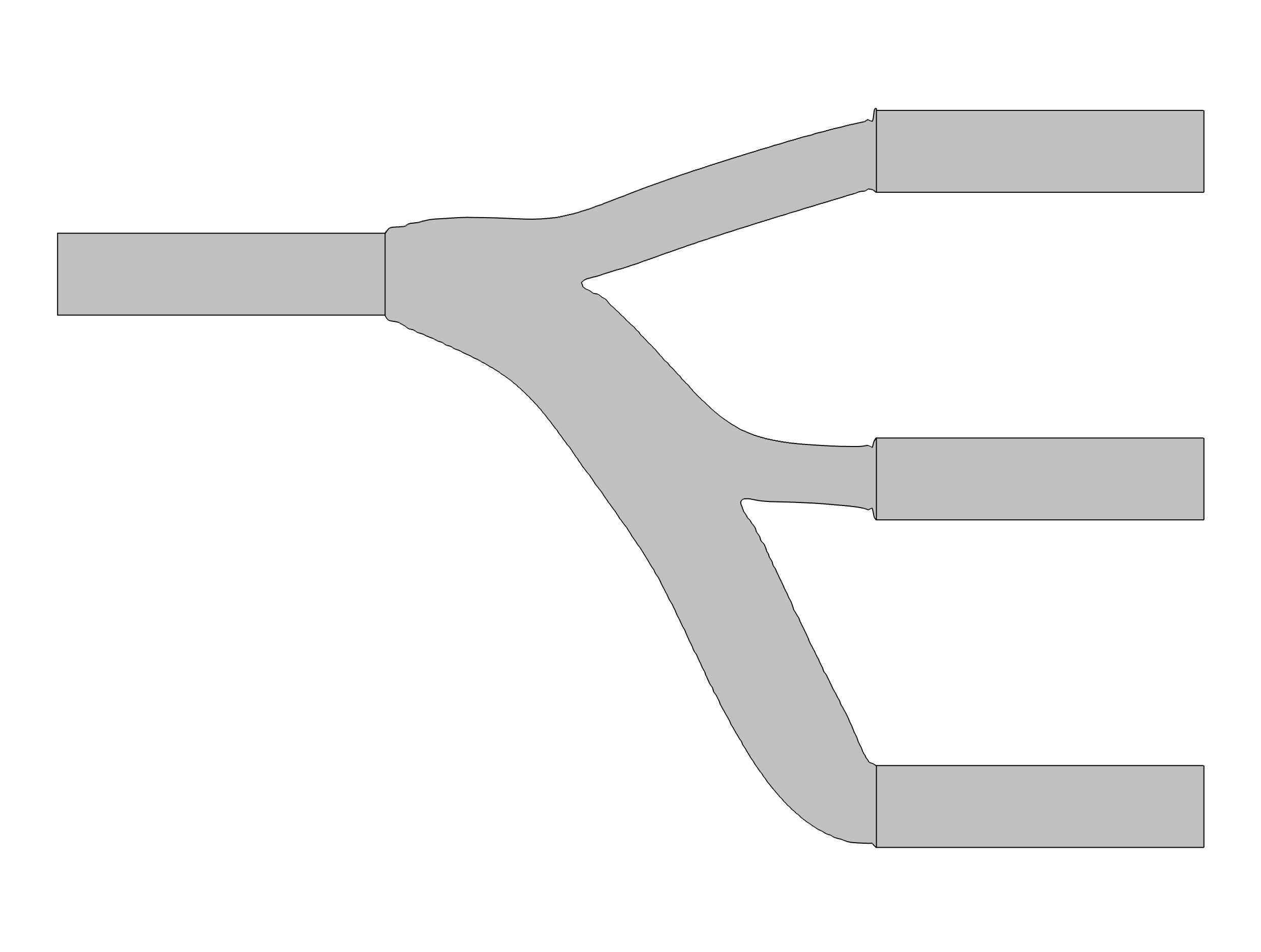}
         \caption{2D}
         \label{fig:example2_isovol-b}
    \end{subfigure}
    \begin{subfigure}{0.3\textwidth}
         \centering
         \includegraphics[width=\textwidth]{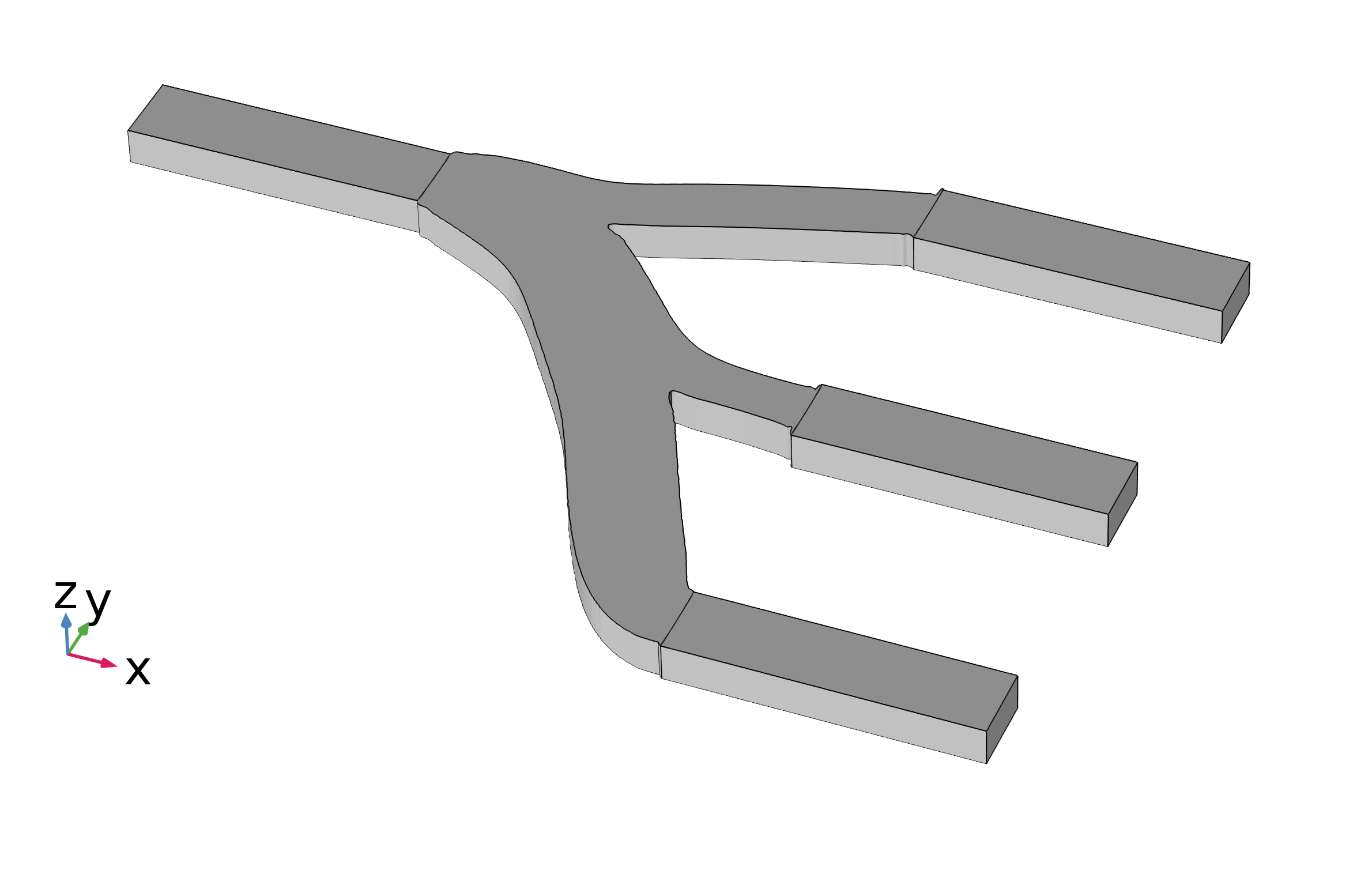}
         \caption{3D}
         \label{fig:example2_isovol-c}
    \end{subfigure}
    \caption{Isocontour geometries for export to body-fitted models: (a) proposed (black) vs. traditional (green); (b) two-dimensional model (proposed); (c) three-dimensional model (proposed).}
    \label{fig:example2_isovol}
\end{figure*}

In order to verify the performance of the optimised topologies produced by the two models, they will be evaluated using body-fitted models in both two and three dimensions. In order to ensure a fair comparison with respect to the pressure drop, the designs produced using the two models are exported at separate isovalues to ensure the exact same final used fluid area, $A_{f} = f_{a}A_\text{des}$. Thus, the traditional model design is exported at $\tilde{\gamma} = 0.569$ and the proposed model design is exported at $\tilde{\gamma} = 0.505$, for which the designs are shown in Figure \ref{fig:example2_isovol}.

The two-dimensional model includes the out-of-plane resistance term as previously discussed and the three-dimensional model only includes the upper half of the geometry.
\begin{table}
    \centering
    \begin{tabular}{c||c|c|c}
        Design & Proposed & 2D & 3D \\
        \hline
        Traditional & 1.017  & 1.129  & 1.175 \\
        Proposed    & 0.994  & 1.100  & 1.158 \\
        \hline
        Improvement & -2.2\% & -4.1\% & -3.9\%
    \end{tabular}
    \caption{Pressure drops (in mPa) for the two optimised flow manifold designs evaluated using the proposed planar continuous model, the two-dimensional body-fitted model, and the three-dimensional body-fitted model.}
    \label{tab:example2_bodyfit}
\end{table}
Table \ref{tab:example2_bodyfit} lists the pressure drop for the two designs evaluated using the proposed planar continuous model, the two-dimensional body-fitted model, and the three-dimensional body-fitted model. It can be seen that no matter the model they are evaluated using, the proposed \textit{topographical} plane model actually delivers a better performing \textit{topological} design.
This may not be a general conclusion, but it definitely deserves further investigation whether the \textit{topographical} design parametrisation generally yields better performing topologies. This will be explored in future work, also for three-dimensional problems although the parametrisation loses physical meaning.

\section{Concluding remarks} \label{sec:conclusion}

The origins of topology optimisation for fluid flow problems has been revisited in this paper. It has been shown that if the channel height between two parallel plates is varied, the traditional approach, using only a Poiseuille-based friction term, is no longer valid. The model also does not hold if the minimum channel height is relatively large or meaning is attributed to intermediate design field values.
In order to remedy this, an augmentation of the mass conservation equation has been introduced. This augmentation ensures that the change in control volume size is taken into account in mass conservation, ensuring accurate description of fully-developed flow between two plates of varying channel height.

The proposed model is applied to the design of a flow distribution problem. The height of the fluid channel between two parallel plates is optimised and, thus, the surface \textit{topography} of the channel. The model provides a significant reduction in the number of degrees-of-freedom, while ensuring reasonable accuracy for low-to-moderate Reynolds numbers in the laminar regime. Common to all planar approximations, the model is not able to capture three-dimensional flow effects. Further, due to the assumption of fully-developed flow everywhere, separation of the flow near sudden expansions are not captured. This is the largest limitation of the model, but an extension improving this behaviour is currently being developed.

Through in-depth parametric studies, it has been shown that accuracy of the proposed model is generally better than or equal to the traditional model, even for decreasing minimum channel height moving towards the limit of \textit{topology} optimisation. Furthermore, when applied to the \textit{topology} optimisation of a flow manifold, the proposed \textit{topographical} model outperforms the traditional \textit{topological} model.
Therefore, the proposed model bridges the gap between \textit{topology} and \textit{topography} optimisation for planar fluid flow problems.

It is interesting to note that the linear interpolation of the channel height used herein only provides a slight penalisation of intermediate design field values. This is in stark contrast to the linear interpolation of the flow resistance as originally discussed by \cite{Borrvall2003}, which strongly penalises intermediate design field values and yields near discrete solutions. This contrast will be explored in future work, where the \textit{topographical} design parametrisation will also be applied to three-dimensional problems, although the parametrisation loses physical meaning.

Lastly, in order to solve the original problem serving as motivation for this work, namely the optimisation of plate heat exchangers, the flow model is currently being coupled to a thermal model similar to those of the pseudo-3D models of the literature.

\begin{acknowledgements}
This work was partly sponsored through the ``NeGeV: Next Generation Ventilation'' project funded by the Danish Energy Agency under the Energy Technology Development and Demonstration Program (EUDP project number 64017-05117).
\end{acknowledgements}

\section*{Conflict of interest}
The author has no conflict of interest.

\section*{Reproduction of results}
The COMSOL file to reproduce Example 1 from Section \ref{sec:results_distribution} are provided as supplementary material.

\bibliographystyle{spbasic}      
\bibliography{main_smo}   

\clearpage
\appendix

\section{Derivations for momentum conservation equations} \label{app:bp_derivation}

Based on the separation of variables defined by Equations \ref{eq:u_sepVar} and \ref{eq:u_zDep}, the derivative of the full dimensional velocity gradient is defined as:
\begin{equation} \label{eq:dudx_sepVar}
    \dpart{u_i}{x_j} = \dpart{\bar{u}_i}{x_j}\eta + \bar{u}_i \dpart{\eta}{x_j} = 
    \begin{cases}
        \dpart{\bar{u}_i}{x_j}\eta & \text{ for } i=1,2 \text{ and } j=1,2 \\
        \bar{u}_i\dpart{\eta}{x_j} & \text{ for } i=1,2 \text{ and } j=3 \\
        0 & \text{ for } i=3 \text{ and } j=1,2,3
    \end{cases}
\end{equation}

The Galerkin weak form of the governing equation for conservation of moment, Equation \ref{eq:navierstokes-a}, is:
\begin{multline}
    \mathcal{R} = \rho\intO{\wi\uj\dpart{\ui}{\xj}} - \mu\intO{\dpart{\wi}{\xj}\bof{\dpart{\ui}{\xj} + \dpart{\uj}{\xi}}} \\+ \intO{p \dpart{\wi}{\xi}} = 0
\end{multline}
with $i,j = 1,2,3$.

By introducing the following assumptions for both the velocity and test function fields: there is no flow in the $x_3$-direction, that is $u_{3} = 0$; separation of variables, Equations \ref{eq:u_sepVar} and \ref{eq:u_zDep}; simplified velocity gradient, Equation \ref{eq:dudx_sepVar}, the weak form can be rewritten to:
\begin{multline}
    \mathcal{R} = \rho \intO{\eta^{3} \wba\ubb\dpart{\uba}{\xb}} - \mu \intO{\eta^{2} \dpart{\wba}{\xb}\bof{\dpart{\uba}{\xb} + \dpart{\ubb}{\xa}}} \\
    - \mu \intO{\bof{\dpart{\eta}{x_{3}}}^{2} \wba\uba} + \intO{\eta p \dpart{\wba}{\xa}}
\end{multline}
with $a,b = 1,2$.

By integrating analytically in the $x_3$-direction, that is applying Equation \ref{eq:int_sepVar}, finally yields the following after a slight rearrangement:
\begin{multline}
    \mathcal{R} = \frac{6}{7}\rho \into{\wba\ubb\dpart{\uba}{\xb}} - \mu \into{\dpart{\wba}{\xb}\bof{\dpart{\uba}{\xb} + \dpart{\ubb}{\xa}}} \\
    - \frac{10\mu}{h^2} \into{\wba\uba} + \frac{5}{4} \into{ p \dpart{\wba}{\xa}}
\end{multline}
with $a,b = 1,2$. This equation can finally be reduced to the equivalent strong form presented in Equation \ref{eq:aug_navierstokes}.

\section{Detailed results of 2D-to-1D parameter study} \label{app:parameterStudies}

\begin{figure}
    \begin{subfigure}{\columnwidth}
         \centering
         \includegraphics[width=\textwidth]{1Dvs2D_smoothContract2.png}
         \caption{$\beta = 1$}
         \label{fig:channel2D_midline_app-a}
    \end{subfigure}
    \\
    \begin{subfigure}{\columnwidth}
         \centering
         \includegraphics[width=\textwidth]{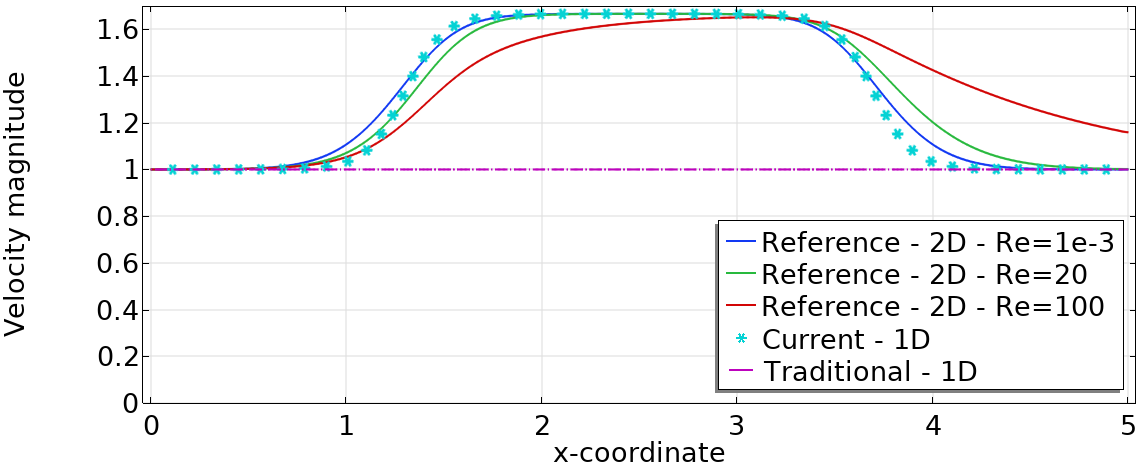}
         \caption{$\beta = 8$}
         \label{fig:channel2D_midline_app-c}
    \end{subfigure}
    \\
    \begin{subfigure}{\columnwidth}
         \centering
         \includegraphics[width=\textwidth]{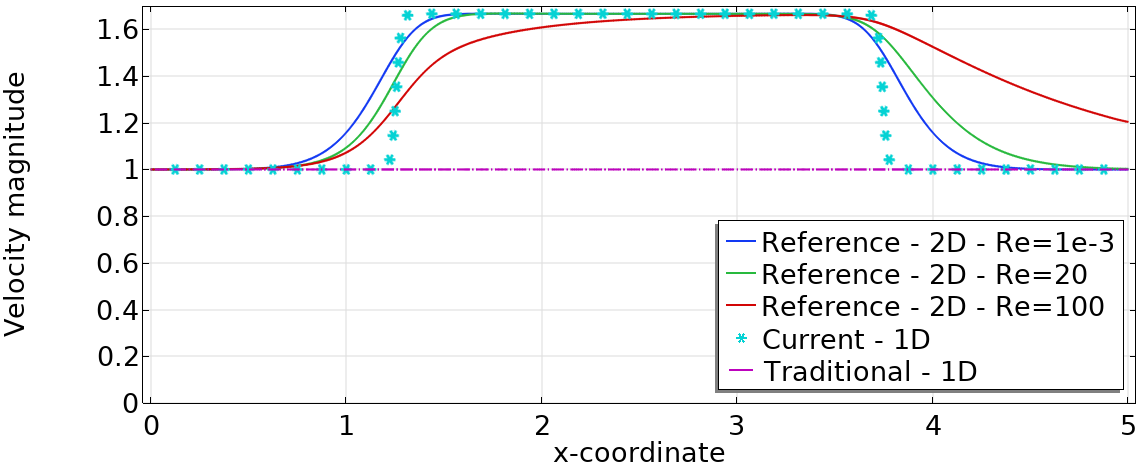}
         \caption{$\beta = 64$}
         \label{fig:channel2D_midline_app-e}
    \end{subfigure}
    \caption{Comparison of velocity magnitude along the mid-line for the different models applied to contraction channel geometries with $h_\text{mid}=0.6$ for different transition sharpness.}
    \label{fig:channel2D_midline1_app}
\end{figure}
\begin{figure}
    \begin{subfigure}{\columnwidth}
         \centering
         \includegraphics[width=\textwidth]{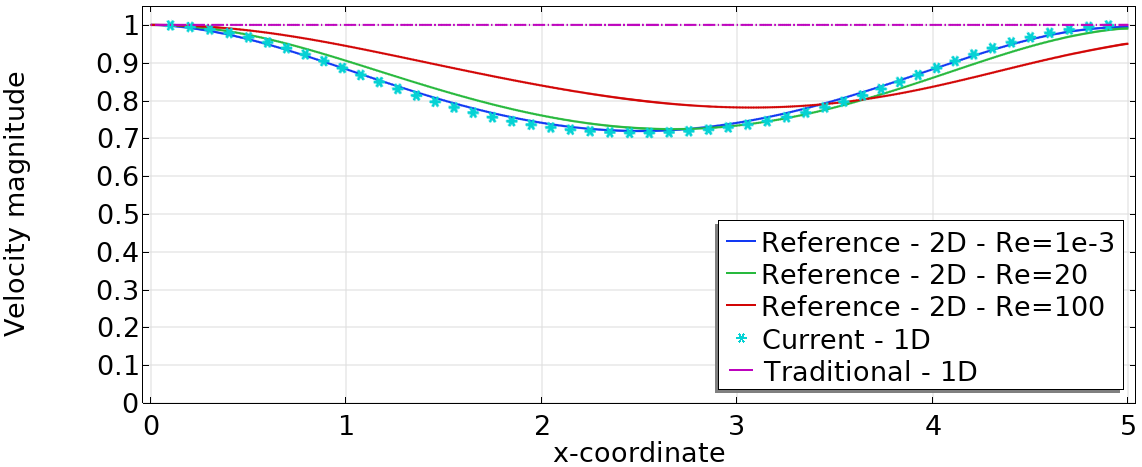}
         \caption{$\beta = 1$}
         \label{fig:channel2D_midline_app-b}
    \end{subfigure}
    \\
    \begin{subfigure}{\columnwidth}
         \centering
         \includegraphics[width=\textwidth]{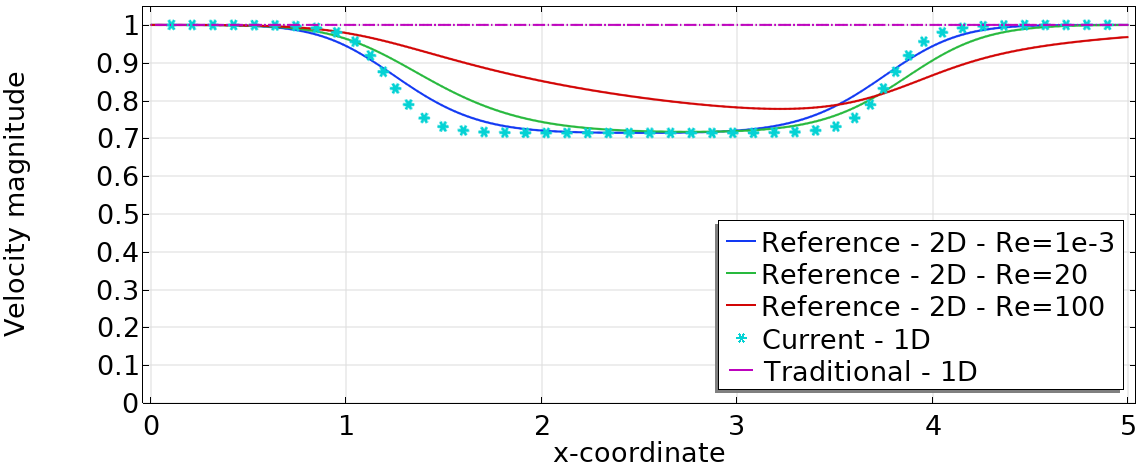}
         \caption{$\beta = 8$}
         \label{fig:channel2D_midline_app-d}
    \end{subfigure}
    \\
    \begin{subfigure}{\columnwidth}
         \centering
         \includegraphics[width=\textwidth]{1Dvs2D_sharpExpand2.png}
         \caption{$\beta = 64$}
         \label{fig:channel2D_midline_app-f}
    \end{subfigure}
    \caption{Comparison of velocity magnitude along the mid-line for the different models applied to expansion channel geometries with $h_\text{mid}=1.4$ for different transition sharpness.}
    \label{fig:channel2D_midline2_app}
\end{figure}
Figures \ref{fig:channel2D_midline1_app} and \ref{fig:channel2D_midline2_app} show the velocity magnitude along the mid-line, comparing the traditional and proposed one-dimensional models with the full two-dimensional model for contraction and expansion channel geometries, respectively.
Firstly, it is observed that the traditional one-dimensional model is completely wrong. It simply does not capture the change in channel height has on the velocity field, since the conservation of mass is not adapted to accommodate this. So although a larger friction term exists when the channel height is reduced, it has not effect on the velocity field what so ever in this one-dimensional case.
Secondly, for both one-dimensional models, the solution does not vary with a change in the Reynolds number. This is due to a combination of the restrictions that a one-dimensional problem presents, as well as a fully-developed laminar (through-thickness) profile being the same independent of Reynolds number. 
Thirdly, the agreement between the models is reasonably good for many parameter values. From Figures \ref{fig:channel2D_midline_app-a} and  \ref{fig:channel2D_midline_app-b}, it is clear that for a slow geometric transition, $\beta=1$, and low Reynolds numbers, the proposed one-dimensional model captures the flow very well as expected. When the Reynolds number increases to $Re=100$, some deviations are observed in the post-contraction and -expansion areas. When increasing the sharpness of the geometric transition slightly, $\beta=8$, Figures \ref{fig:channel2D_midline_app-c} and  \ref{fig:channel2D_midline_app-d} show that there is still reasonable agreement for lower Reynolds numbers. However, when the geometric transition is abrupt, $\beta=64$, it can be seen from Figures \ref{fig:channel2D_midline_app-e} and  \ref{fig:channel2D_midline_app-f} that the agreement deteriorates even for low Reynolds number flows. This is because the proposed one-dimensional model exhibits instantaneous expansion and contraction, due to the assumption of a fully-developed flow profile at all points along the channel. This assumption allows us to dimensionally reduce the problem, but like most assumptions it also introduces errors and limitations.

\begin{figure}
    \begin{subfigure}{\columnwidth}
         \centering
         \includegraphics[width=\textwidth]{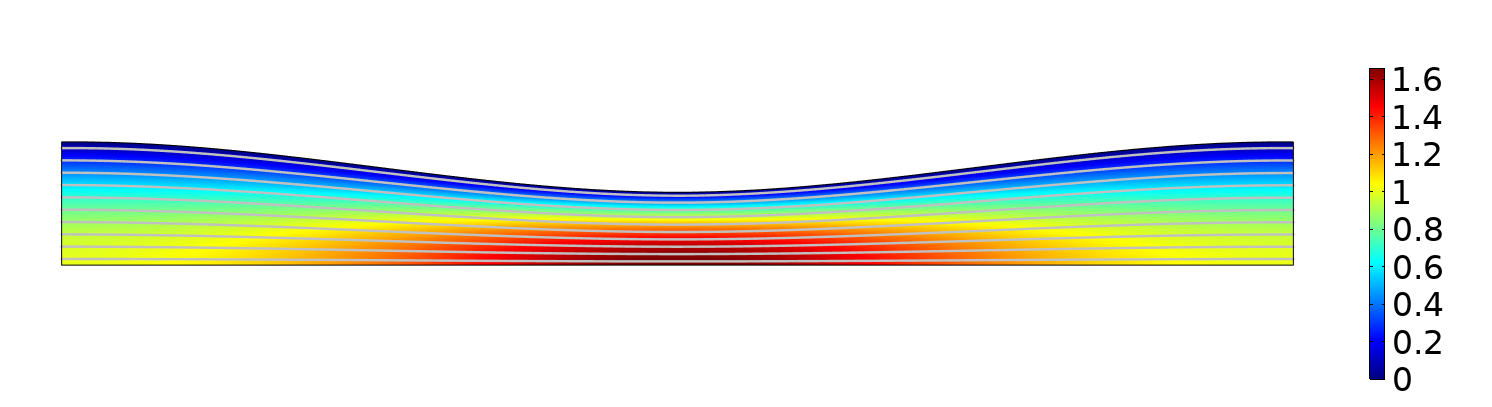}
         \caption{$Re = 10^{-3}, \beta = 1$}
         \label{fig:channel2D_contraction-a}
    \end{subfigure}
    \\
    \begin{subfigure}{\columnwidth}
         \centering
         \includegraphics[width=\textwidth]{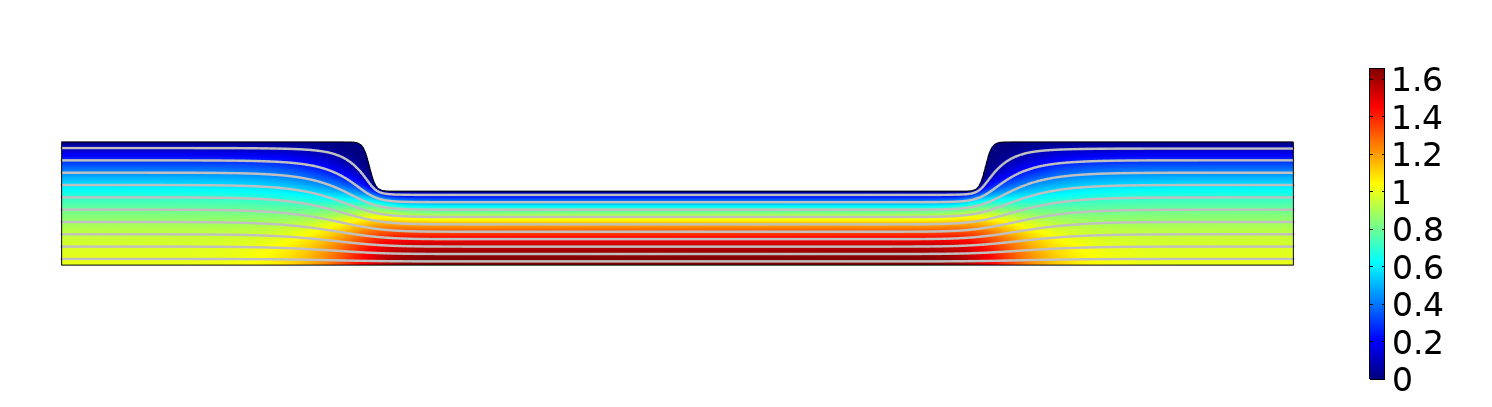}
         \caption{$Re = 10^{-3}, \beta = 64$}
         \label{fig:channel2D_contraction-b}
    \end{subfigure}
    \caption{Velocity magnitude field and streamlines for two-dimensional contraction channel geometries for different transition sharpness with $h_\text{mid}=0.6$ and $Re = 10^{-3}$.}
    \label{fig:channel2D_contraction1}
\end{figure}

\begin{figure}
    \begin{subfigure}{\columnwidth}
         \centering
         \includegraphics[width=\textwidth]{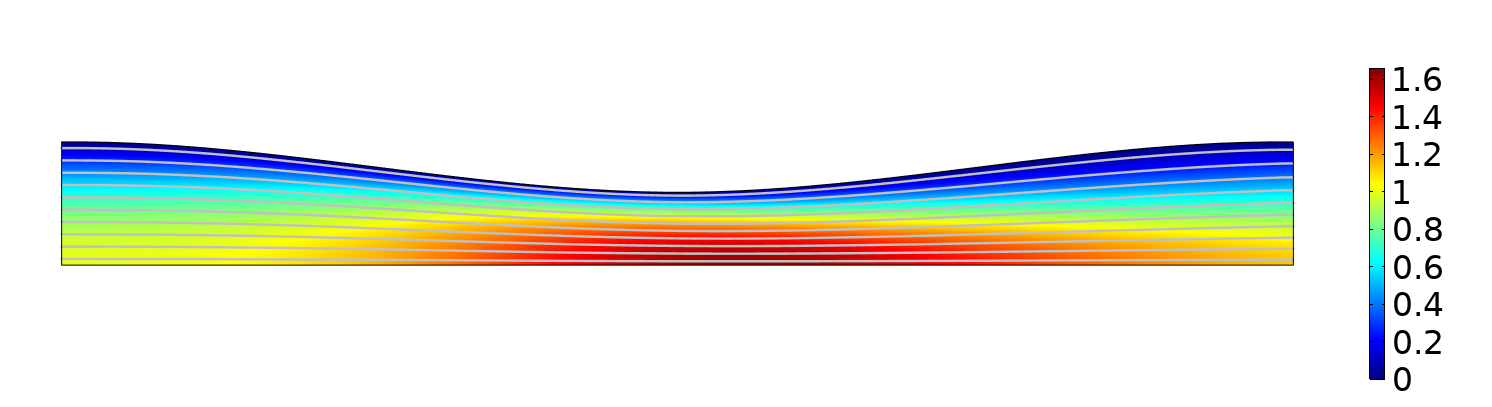}
         \caption{$\beta = 1$}
         \label{fig:channel2D_contraction-e}
    \end{subfigure}
    \\
    \begin{subfigure}{\columnwidth}
         \centering
         \includegraphics[width=\textwidth]{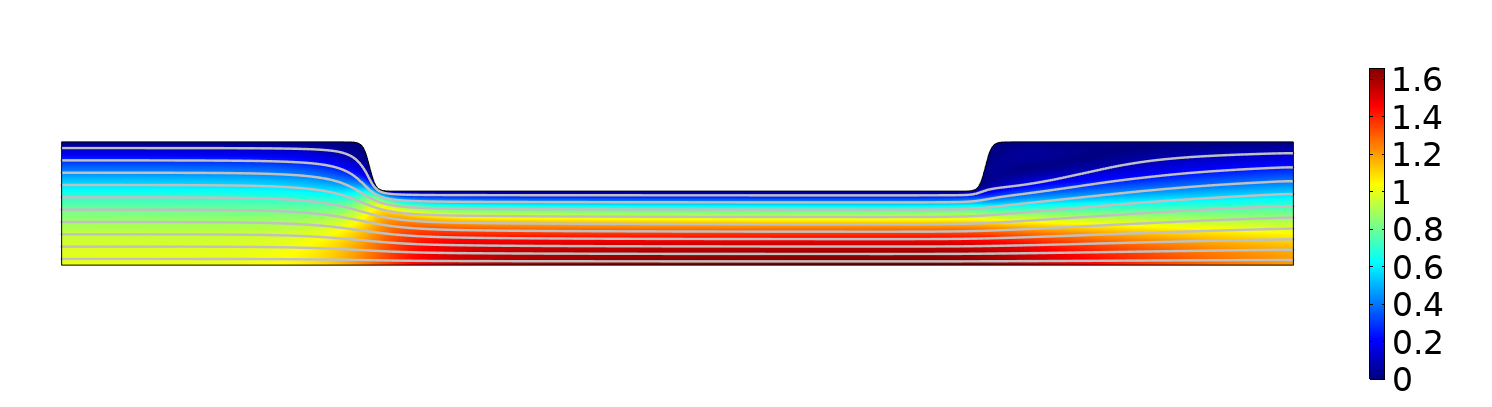}
         \caption{$\beta = 64$}
         \label{fig:channel2D_contraction-f}
     \end{subfigure}
    \caption{Velocity magnitude field and streamlines for two-dimensional contraction channel geometries for different transition sharpness with $h_\text{mid}=0.6$ and $Re = 100$.}
    \label{fig:channel2D_contraction2}
\end{figure}

\begin{figure}
    \begin{subfigure}{\columnwidth}
         \centering
         \includegraphics[width=\textwidth]{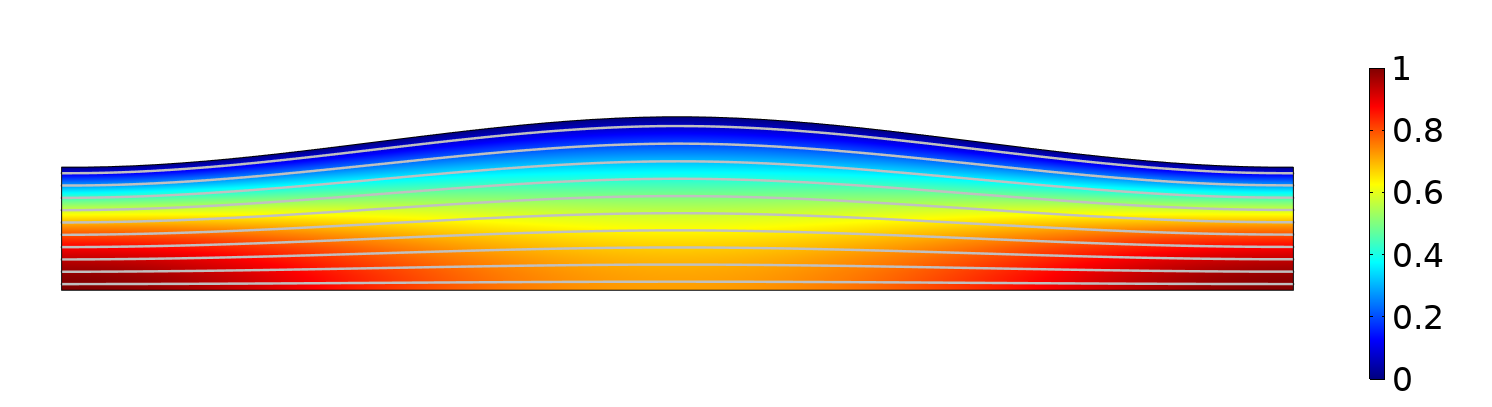}
         \caption{$Re = 10^{-3}, \beta = 1$}
         \label{fig:channel2D_expansion-a}
    \end{subfigure}
    \\
    \begin{subfigure}{\columnwidth}
         \centering
         \includegraphics[width=\textwidth]{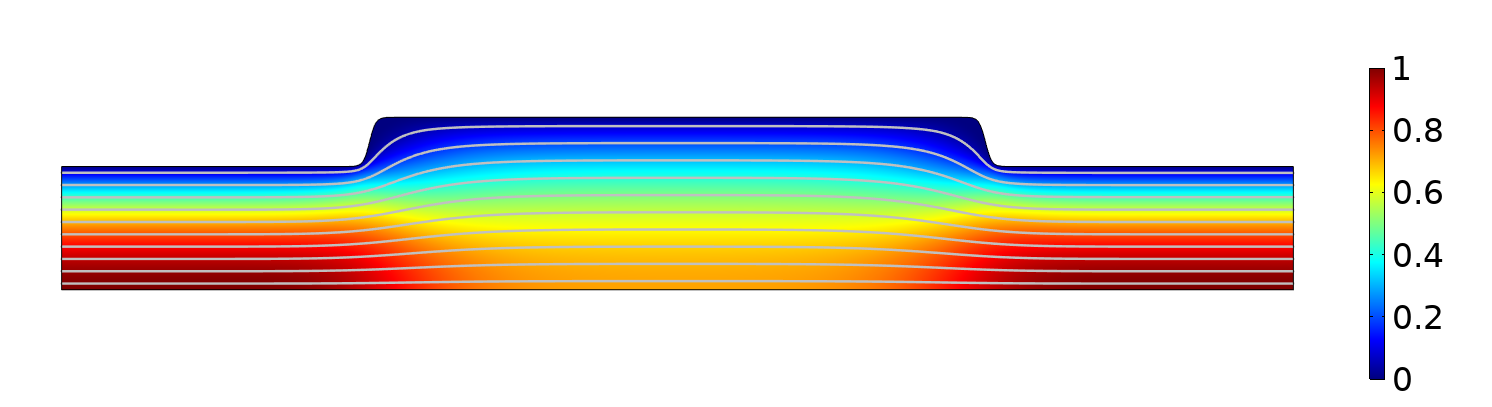}
         \caption{$Re = 10^{-3}, \beta = 64$}
         \label{fig:channel2D_expansion-b}
    \end{subfigure}
    \caption{Velocity magnitude field and streamlines for two-dimensional expansion channel geometries for different transition sharpness with $h_\text{mid}=1.4$ and $Re = 10^{-3}$.}
    \label{fig:channel2D_expansion1}
\end{figure}

\begin{figure}
    \begin{subfigure}{\columnwidth}
         \centering
         \includegraphics[width=\textwidth]{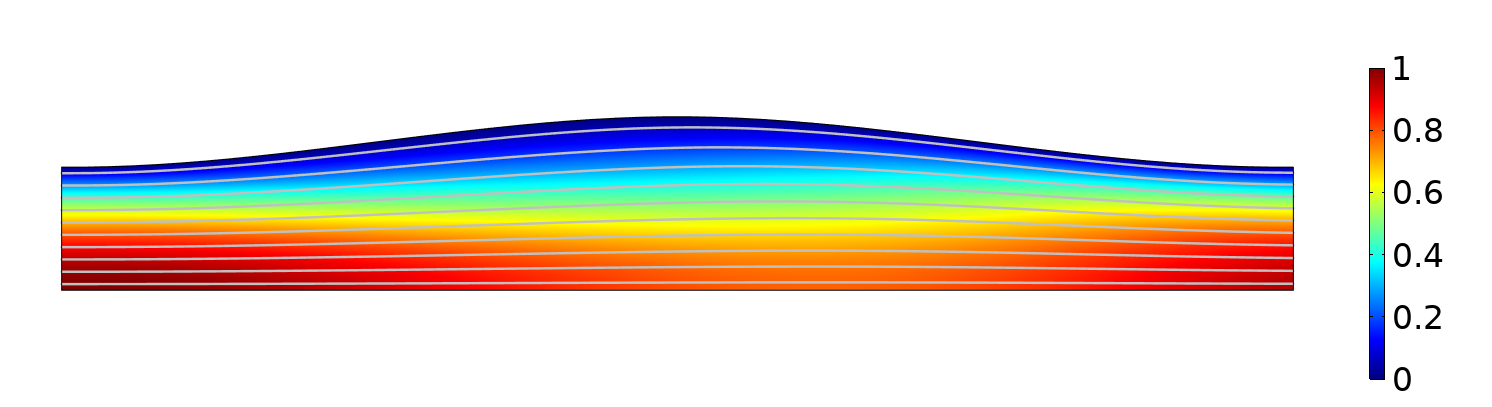}
         \caption{$Re = 100, \beta = 1$}
         \label{fig:channel2D_expansion-e}
    \end{subfigure}
    \\
    \begin{subfigure}{\columnwidth}
         \centering
         \includegraphics[width=\textwidth]{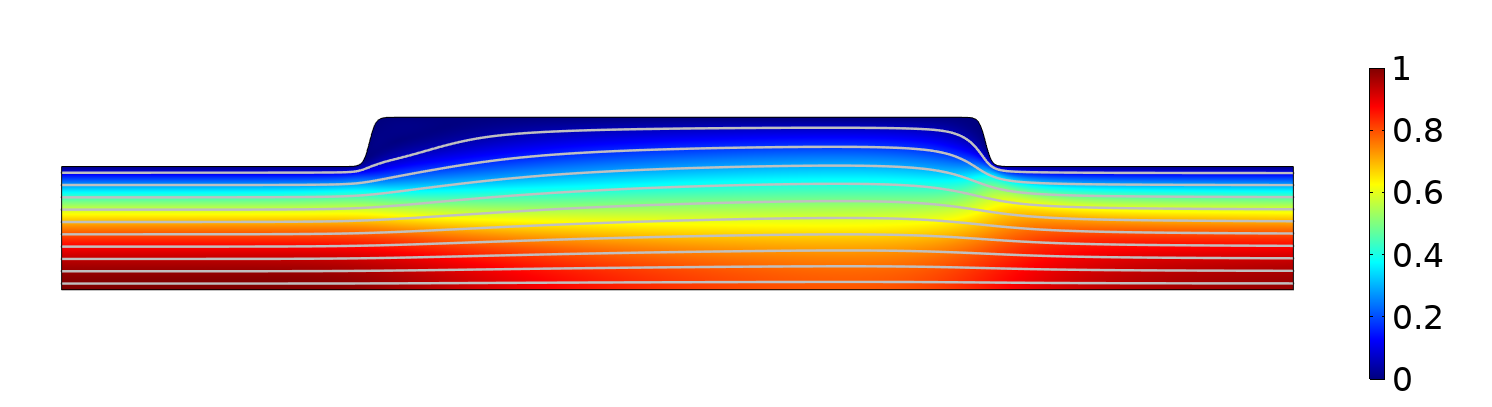}
         \caption{$Re = 100, \beta = 64$}
         \label{fig:channel2D_expansion-f}
     \end{subfigure}
    \caption{Velocity magnitude field and streamlines for two-dimensional expansion channel geometries for different transition sharpness with $h_\text{mid}=1.4$ and $Re = 100$.}
    \label{fig:channel2D_expansion2}
\end{figure}

Figures \ref{fig:channel2D_contraction1}-\ref{fig:channel2D_expansion2} show the velocity fields and streamlines for contraction (Figures \ref{fig:channel2D_contraction1} and \ref{fig:channel2D_contraction2}) and expansion (Figures \ref{fig:channel2D_expansion1} and \ref{fig:channel2D_expansion2}) channels, respectively, with different Reynolds number and transition sharpness.
For a slowly changing channel height (Figures \ref{fig:channel2D_contraction-a},
\ref{fig:channel2D_contraction-e} and Figures \ref{fig:channel2D_expansion-a},
\ref{fig:channel2D_expansion-e}), it can be seen that an increase in Reynolds number, or increase in inertia, does not have a large impact on the flow field. For $Re=100$, a slight degree of separation (or rather the transition thereto) is observed during the parts of the channel where the height is increasing.
For a sharp transition in the channel height (Figures \ref{fig:channel2D_contraction-b},
\ref{fig:channel2D_contraction-f} and Figures \ref{fig:channel2D_expansion-b},
\ref{fig:channel2D_expansion-f}), it can be seen that an increase in Reynolds number, has a significantly larger effect. For $Re=100$, actual flow separation is experienced after the expansions, where small recirculation zones exist.

\begin{figure}
    \begin{subfigure}{\columnwidth}
         \centering
         \includegraphics[width=\textwidth]{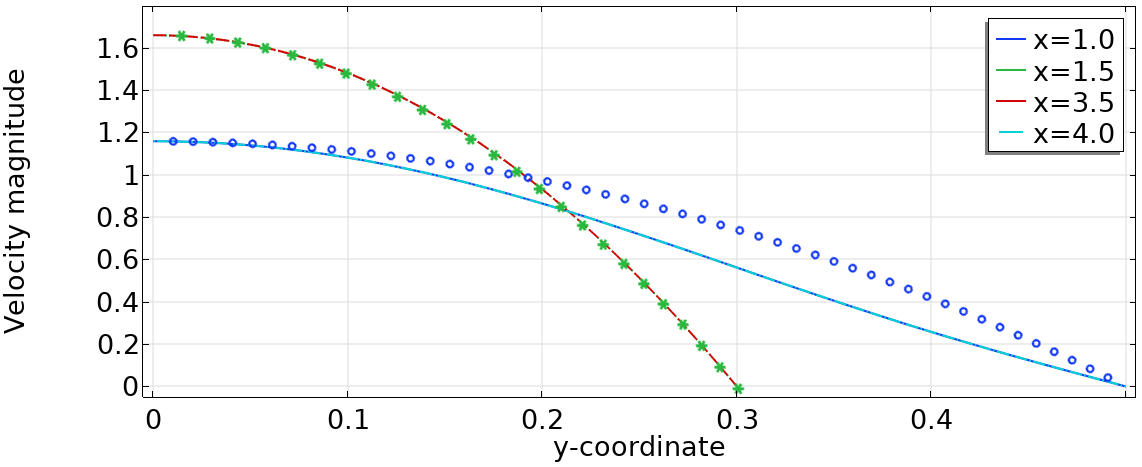}
         \caption{$Re=10^{-3}$}
         \label{fig:channel2D_cross-a}
    \end{subfigure}
    \\
    \begin{subfigure}{\columnwidth}
         \centering
         \includegraphics[width=\textwidth]{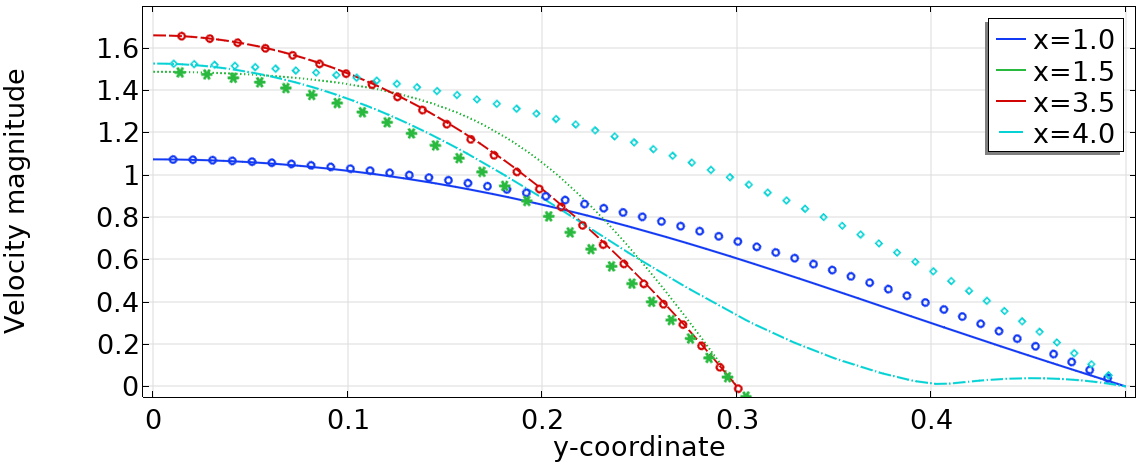}
         \caption{$Re=100$}
         \label{fig:channel2D_cross-b}
     \end{subfigure}
    \caption{Through-thickness velocity profiles at various axial locations for the full two-dimensional abrupt contraction channels ($\beta = 64$) shown in Figures \ref{fig:channel2D_contraction-b} and \ref{fig:channel2D_contraction-f}. For comparison, symbols show quadratic profile with the same maximum velocity.}
    \label{fig:channel2D_cross}
\end{figure} 
Figure \ref{fig:channel2D_cross} shows the through-thickness velocity profiles at various axial locations for the full two-dimensional abrupt contraction channels ($\beta = 64$) shown in Figures \ref{fig:channel2D_contraction-b} and \ref{fig:channel2D_contraction-f}. The positions $x=1.0$ and $x=1.5$ are located just before and after the initial contraction, respectively, whereas $x=3.5$ and $x=4.0$ are located just before and after the subsequent expansion, respectively. For the low Reynolds number, $Re = 10^{-3}$, it can be seen from Figure \ref{fig:channel2D_cross-a} that there are smaller disagreements between the actual velocity profile and an equivalent quadratic one. However, for the high Reynolds number, $Re = 100$, it can be seen from Figure \ref{fig:channel2D_cross-b} that there are significant deviations from the assumed quadratic profiles. This is exactly what leads to the discrepancies between the full two-dimensional model and the proposed one-dimensional model.

\end{document}